\documentclass[prl,a4paper,11pt]{article}
\pdfoutput=1

\usepackage{bm}        
\usepackage{amsmath}						
\usepackage{amssymb}						
\usepackage{amsthm}						
\usepackage{array}							
\usepackage[english]{babel}					
\usepackage{cancel}						
\usepackage{eufrak}							
\usepackage{fullpage}						
\usepackage{graphicx}						
\usepackage{multirow}						
\usepackage{nccmath}						
\usepackage{nicefrac}						
\usepackage[section]{placeins}                         
\usepackage{booktabs}						
\usepackage[format=hang]{caption}				
\usepackage[hang,flushmargin]{footmisc}		
\usepackage{hyperref}						
\usepackage{mathtools}						
\usepackage[usenames]{color}

\numberwithin{equation}{section}				
\def\scalar{{\mathbf S}}

\newcommand{\be}{\begin{equation}}
\newcommand{\ee}{\end{equation}}		
\newcommand{\DD}{\mathcal{D}}

\begin{document}

\thispagestyle{empty}

\vspace*{2cm}

\begin{center}
{\bf \LARGE Holographic thermalization, quasinormal modes and superradiance in Kerr-AdS}\\
\vspace*{2.5cm}

{\bf Vitor Cardoso$^{1,2}$, } {\bf\'Oscar J. C.~Dias$^{3,4}$, } {\bf Gavin S.~Hartnett$^5$, } \\ {\bf Luis Lehner$^2$, } {\bf Jorge E.~Santos$^{6,7}$}

\vspace*{1cm}

{\it $1$ CENTRA, Departamento de F\'{\i}sica, Instituto Superior T\'ecnico, \\
Universidade de Lisboa, Av.~Rovisco Pais 1, 1049 Lisboa, Portugal}\\ \vspace{0.3cm}
{\it $2$ Perimeter Institute for Theoretical Physics, Waterloo, Ontario N2J 2W9, Canada.}\\ \vspace{0.3cm}
{\it $3$ CAMGSD, Departamento de Matem\'atica and LARSyS, Instituto Superior T\'ecnico, 1049-001 Lisboa, Portugal}\\ \vspace{0.3cm}
{\it $4$ Institut de Physique Theorique, CEA Saclay, CNRS URA 2306, F-91191 Gif-sur-Yvette, France}\\ \vspace{0.3cm}
{\it $5$ Department of Physics, UCSB, Santa Barbara, CA 93106}\\ \vspace{0.3cm}
{\it $6$ Department of Physics, Stanford University, Stanford, CA 94305-4060, U.S.A.}\\ \vspace{0.3cm}
{\it $7$ Department of Applied Mathematics and Theoretical Physics, University of Cambridge, Wilberforce Road, Cambridge CB3 0WA, UK}

\vspace*{0.5cm} {\tt vitor.cardoso@ist.utl.pt, oscar.dias@ist.utl.pt, hartnett@physics.ucsb.edu, llehner@perimeterinstitute.ca, jss55@stanford.edu}

\end{center}

\vspace*{1cm}

\begin{abstract}
Black holes in anti-de Sitter (AdS) backgrounds play a pivotal role in the gauge/gravity duality where they determine, among other things, the approach to equilibrium of the dual field theory. We undertake a detailed analysis of perturbed Kerr-AdS black holes in four- and five-dimensional spacetimes, including the computation of its quasinormal modes, hydrodynamic modes and superradiantly unstable modes. Our results shed light on the possibility of new black hole phases with a single Killing field, possible new holographic phenomena and phases in the presence of a rotating chemical potential, and close a crucial gap in our understanding of linearized perturbations of black holes in anti-de Sitter scenarios.
\end{abstract}
\noindent

\newpage
\thispagestyle{empty}
\tableofcontents

\setcounter{page}{1} \setcounter{footnote}{0}

\section{Introduction and summary\label{sec:IntroSum}}
The behavior of perturbed black holes (BHs) in asymptotically anti de-Sitter (AAdS) spacetimes is
of central importance in both current fundamental and practical research endeavors. Since these spacetimes
contain a timelike boundary, exploring such behavior requires taking into account the role of boundary conditions.
Of particular relevance are physically motivated conditions implying the absence of dissipation at infinity. 
This introduces new features and challenges in the analysis of fluctuations in AAdS scenarios:
generic perturbations ``bounce back'' off infinity and come back to interact, in the core region of AdS or with the black hole, in finite time. 
Such interaction dissipates the quasinormal modes (QNMs) only at the horizon and can trigger superradiant instabilities at the linear level, and even induce other nonlinear phenomena.
Additionally, BHs in AAdS play a central role in the formulation and applications of the AdS/CFT correspondence.
This correspondence~\cite{Maldacena:1997re,Aharony:1999ti} provides a
remarkable framework for studying certain strongly coupled gauge theories in
$d$ dimensions by mapping them to weakly coupled quantum gravitational systems in $d+1$ dimensions. In a certain limit (namely in the large `t Hooft coupling and planar limit), quantum gravity in the bulk reduces to classical general relativity. Within this holographic framework, a black hole is dual to a thermal state and the question of thermalization in the boundary gauge theory translates, in the gravitational bulk, to understanding the ``return to equilibrium'' behavior of perturbed black holes \cite{Horowitz:1999jd,Danielsson:1999fa,Birmingham:2001pj,Son:2002sd,Kovtun:2005ev,Policastro:2002se,Friess:2006kw,Michalogiorgakis:2006jc}. 

Here, we will be interested in the original gauge/gravity duality, namely the AdS$_{d+1}$/CFT$_d$ correspondence  (for the $d=3,4$ cases for which Super-Yang-Mills theory is dual to string theory on  $AdS_4 \times S^7$ and  $AdS_5 \times S^5$, respectively). Moreover, we are particularly interested in systems with a rotating chemical potential. This requires looking to CFTs formulated on a sphere (since a rotational shift is a pure gauge transformation on a plane), i.e. for bulk solutions that asymptote to global AdS (which is conformal to the static Einstein Universe $R_t \times S^{d-1}$). Henceforward, when we refer to AAdS spacetimes it is implicitly assumed that we mean global AdS (although some of the discussions are also valid for planar AdS i.e. the Poincar\'e patch of AdS that asymptotes to $R_t \times R^{d-1}$). We will often use the notation $D=d+1$ for the bulk spacetime dimension; Greek indices denote the bulk dimensions while Latin indices describe boundary coordinates.

Certainly, important headways on thermalization can be made by studying the behavior of perturbed
black hole spacetimes at a {\em linearized level}. Naturally, the applicability of such analysis depends on the strength of the perturbation off the stationary black hole and the behavior obtained can hint of possible instabilities~\cite{Holzegel:2011rk,Holzegel:2011uu,Holzegel:2012wt,Holzegel:2013kna}.

The analog problem in asymptotically flat spacetimes is, to a large degree, understood.
The approach to equilibrium depends sensitively on the character of the perturbation:
massless fields (scalar, vector or tensor type) die off through their quasinormal modes (QNMs), with a time dependence of the form $e^{-i \omega t}$ with $\mathrm{Im}(\omega)<0$ (for
a review see~\cite{Berti:2009kk})\footnote{Exceptions exist however, as QNMs do not constitute a basis for perturbations, nevertheless cases where QNMs are known to fail to describe the solution in linearized perturbative regimes are either finely-tuned or, like tails, arise after
a QNM epoch can be identified).}. Massive fields on the other hand, have a much richer phenomenology, tied to the fact that they can be trapped inside a cavity with size of the order of the Compton wavelength. This trapping causes the field to decay much slower, or can even become unstable for large black hole rotation (see~\cite{Cardoso:2013krh} and references therein). The linear behavior of massive fields around rotating black holes is not fully understood yet (and certainly not the nonlinear regime), but it is akin to that of massless fields in AAdS backgrounds in that both can develop trapping potentials. However, an important difference is that the height of the potential barrier is unbounded in the AdS case while it is finite for a massive scalar in a flat background.

Accurate expressions for the QNMs for generic black holes in the asymptotically flat case are known for both static and stationary black holes (see the review~\cite{Berti:2009kk}); remarkably this is not the case in the AdS background as they are not known for the Kerr-AdS BH. This status of affairs is, at first sight, surprising given the central role they play in holographic dualities as well as in studies of AAdS black hole stability. 

It is thus worth discussing in detail the reason for this gap in our knowledge.
Since an AAdS spacetime is a non-globally hyperbolic spacetime (i.e., spatial infinity is a timelike boundary in the Carter-Penrose diagram and thus null rays can reach it in finite time), in order to predict the future time evolution of the system we need to give not only initial data but also to specify the choice of boundary conditions (BCs). At the inner boundary (origin or horizon) regularity fixes the BC. However, at the asymptotic boundary this choice is {\it \`a priori} arbitrary, being fixed by a physical motivation. 
From a pure gravitational perspective it is often stated in the literature that one is interested in ``reflecting BCs" which suggests the idea that we want vanishing flux of energy and angular momentum across the asymptotic boundary. On the other hand, in the context of the AdS/CFT duality we typically want to choose BCs that preserve the asymptotic boundary (conformal) metric. Next, and in Appendix \ref{sec:Fluxes}, we 
emphasize that these two perspectives require exactly the very same BCs.  
Formally, the discussion of the asymptotic BCs is more clear if we write the total metric (background plus perturbations, if present) in Fefferman-Graham  coordinates (this frame is defined such that $g_{zz}=L^2/z^2$ and $g_{zb}=0$, where $z$ is the radial distance with boundary at $z=0$, and $x^b$ are the coordinates on the boundary),  and  looking at its boundary expansion (see \cite{Haro:2000xn} and references therein). For odd boundary dimension $d$ this reads\footnote{For even $d$, the asymptotic expansion \eqref{introFG} contains also a logarithmic term $z^d \log z^2 \tilde{g}_{ab}^{(d)}$ and the holographic stress tensor has an extra contribution proportional to the conformal anomalies of the boundary CFT \cite{Haro:2000xn}. These details are not essential for the present discussion.} 
\begin{eqnarray}\label{introFG}
&&ds^2 = \frac{L^2}{z^2}\left[dz^2+g_{ab}(z,x) dx^{a}dx^{b}\right]\,, \nonumber\\
&& g_{ab}{\bigl |}_{z\to 0}=g_{ab}^{(0)}(x)+\cdots+z^d g_{ab}^{(d)}(x)+\cdots 
\qquad \hbox{with} \quad \langle T_{ab}(x)\rangle\equiv \frac{d}{16\pi G_N}\,g_{ab}^{(d)}(x),
\end{eqnarray}
where $g^{(0)}(x)$ and $g^{(d)}(x)$ are the two integration ``constants" of the expansion; the first dots include only even powers of $z$ (smaller than $d$) and depend only on $g^{(0)}$ (thus being the same for {\it any} solution that asymptotes to global AdS$_d$) while the second dots depend on the two independent terms $g^{(0)}, g^{(d)} $ (we will fix Newton's constant as $G_N\equiv 1$).
Within the AdS/CFT duality we are (typically) interested in Dirichlet BCs that do not deform the conformal metric $g^{(0)}$. Indeed, this defines the gravitational background where the CFT is formulated and we want to keep it fixed; in our case this is the static Einstein Universe. Stated in other words, we allow perturbations in the bulk that only deform the expectation value of holographic stress tensor $ \langle T_{ab}(x)\rangle $ (that specifies and describes the boundary CFT) \footnote{Note that  $g^{(d)}$ is an integration ``constant" but not a free function; it is fixed solving the Einstein equation in the bulk subject to regular BCs at the horizon or radial origin.} but that preserve the asymptotic structure of the original background that we perturb.\footnote{Other BCs that might be called asymptotically globally AdS (and that promote the boundary graviton to a dynamical field) were proposed in \cite{Andrade:2011dg}. However, they turn out to lead to ghosts (modes with negative kinetic energy) and thus make the energy unbounded below \cite{Compere:2008us}.} 
As discussed in Appendix \ref{sec:Fluxes} these BCs do not allow asymptotic dissipation of energy or angular momentum. In other words, everything that hits the asymptotic boundary is reflected back to the bulk core allowing for a non-trivial interplay between the asymptotic and inner (e.g. horizon) boundaries. We have now growing evidence that these conditions favour the development of instabilities. For instance, it has been shown recently that even arbitrarily small  perturbations can trigger black hole formation in global AdS~\cite{Bizon:2011gg,Dias:2011ss}, indicating that global AdS is nonlinearly unstable to a weak perturbative turbulent mechanism (note however the existence of ``islands'' of stability~\cite{Buchel:2013uba,Dias:2012tq,Maliborski:2013jca}). Additionally, it has recently
been shown that turbulent behavior \footnote{As well, in planar AdS backgrounds, turbulent behavior of gravity has also been uncovered for (sufficiently) long-wavelength perturbations of black holes in~\cite{Adams:2013vsa,Carrasco:2012nf}.}
(akin to the one displayed by hydrodynamics) arises in long-wavelength perturbations of 3+1 Kerr-AdS~\cite{Carrasco:2012nf,Green:2013zba}. 

The BCs we need to impose to study AAdS perturbations of  global AdS BHs are therefore well known. Yet, we still need to understand why the study of QNMs and superradiant instabilities of global AdS BHs is not a closed chapter.  For that, we need to look to the perturbation equations. Studying linearized gravitational perturbations requires solving the linearized Einstein equations for the metric perturbation. For generic perturbations this is a coupled nonlinear system of PDEs. Solving this PDE system directly with the above BCs is a hard problem, even numerically. In certain special cases, however, drastic simplifications occur. Fortunately, and quite remarkably, in four dimensions it has been shown that if we use certain gauge invariant scalar variables we can reduce the problem of looking for the most generic perturbations of the above AAdS BHs  to solving a single PDE. Moreover, using the harmonic decomposition of the system, the later reduces to solving two ODEs. This remarkable reformulation of the linearised perturbation problem is known as the Regge-Wheeler$-$Zerilli or Kodama-Ishisbashi formalism for perturbations of Schwarzchild BHs \cite{Regge:1957td,Zerilli:1970se,Kodama:2003jz}, and Teukolsky formalism  for perturbations of Kerr BHs \cite{Teukolsky:1972my,Teukolsky:1973ha}. We ask the reader to see the companion paper \cite{Dias:2013sdc} for a detailed discussion of these two formalisms and for the map relating them when the background rotation vanishes. Once the solution for the gauge invariant scalars is known a simple differential map generates the corresponding metric perturbation tensors (in a particular gauge). 

At this point, to find QNMs or instability timescales of AAdS BHs we just need to take the above BCs, discussed for the metric perturbations, and translate them to get the corresponding BCs that need to be imposed on the gauge invariant scalars. On general grounds we should expect the Dirichlet BCs on the metric to translate to Robin BCs (which relate the field with its derivative) on the gauge invariant scalars.  
In the static background case, this dictionary was found by \cite{Friess:2006kw,Michalogiorgakis:2006jc}. There are two families of perturbations: scalar (also called even or polar) and vector (odd or axial) sectors. 
The associated QNMs of the global AdS Schwarzchild BH were then computed \cite{Friess:2006kw,Michalogiorgakis:2006jc,Dias:2013sdc}:  vector QNMs agree with those first computed in
 \cite{Cardoso:2001bb,Berti:2003ud,Natario:2004jd} (the scalar modes of  \cite{Cardoso:2001bb,Berti:2003ud,Natario:2004jd} do not preserve the asymptotic AdS structure).
 In the stationary case, the BC map was constructed only recently in the companion paper \cite{Dias:2013sdc}. With it at hand, we can finally compute the gravitational QNM spectrum and superradiant instability growth rates of the Kerr-AdS BH. This is one of main aims of the work here reported. (Previous work on gravitational QNMs \cite{Giammatteo:2005vu} and superradiant instability of  Kerr-AdS \cite{Cardoso:2006wa} imposed BCs that do not keep the boundary metric fixed).
While many of the methods presented here are readily applicable to arbitrary dimensions we concentrate in dimensions $d=3$ and $d=4$ because of their interest for the AdS$_{d+1}$/CFT$_d$ holographic dualities. 

The interest on the superradiant instability is not restricted to its growth rate. Indeed, the onset curve of this instability (where the imaginary part of the frequency vanishes) is an exact zero mode that is invariant under the horizon-generating Killing field of Kerr-AdS. Therefore we will argue that, in a phase diagram of stationary solutions,  this onset curve signals a bifurcation curve to a new family of BHs that have a single Killing vector field (KVF), i.e. that are periodic but not time dependent neither axisymmetric. A far reaching consequence of this statement is that Kerr-AdS BHs are not the only stationary BHs of Einstein-AdS gravity. These BHs can exist because they evade a main assumption of the rigidity theorems \cite{Hawking:1971vc,Hollands:2006rj,Moncrief:2008mr}. We will give the explicit perturbative construction of the leading order thermodynamics and properties of these BHs. These ideas were first proposed in \cite{Kunduri:2006qa} and further developed in \cite{Dias:2011at,Dias:2011ss}. Now that we have the precise onset curve of superradiance, we have the opportunity to expand their discussion. 

Another aim of the present work is to confirm that long wavelength gravitational QNM frequencies agree with the hydrodynamic relaxation timescales that we obtain when we consider the near-equilibrium and long wavelength effective description of the $CFT_d$. This will provide the first explicit confirmation that the match between the QNM spectrum and the CFT thermalization timescales also holds in the presence of a rotating chemical potential. Incidentally, it provides the first non-trivial confirmation that the Robin boundary conditions for the Teukolsky gauge-invariant variable derived in the companion paper [59] are indeed the ones that we must impose if we want the perturbations to preserve the conformal metric.

This work is divided as follows. Section \ref{sec:KerrAdS} reviews relevant properties of $D=4$ Kerr-AdS spacetime and the equations of motion and the BCs \cite{Dias:2013sdc} governing the behavior of perturbations at the linear level. Section \ref{sec:num} describes the numerical methods employed to solve them. One of these numerical approaches is novel and we expect it to be of interest for other applications.
Section \ref{sec:QNMs} presents our results for the full spectrum of gravitational QNMs and superradiant instability timescales of the Kerr-AdS BH. In Section \ref{sec:singleKVFBH} we construct and discuss the novel single Killing vector field BHs that merge with the Kerr-AdS BH at the onset of superradiance. 
In section \ref{sec:Hydro} we use the fluid/gravity duality to confirm the match between the gravitational long-wavelength QNM spectrum and the $CFT_3$ hydrodynamic modes even in the presence of a rotating chemical potential. Section \ref{sec:5d} repeats the previous section computations and discussions but this time for the $D=5$ rotating system that is of interest for the AdS$_5$/CFT$_4$ duality. It will also contribute to identify universal properties of systems with a rotating chemical potential. We work in a particularly clean environment where we study perturbations around the equal angular momentum Myers-Perry BH. Indeed, this background has enhanced symmetry $-$ it only depends non-trivially on the radial direction $-$ and its perturbations have an exact harmonic decomposition. 
The present study fills important gaps in our knowledge but confirms and opens some interesting questions. In Section \ref{sec:Conc} we discuss these open questions in what can be viewed as a roadmap  in the subject from our viewpoint.

\section{Gravitational perturbations \& boundary conditions of Kerr-AdS black hole\label{sec:KerrAdS}}
In this section we review the basic properties of Kerr-AdS black holes and their gravitational perturbations.

\subsection{Kerr-AdS black hole  \label{sec:KerrAdSbh}}

The Kerr-AdS geometry was originally discovered by Carter in the Boyer-Lindquist coordinate system $\{ T,r,\theta,\phi\}$ \cite{Carter:1968ks}.  For our purposes, it is convenient here, 
to follow Chambers and Moss \cite{Chambers:1994ap} and introduce the new time and polar 
coordinates $\{t,\chi\}$, which are related to the  Boyer-Lindquist coordinates $\{T,\theta\}$ by
\begin{equation}\label{CMtoBL}
t=\Xi\,  T \,,\qquad \chi=a \cos\theta\,,
\end{equation}
where $a$ is the rotation parameter of the solution and $\Xi$ is to be defined in \eqref{metricAux}. In this coordinate system the Kerr-AdS black hole line element  reads \cite{Chambers:1994ap}
\begin{eqnarray}\label{metric}
&& ds^2=-\frac{\Delta _r}{\left(r^2+\chi ^2\right)\Xi^2}\left(dt-\frac{a^2-\chi ^2}{a}\,d\phi \right)^2+\frac{\Delta _{\chi }}{\left(r^2+\chi ^2\right)\Xi ^2}\left(dt-\frac{a^2+r^2}{a}\,d\phi\right)^2 \nonumber \\
&&\hspace{1.1cm}
+\frac{\left(r^2+\chi ^2\right)}{\Delta _r}\,dr^2+\frac{\left(r^2+\chi ^2\right)}{\Delta _{\chi}}\,d\chi^2\,,
\end{eqnarray}
where 
\begin{equation} \label{metricAux}
\Delta _r=\left(a^2+r^2\right)\left(1+\frac{r^2}{L^2}\right)-2M r\,,\qquad \Delta _{\chi }=\left(a^2-\chi ^2\right)\left(1-\frac{\chi ^2}{L^2}\right)\,,\qquad \Xi =1-\frac{a^2}{L^2} \,.
\end{equation}
The Chambers-Moss (CM) coordinate system $\{t,r,\chi,\phi\}$ has the appealing property that the line element treats the radial $r$ and polar $\chi$ coordinates on an almost equal footing. This property extends to the radial and angular equations describing gravitational perturbations in the Kerr-AdS background. In this frame, the horizon angular velocity and temperature are given by 
\begin{equation} \label{OmegaT}
\Omega_H=\frac{a}{r_+^2+a^2}\,,\qquad 
 T_H=\frac{1}{\Xi}\left[ \frac{r_+}{2 \pi}\left(1+\frac{r_+^2}{L ^2}\right)\frac{1}{r_+^2+a^2}-\frac{1}{4\pi r_+}\left(1-\frac{r_+^2}{L ^2}\right)\right].
\end{equation}
The Kerr-AdS black hole obeys $R_{\mu\nu} =-3L^{-2}g_{\mu\nu}$, and asymptotically approaches global AdS space with radius of curvature $L$.  This asymptotic structure is not manifest in \eqref{metric}, one of the reasons being that the coordinate frame $\{t,r,\chi,\phi\}$  rotates at infinity with angular velocity $\Omega_{\infty}=-a/(L^2 \Xi)$.  However, if we introduce the coordinate change 
\begin{eqnarray}\label{globalAdScoordtransf}
&& T=\frac{t}{\Xi }\,, \qquad \Phi =\phi +\frac{a}{L^2}\, \frac{t}{\Xi }\,,\nonumber \\
&& R=\frac{\sqrt{L^2 \left(a^2+r^2\right)-\left(L^2+r^2\right) \chi ^2}}{L\sqrt{\Xi }},\qquad \cos\Theta=\frac{L\sqrt{\Xi } \,r \, \chi }{a\sqrt{L^2 \left(a^2+r^2\right)-\left(L^2+r^2\right) \chi ^2}}\,, 
\end{eqnarray}
we find that as $r\to\infty$ (i.e. $R\to \infty$), the Kerr-AdS geometry  \eqref{metric} approaches
\begin{eqnarray}\label{metricAdS}
ds^2_{AdS}=-\left(1+\frac{R^2}{L^2}\right)dT^2+\frac{dR^2}{1+\frac{R^2}{L^2}}+R^2\left(d\Theta^2+\sin^2\Theta\, d\Phi^2\right),
\end{eqnarray}
which we recognize as the line element of global AdS. In other words, the conformal boundary of the bulk spacetime is the static Einstein universe $R_t \times S^2$:  $ds^2_{\partial}=\lim_{R\to\infty} \frac{L^2}{R^2}\, ds^2_{AdS}=-dT^2+d\Theta^2+\sin^2\Theta\, d\Phi^2$. This is the boundary metric where the CFT lives in the context of the AdS$_4$/CFT$_3$ correspondence.

\begin{figure}[ht]
\centering
\includegraphics[width=.9\textwidth]{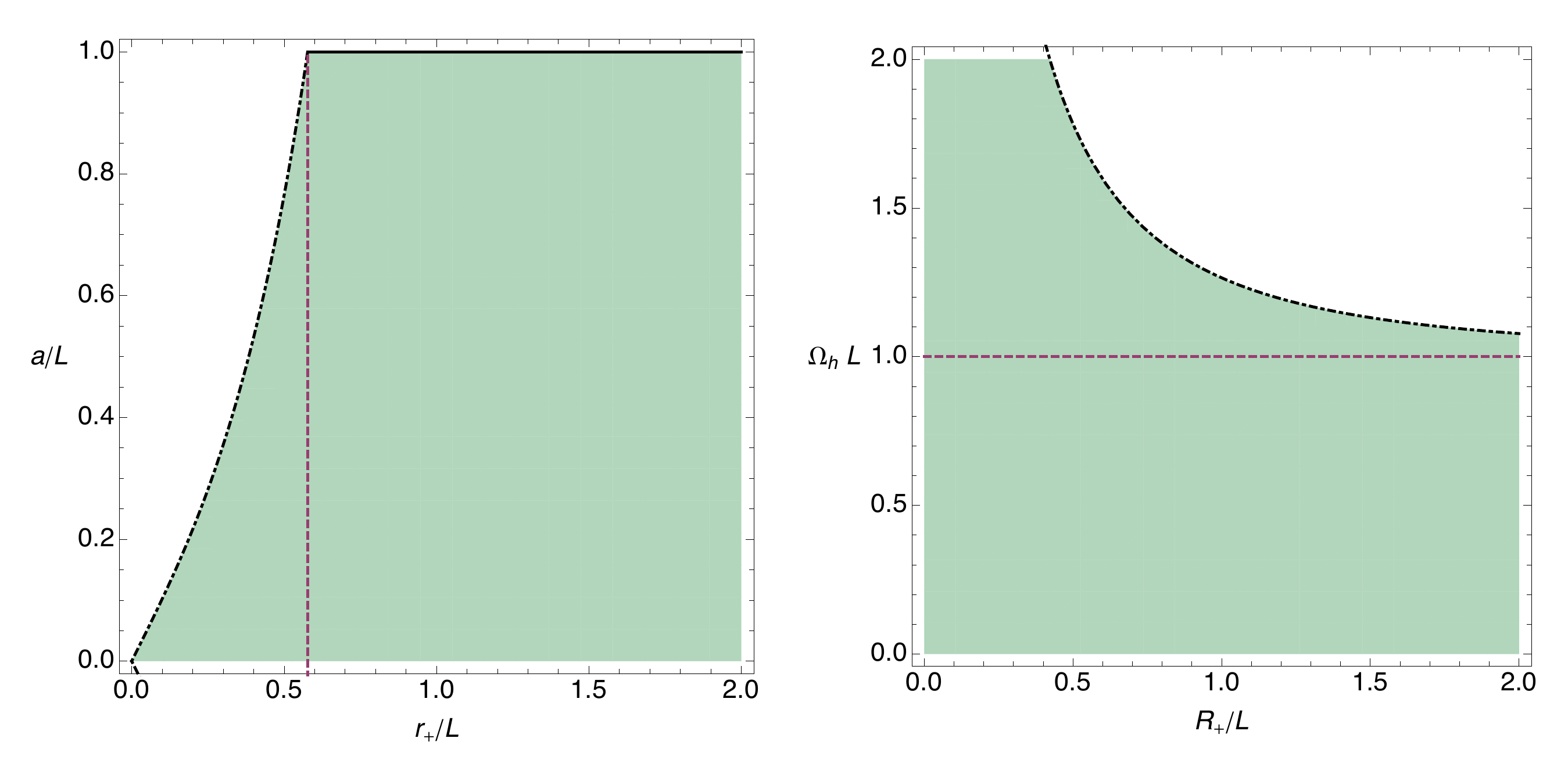}
\caption{\emph{Left panel:} Allowable region for $a/L$ and $r_+/L$: the vertical dashed line is given by $r_+/L=1/\sqrt{3}$, the dashed dotted lines indicate extremality, and the horizontal solid lines indicate $|a|=L$.\\
\emph{Right panel:} Allowable region for $\Omega_h L$ and $R_+/L$: the horizontal dashed line marks the onset of superradiance, the dashed dotted lines indicate extremality.}
\label{Fig:domain}
\end{figure}

The ADM mass and angular momentum of the black hole are related to the mass $M$ and rotation $a$ parameters through $M_{ADM}=M/\Xi^2$
and $J_{ADM}=M a/\Xi^2$, respectively \cite{Caldarelli:1999xj,Gibbons:2004ai}. 
The horizon angular velocity and temperature relevant for the thermodynamic analysis are the ones measured with respect to the non-rotating frame at infinity \cite{Caldarelli:1999xj,Gibbons:2004ai} and are given in terms of \eqref{OmegaT} by $T_h= \Xi \,T_H$ and $\Omega_h=\Xi  \,\Omega _H+\frac{a}{L^2}$.
The event horizon is located at $r=r_+$ (the largest real root of $\Delta_r$), and it is a Killing horizon generated by the Killing vector $K=\partial_T+\Omega_h \partial_\Phi $. We discuss our results in terms of the horizon radius and rotation parameter, which uniquely determine a given Kerr-AdS black hole. The mass parameter is given in terms of these by  $M=\left(r_+^2+a^2\right) \left(r_+^2+L^2\right)/\left(2 L^2 r_+\right)$. All regular black hole solutions must obey $T_H\geq0$ and $a/L<1$. This translates into the following conditions for $r_+/L$ and $a/L$:
\begin{align}
&\frac{a}{L}\leq \frac{r_+}{L}\sqrt{\frac{L^2+3 r_+^2}{L^2-r_+^2}},\quad\text{for}\quad \frac{r_+}{L}<\frac{1}{\sqrt{3}}, \nonumber
\\
\\
&\frac{a}{L}< 1,\quad\text{for}\quad \frac{r_+}{L}\geq\frac{1}{\sqrt{3}}\,.\nonumber
\end{align}
The first inequality is saturated for a degenerate extremal regular horizon. On the left panel Fig.~\ref{Fig:domain}, we show the allowable domain for $a/L$ and $r_+/L$. Further properties of the Kerr-AdS spacetime are discussed in Appendix A of~\cite{Dias:2010ma}.

We will find it useful to parametrize the black hole in variables that are naturally related to the onset of superradiance, and that are gauge invariant. Here we choose the pair $(R_+,\Omega_h)$, with $R_+$ given by:
\begin{equation}
R_+ = \frac{\sqrt{r_+^2+a^2}}{\sqrt{\Xi}}\,.
\end{equation}
Extremality is attained at
\begin{equation}
\left|\Omega^{\text{ext}}_h\right| = \frac{1}{L R_+} \sqrt{\frac{(L^2+R_+^2)(L^2+3R_+^2)}{2L^2+3 R_+^2}}\,.
\end{equation}

Note that $R_+$ is just the square root of the area of the spatial section of the event horizon, divided by $4\pi$, often denominated areal horizon radius. The allowed values of $R_+/L$ and $\Omega_h L$ are depicted on the right panel of Fig.~\ref{Fig:domain}.

\subsection{Gravitational master equation and global AdS boundary conditions  \label{sec:Teukolsky}}

In the Newman-Penrose$-$Teukolsky formalism, all the information about (non-trivial) gravitational perturbations with spin $s=-2$ is encoded in the single variable $\delta \Psi_4$ which describes the perturbation of the Weyl scalar $ \Psi_4=C_{abcd} n^a \bar m^b n^c \bar m^d$. The equation of motion for this perturbation $\delta \Psi_4$ is described by the $s=-2$ Teukolsky master  equation \cite{Teukolsky:1972my,Teukolsky:1973ha}.  Introducing the separation ansatz 
\begin{equation}\label{TeukAnsatz}
\delta \Psi_4=(r-i \chi )^{-2}\,e^{-i \tilde{\omega}  t}\,e^{i m \phi }\, R_{\tilde{\omega} \ell m}^{(-2)}(r)\, S_{\tilde{\omega} \ell m}^{(-2)}(\chi )\,,
\end{equation}
the spin $s=-2$ Teukolsky master equation separates into angular and  radial equations  \cite{Chambers:1994ap,Dias:2013sdc}:
\begin{equation}
\label{MasterAng}
\partial _{\chi }\left(\Delta _{\chi }\partial _{\chi } S_{\tilde{\omega} \ell m}^{(-2)} \right)+ \left[-\frac{\left(K_{\chi }+\Delta _{\chi }'\right)^2}{ \Delta _{\chi }}+ \left(\frac{6\chi ^2}{L^2}+4K_{\chi }'+\Delta _{\chi }''\right)+\lambda \right]S_{\tilde{\omega} \ell m}^{(-2)}=0   \,, 
\end{equation}
\begin{equation}\label{MasterRad}
\partial _r\left(\Delta _r\partial _r R_{\tilde{\omega} \ell m}^{(-2)}\right)+\left[\frac{\left(K_r-i \Delta _r'\right)^2}{ \Delta _r}+\left(\frac{6 r^2}{L^2}+4i K_r'+ \Delta _r''\right)-\lambda \right]R_{\tilde{\omega} \ell m}^{(-2)}=0\,,
\end{equation}
where we have defined
\begin{equation} \label{KrKchi}
K_r=\Xi \left[m a-\tilde{\omega} \left(a^2+r^2\right)\right],\qquad K_{\chi }=\Xi \left[m a -\tilde{\omega} \left(a^2-\chi ^2\right)\right].
\end{equation}
The eigenfunctions $S_{\tilde{\omega} \ell m}^{(-2)}(\chi)$ are the spin-weighted $s=-2$ AdS-spheroidal harmonics.  The positive integer $\ell$ specifies the number of zeros of the eigenfunction along the polar direction which are given by $\ell-{\rm max}\{|m|,|s|\}$ (so the smallest  $\ell$ is $\ell=|s|=2$). The associated eigenvalues $\lambda$ are functions of $\tilde{\omega},\ell,m$ which can be  computed numerically. Regularity imposes the constraints that $-\ell\leq m\leq \ell$ must be an integer and $\ell\geq |s|$. This equation has been studied in \cite{Dias:2013sdc} in the limit where the rotation vanishes.

If we solve (simultaneously) the angular and radial equations, which are coupled through the two eigenvalues $\tilde{\omega}$ and $\lambda$, we get information about the most general perturbation of the Kerr-AdS black hole. In particular, the Teukolsky equation and its solution for the spin $s=+2$ perturbations, described by the variable $\delta \Psi_0=(r-i \chi )^{-2}\,e^{-i \tilde{\omega}  t}\,e^{i m \phi }\, R_{\tilde{\omega} \ell m}^{(2)}(r)\, S_{\tilde{\omega} \ell m}^{(2)}(\chi )$, follow trivially from the spin $s=-2$ solution. Namely,  $R_{\tilde{\omega} \ell m}^{(2)}$ is the complex conjugate of  $R_{\tilde{\omega} \ell m}^{(-2)}$  and $S_{\tilde{\omega} \ell m}^{(2)}(\chi)=S_{\tilde{\omega} \ell m}^{(-2)}(-\chi)$. The later statement implies that the separation constants are such that $\lambda_{\tilde{\omega}\ell m}^{(-2)}=\lambda_{\tilde{\omega}\ell m}^{(2)}\equiv \lambda$. The only exceptions to the above are the  trivial perturbations, the ``$\ell=0$" and ``$\ell=1$" modes, which shift, respectively, the mass and angular momentum of the solution along the original Kerr-AdS family, and to which the Teukolsky formalism is blind  \cite{WaldL0L1,Kodama:2003jz,Dias:2013hn,Dias:2013sdc}. 

Quasinormal modes and unstable modes of the Kerr-AdS black hole are solutions of \eqref{MasterAng}-\eqref{MasterRad}  obeying physically relevant boundary conditions (BCs) \cite{Dias:2013sdc}. 
At the horizon, the BCs must be such that only ingoing modes are allowed. A Frobenius analysis at this boundary gives two independent solutions,
\begin{equation}\label{bcH0}
R_{\tilde{\omega} \ell m}^{(-2)}\sim  A_{\rm{in}}\left(r-r_+\right)^{1-i\frac{ \tilde{\omega} -m \Omega _H}{4\pi  T_H}}\left[1+\mathcal{O}\left(r-r_+\right)\right]+A_{\rm{out}}\left(r-r_+\right)^{-1+i\frac{ \tilde{\omega} -m \Omega _H}{4\pi  T_H}}\left[1+\mathcal{O}\left(r-r_+\right)\right],
\end{equation}
where $A_{\rm{in}}, A_{\rm{out}}$ are arbitrary amplitudes and $\Omega_H,T_H$ are the angular velocity and temperature defined in \eqref{OmegaT}.
To impose the correct BC, we introduce the ingoing Eddington-Finkelstein coordinates $\{v,r,\chi,\widetilde{\phi} \}$, which are appropriate to extend the solution through the horizon. These are defined via
\begin{equation}\label{EF}
t= v-\Xi\int \frac{r^2+a^2}{\Delta _r}\,dr ,\qquad \phi =\widetilde{\phi }-\int \frac{a\, \Xi }{\Delta _r} \,dr\,.
\end{equation}
The BC is determined by the requirement that the metric perturbation is regular in these ingoing Eddington-Finkelstein coordinates, where the metric tensor is constructed applying a differential operator to $\delta \Psi _4$ (this is known as the Hertz map; see the companion paper \cite{Dias:2013sdc}). It follows that the metric perturbation is regular at the horizon if and only if  $R(r)|_H$ behaves as $R(r)|_H \sim R_{IEF}(r)|_H\left(r-r_+\right)^{1-i \frac{\tilde{\omega} -m \Omega _H}{4\pi  T_H}}$  where $R_{IEF}(r)|_H$ is a smooth function\,\footnote{This analysis misses the special case in which $2i \frac{\tilde{\omega} -m \Omega _H}{4\pi  T_H}$ is a positive integer. For this special value, our boundary conditions still allow for outgoing modes at the horizon. However, by inspecting our numerical data we can \emph{a posteriori} test if this condition is satisfied. In all our simulations, this never seems to be the case.}.
Therefore, the appropriate BC at the horizon demands we set $A_{\rm{out}}=0$ in \eqref{bcH0}:
\begin{equation}\label{bcH}
R_{\tilde{\omega} \ell m}^{(-2)}{\biggl |}_{r\to r_+}=A_{\rm{in}}\left(r-r_+\right)^{1-i\varpi}\left[1+\mathcal{O}\left(r-r_+ \right)\right]\,
\end{equation}
where
\begin{equation}
\label{superradiance factor}
\varpi=\frac{\tilde{\omega}-m \Omega _H}{4\pi  T_H}\,,
\end{equation}
is what we might call the superradiant factor. 
Less formally, but perhaps more intuitively, when $\tilde{\omega}$ is real and non-zero we can understand this horizon BC by noting that the wave solution $e^{-i \tilde{\omega} t}\left(r-r_+\right)^{-i \varpi}=
e^{-i \left(\tilde{\omega} t+ \varpi\,\hbox{ln}\left(r-r_+\right)\right)}$ is the one that describes ingoing modes at the horizon since $r$ must decrease when $t$ grows to keep the phase constant  (classically, we cannot have outgoing modes leaving the horizon). 

Consider now the asymptotic boundary. A  Frobenius analysis of the radial Teukolsky equation\eqref{MasterRad} at infinity yields the two independent asymptotic decays 
\begin{equation}\label{FrobRinf}
R_{\tilde{\omega} \ell m}^{(- 2)}{\bigl |}_{r\to \infty}= B_{+}^{(- 2)}\,\frac{L}{r}+ B_{-}^{(-2)}\,\frac{L^2}{r^2}+\mathcal{O}\left(  \frac{L^3}{r^3} \right)\,,
\end{equation}
for arbitrary  amplitudes $B_{\pm}^{(-2)}$.
We want the perturbations to preserve the asymptotic global AdS structure of the background Kerr-AdS black hole, i.e. we want the deformation to preserve the asymptotic line element \eqref{metricAdS}. In the companion paper \cite{Dias:2013sdc} we found that  this requirement yields the following Robin BC,   
\begin{equation}\label{TheBCs}
B_-^{(-2)}=i \,\beta B_+^{(-2)} \,,
\end{equation}
with two possible solutions for $\beta$, that we call $\beta_{\rm \bf s}$ and $\beta_{\rm \bf v}$,
\begin{eqnarray}
&& 1)   \quad  \beta=\beta_{\rm \bf s}= \frac{\Lambda_0-\sqrt{\Lambda_1}}{\Lambda_2}\,,\qquad \hbox{or} \label{TheBCsS} \\
&& 2) \quad \beta=\beta_{\rm \bf v}= \frac{\Lambda_0+\sqrt{\Lambda_1}}{\Lambda_2}\,,\label{TheBCsV}
\end{eqnarray}
where we have introduced  
\begin{eqnarray}  \label{TheBCaux}
&& \Lambda_0 \equiv 2 a^2 (\lambda -6)-8 (\lambda +1) L^4 \tilde{\omega} ^2\Xi ^2+8 L^6 \tilde{\omega} ^4\Xi ^4+L^2 \left[\lambda  (\lambda +2)-4\text{  }\Xi ^2a \tilde{\omega}  \left[5 (m-a \tilde{\omega} )+2 a \tilde{\omega} \right]\right], \nonumber \\
&&  \Lambda_1 \equiv  4 a^4 (\lambda -6)^2+L^4\lambda ^2 (\lambda +2)^2+48 (\lambda +6)a^3 \Xi ^2L^2 \tilde{\omega} (m-a \tilde{\omega} )
+8\lambda  (5 \lambda +6)(m-a \tilde{\omega} ) L^4\Xi ^2 a \tilde{\omega} 
\nonumber\\
&& \hspace{0.9cm} +4 a^2 L^2 {\biggl [} \lambda  \left[-12+(\lambda -4) \lambda +24 (m-a \tilde{\omega} )^2 \Xi ^2\right] 
+12  \Xi ^2 L^2 \tilde{\omega} ^2 \left[2 \lambda +3 (m-a \tilde{\omega} )^2 \Xi ^2\right] {\biggr ]}, \nonumber\\
&&\Lambda_2 \equiv 4 L \Xi \left[2 a m+L^2 \tilde{\omega}  \left(2+\lambda -2 L^2 \tilde{\omega} ^2\Xi ^2\right)\right].
\end{eqnarray}
Perturbations obeying the BCs \eqref{TheBCs}-\eqref{TheBCsS} preserve the asymptotically global AdS behavior of the background. These are also natural BCs in the context of the AdS/CFT correspondence: they allow a non-zero expectation value for the CFT stress-energy tensor while keeping fixed the boundary metric. 

The BC \eqref{TheBCs},\eqref{TheBCsS} generates what we might call the ``rotating sector of scalar modes", in the sense that when the rotation vanishes, these perturbations reduce continuously to the Kodama-Ishibashi scalar modes  \cite{Dias:2013sdc}.\footnote{The Kodama-Ishibashi  vector master equation is the Regge-Wheeler master equation for odd (also called axial) perturbations \cite{Regge:1957td}, and the Kodama-Ishibashi scalar master equation is the Zerilli master equation for even (also called polar) perturbations \cite{Zerilli:1970se}.} By a similar reasoning  the BC \eqref{TheBCs}, \eqref{TheBCsV} selects the ``rotating vector modes" of the spectrum. Having this in mind we will often use the nomenclature ``scalar/vector'' modes when discussing our results  \cite{Dias:2013sdc}.

As discussed previously, the Chambers-Moss coordinate system $\{t,r,\chi,\phi\}$  rotates at infinity.  However, the coordinate transformation \eqref{globalAdScoordtransf} 
introduces the coordinate frame $\{T,R,\Theta,\Phi\}$ appropriate to discuss the asymptotic global AdS$_4$ structure of the geometry and the boundary metric where the dual CFT$_3$ and its hydrodynamic limit are formulated. 
Consider  a generic linear perturbation in Kerr-AdS written in the Chambers-Moss frame $\{t,r,\chi,\phi\}$. Since $\partial_t$ and $\partial_\phi$ are isometries of the background geometry we can Fourier decompose the perturbation in these directions as $e^{-i \tilde{\omega}  t}\,e^{i m \phi }$ as we did in  \eqref{TeukAnsatz}. The frequency $\tilde{\omega}$ measured in the frame $\{t,r,\chi,\phi\}$ differs from the frequency measured in the frame $\{T,R,\Theta,\Phi\}$. It follows from the coordinate transformation \eqref{globalAdScoordtransf} that they are related by
\begin{equation}\label{freqshift}
e^{-i \tilde{\omega}  t}\,e^{i m \phi } \equiv e^{-i \,\omega\,  T}\,e^{i m \Phi }, \qquad \hbox{with} \quad 
\omega \equiv \tilde{\omega}\, \Xi+m \frac{a}{L^2}.
\end{equation}
The quantity $\omega$ can be viewed as the natural or fundamental frequency since it measures the frequency with respect to a frame that does not rotate at infinity. This is also the natural frequency measured by a CFT$_3$ and associated fluid rest frame observer.  
Therefore, although we will use the frame $\{t,r,\chi,\phi\}$ and $\tilde{\omega}$ to discuss many of our results, we choose to plot our  results in terms of $\omega$. Note that the superradiant factor defined 
in \eqref{bcH} can equally be written as $\varpi=\frac{ \omega -m \Omega _h}{4\pi  T_h}$ where the angular velocity $\Omega _h$ and temperature $ T_h$ as measured in the $\{T,R,\Theta,\Phi\}$ frame are given below \eqref{metricAdS}.

\section{Numerical procedures \label{sec:num}}
In this section we discuss the numerical procedures applied to solve for the characteristic frequency $\omega$ and separation constant $\lambda$. We present three such methods based on: shooting, series expansion, and Newton-Raphson. The first two methods are typically used in studies of QNMs and the latter we introduce here and have found it to be the most robust when exploring limiting cases. As a powerful check, we find excellent agreement between different methods when more than one is applicable.

\vskip 1cm
\noindent {\bf Shooting.} The first method ``shoots'' for the correct answer in both the angular and radial component.
Regularity of the angular eigenfunctions require that they admit the following expansion
\begin{eqnarray}
S(\theta)&\sim& \theta^{|m-2|}\sum_{n=0} B^L_n(\tilde{\omega},\lambda) \theta^n\,,\quad \theta\sim 0\,,\\
         &\sim&(\pi-\theta)^{|m+2|}\sum_{n=0}B^R_n(\tilde{\omega},\lambda)(\pi-\theta)^n \,,\quad \theta \sim \pi\,,
\end{eqnarray}
at the left- and right-boundaries respectively. The coefficients $B^L_n,B^R_n$ can be extracted from the angular equation
and are functions of the frequency and the separation constant. We typically keep the first six terms in the expansion, 
numerically integrate the solutions towards each other where we match the logarithmic derivative at an
intermediate point. We proceed identically with the radial equation, by imposing conditions~\eqref{bcH} and~\eqref{TheBCs} at the boundaries.
Due to well-known divergences of QNMs at the horizon (stable modes diverge exponentially), we use an {\it analytical},
series expansion close to the horizon and a similar expansion close to spatial infinity.
An example notebook of how the radial equation is dealt with can be found online \cite{Berti:2009kk}.
The method gives stable, convergent results for small black holes, but becomes less accurate for large black holes.
\\ \\
\noindent {\bf Series expansion.} A powerful alternative is based on a series solution of the radial equation which avoids the divergent nature of QNMs at the horizon altogether by factoring the relevant terms~\cite{Horowitz:1999jd,Berti:2009kk}.
For simplicity let us focus on non-rotating BHs in this brief description, the extension to rotating BHs is straightforward. Let us start by re-expressing the boundary condition \eqref{TheBCs} as $-r(r/LR^{(-2)}_{\tilde{\omega}\ell m})'=i\beta R^{(-2)}_{\tilde{\omega}\ell m}$, where primes denote derivative with respect to $r$ and all quantities are evaluated at spatial infinity. Redefine the wavefunction to $R^{(-2)}_{\tilde{\omega}\ell m}=\frac{\Delta_r}{r^5} e^{-i\omega r_*}Z(r)$, with $dr/dr_*=\Delta_r/r^2$.
Then, make the variable change $z=1/r$ and re-write the radial equation as
\begin{equation}
s(z)\frac{d^2Z}{dz^2}+\frac{t(z)}{z-z_\star}\frac{dZ}{dz}+\frac{u(z)}{(z-z_\star)^2}Z=0\,,
\end{equation}
and the boundary conditions as
\begin{equation}
Z'/L=iZ\left(\beta-L\omega\right)\,,
\end{equation}
where primes now denote derivative with respect to $z$ and $z_\star=1/r_+$.

The idea is now to look for a series solution, $Z=a_n(z-z_\star)^n$, where the coefficients $a_n$
are found through the recurrence relation
\begin{equation}
a_n=-\frac{1}{P_n}\sum_{k=0}^{n-1}\left[k(k-1)s_{n-k}+k\,t_{n-k}+u_{n-k}\right]a_k\,,
\end{equation}
where $P_n=n(n-1)s_0+n\,t_0$ and where $s,t,u$ have been expanded in Taylor series around the horizon. 
The boundary condition then translates into
\begin{equation}
\sum a_n(-z_\star)^n\left[1+\frac{n}{h(i\beta-i\omega)}\right]=0\,,
\end{equation}
where $\beta$ is given by either Eq.~(\ref{TheBCsS}) or Eq.~(\ref{TheBCsV}). Extension to rotating geometries is obtained simply by replacing $\omega$ with the corresponding superradiant factor.
\\ \\
\noindent {\bf Newton-Raphson.} We have also developed a novel numerical procedure based on the Newton-Raphson root-finding algorithm that searches for specific quasinormal modes, once a seed solution is given. In order to proceed we first need to recast Eq.~(\ref{MasterAng}) and Eq.~(\ref{MasterRad}) in a different form. Let us introduce the following auxiliary functions:
\begin{align}
&R_{\tilde{\omega} \ell m}^{(-2)}( r) = \left(1-\frac{r_+}{r}\right)^{1-i\varpi} \frac{L}{r}q_1\left(1-\frac{r_+}{r}\right),\\
&S_{\tilde{\omega} \ell m}^{(-2)} (\chi)= \left(1+\frac{\chi}{a}\right)^{\left|\frac{m}{2}+1\right|}\left(\frac{\chi}{a}\right)^{\left|\frac{m}{2}-1\right|}q_2\left(1+\frac{\chi}{a}\right)\,,
\end{align}
where we have implicitly introduced two new compact 
coordinates $y  = 1-r_+/r$ and $\tilde{x} = 1+\chi/a$, which map the problem to the unit square: $(\tilde{x},y)\in (0,1)\times (0,1)$. The boundary conditions on the $q_I$ simply arise from regularity, and translate into four Robin boundary conditions at each integration boundary, \emph{i.e.}
$$
q_I'(0)=a_I\,q_I(0)\quad\text{and}\quad q_I'(1)=b_I\,q_I(1)\,,
$$
where both $a_I$ and $b_I$ are constants and $I=\{1,2\}$. For $q_2$, both $a_2$ and $b_2$ are determined by solving the equations of motion (\ref{MasterAng}) off the singular points $\tilde{x}=\{0,1\}$. $q_1$ on the other hand, is a little more subtle. At $y=0$, we still get the Robin boundary conditions by solving Eq.~(\ref{MasterRad}) off $y=0$, but the condition at $y=1$ is obtained directly from either Eq.~(\ref{TheBCsS}) or Eq.~(\ref{TheBCsV}).

We are now ready to introduce the new numerical procedure that determines $\{q_1,q_2,\omega,\lambda\}$. For the sake of presentation we will only discuss below the case in which we have a single differential equation to solve. The extension to a coupled system like the one above is straightforward.

Consider the following ``nonlinear St\"urm-Liouville'' problem in $\{\mathfrak{f},\tilde{\lambda}\}$:
\begin{equation}
H(\tilde{\lambda})\mathfrak{f}=0\,\quad\text{with}\quad \mathfrak{f}'(0)=a_0\,\mathfrak{f}(0)\,,\quad \mathfrak{f}'(1)=b_0\,\mathfrak{f}(1)\,,
\label{eq:nonlinearsturm}
\end{equation}
where $H(\tilde{\lambda})$ is nonlinear function in $\tilde{\lambda}$, and a linear differential operator in $\mathfrak{f}$ and both $\{a_0,b_0\}$ are constants. In many circumstances $H$ takes the following simple form: $H(\tilde{\lambda}) \mathfrak{f}= H_0 \mathfrak{f}-\tilde{\lambda} H_1 \mathfrak{f}-\tilde{\lambda}^2 H_2\mathfrak{f}$, where each of the $H_i$ is a second order differential operator independent of $\tilde{\lambda}$. The former differential equation is often called a quadratic eigenvalue problem, so long as the constants $\{a_0,b_0\}$ admit a similar expansion. The method we describe below allows for \emph{any} dependence in $\tilde{\lambda}$.

We discretize our Eq.~(\ref{eq:nonlinearsturm}) by introducing a spatial grid $\{y_i\}$, with $N+1$ grid points. Because we are solving for manifestly analytic functions $q_I$, we can readily use a pseudospectral collocation discretization scheme. We choose the Gauss-Chebyshev-Lobatto grid as our collocation points. 
The nonlinear St\"urm-Liouville problem (\ref{eq:nonlinearsturm}) reduces to a nonlinear eigenvalue problem of the form:
\begin{equation}
H_{i,j}(\tilde{\lambda})\mathfrak{f}_j=0\,\quad\text{with}\quad D_{N+1,i}\mathfrak{f}_i=a_0\,\mathfrak{f}_{N+1}\,,\quad D_{1,i}\mathfrak{f}_i=b_0\,\mathfrak{f}_1\,,
\label{eq:nonlinearlambda}
\end{equation}
where $D_{i,j}$ is a Chebyshev differentiating matrix and $H_{i,j}$ is the discretization of the operator $H$. We now introduce a normalization for the eigenvector $\{\mathfrak{f}_i\}$, using an auxiliary constant vector $\{v_i\}$, such that $v_i \mathfrak{f}_i=1$. In all cases, we choose $\{v_i\}$ to have only one nonzero component, which without loss of generality we choose to be the horizon and the south pole located at $y=0$ and $\tilde{x}=0$, respectively.

The procedure is now clear: we promote $\tilde{\lambda}$ to be a parameter to be determined via the Newton-Raphson method. Recall that we have to solve
$$
f(\mathfrak{f}_j,\tilde{\lambda})=
\left\{
\begin{array}{l}
\hat{H}_{i,j}(\tilde{\lambda})\mathfrak{f}_j \\
v_i \mathfrak{f}_i-1
\end{array}
\right\}\,=\,0\,,
$$
where $\hat{H}_{i,j}$ is obtained from $H_{i,j}$ by removing its first and last lines, and substitute them by the last two conditions in Eq.~(\ref{eq:nonlinearlambda}). The Newton-Raphson method states that the correction to our initial guess for $(\{\mathfrak{f}^{(0)}_i\},\tilde{\lambda}^{(0)})$ can be determined by inverting the following \emph{linear} system of equations:
\begin{equation}
\left[
\begin{array}{cc}
\hat{H}_{i,j}(\tilde{\lambda}^{(0)}) & \left.\frac{\partial \hat{H}_{i,j}}{\partial \tilde{\lambda}}\mathfrak{f}_j\right|_{\mathfrak{f}_j = \mathfrak{f}^{(0)}_j\,,\tilde{\lambda} = \tilde{\lambda}^{(0)}} \\
v_j & 0
\end{array}
\right]
\left[
\begin{array}{l}
\delta{\mathfrak{f}}_j\\
\delta{\tilde{\lambda}}
\end{array}
\right] = \,-\,\left[\begin{array}{l}
\hat{H}_{i,j}(\tilde{\lambda}^{(0)})\mathfrak{f}^{(0)}_j \\
v_j \mathfrak{f}^{(0)}_j-1
\end{array}
\right]\,.
\end{equation}
We then iterate this procedure until $|\delta{\mathfrak{f}}_j|$ and $|\delta\tilde{\lambda}|$ are below some tolerance, which in this manuscript we take to be $10^{-30}$. All computations using this method were performed with octuple precision, which is particularly relevant for small black holes.

Our results have been benchmarked using previous results in the literature, specifically for scalar field perturbations
of Kerr-AdS BHs \cite{Cardoso:2004hs,Cardoso:2006wa,Uchikata:2009zz}. In particular, we recover to all significant digits the numerical results
reported in Ref.~\cite{Uchikata:2009zz}. Furthermore, we recover all known results from gravitational perturbations of Schwarzschild-AdS BHs with the same boundary conditions~\cite{Cardoso:2001bb,Cardoso:2003cj,Michalogiorgakis:2006jc,Dias:2013sdc}.

Finally, we note that an important symmetry of the relevant perturbation equations and boundary conditions for QNMs is that if $(\omega, \lambda)$ is a solution for a given $m$ then $(-\omega^*,\lambda^*)$ is a solution for $-m$. As such, we will only discuss positive real part modes, with the understanding that they come in complex conjugate pairs.

\section{QNMs and superradiance in Kerr-AdS: results \label{sec:QNMs}}
In this section we present the numerical results obtained, make contact with some analytical results, 
and discuss implications with the phenomena of superradiance.

\subsection{Comparison between analytical and numerical results \label{sec:CompareNumAnal}}

The angular \eqref{MasterAng} and radial \eqref{MasterRad} equations constitute a system of ordinary
differential equations 
coupled through the frequency $\tilde{\omega}$ and angular $\lambda$ eigenvalues that cannot be solved analytically when $M,a\neq 0$. For this reason, we solve these equations using the numerical methods outlined in Section \ref{sec:num}.
There is however a regime where we can use a matched asymptotic expansion procedure to get an approximate analytical solution for the QNM and superradiant instability frequency spectra. 
This perturbative analytical computation provides useful physical insights about the system and is valuable to check our numerical results. We leave the details of this analytical construction to Appendix \ref{sec:AnalyticalQNMAppendix} and present here only the main outcome of the computation and its comparison with the numerical results.

As justified in Appendix \ref{sec:AnalyticalQNMAppendix}, the perturbative analytical results are valid  in the regime of parameters where 
\begin{eqnarray}\label{AnalyticApprox}
&&\qquad \frac{r_+}{L} \ll 1\,\quad \Rightarrow \quad \frac{a}{L} \ll 1\,,\quad a \tilde{\omega} \ll 1\,, \quad  r_+\tilde{\omega} \ll 1 \,; \nonumber\\
&&\qquad \frac{a}{r_+}\ll 1\,.
\end{eqnarray}
i.e., for Kerr-AdS black holes with small horizon radius in AdS radius units and even smaller rotation parameter, and for perturbations whose wavelength is much bigger than the black hole lengthscales.

\begin{figure}[ht]
\centering
\includegraphics[width=.47\textwidth]{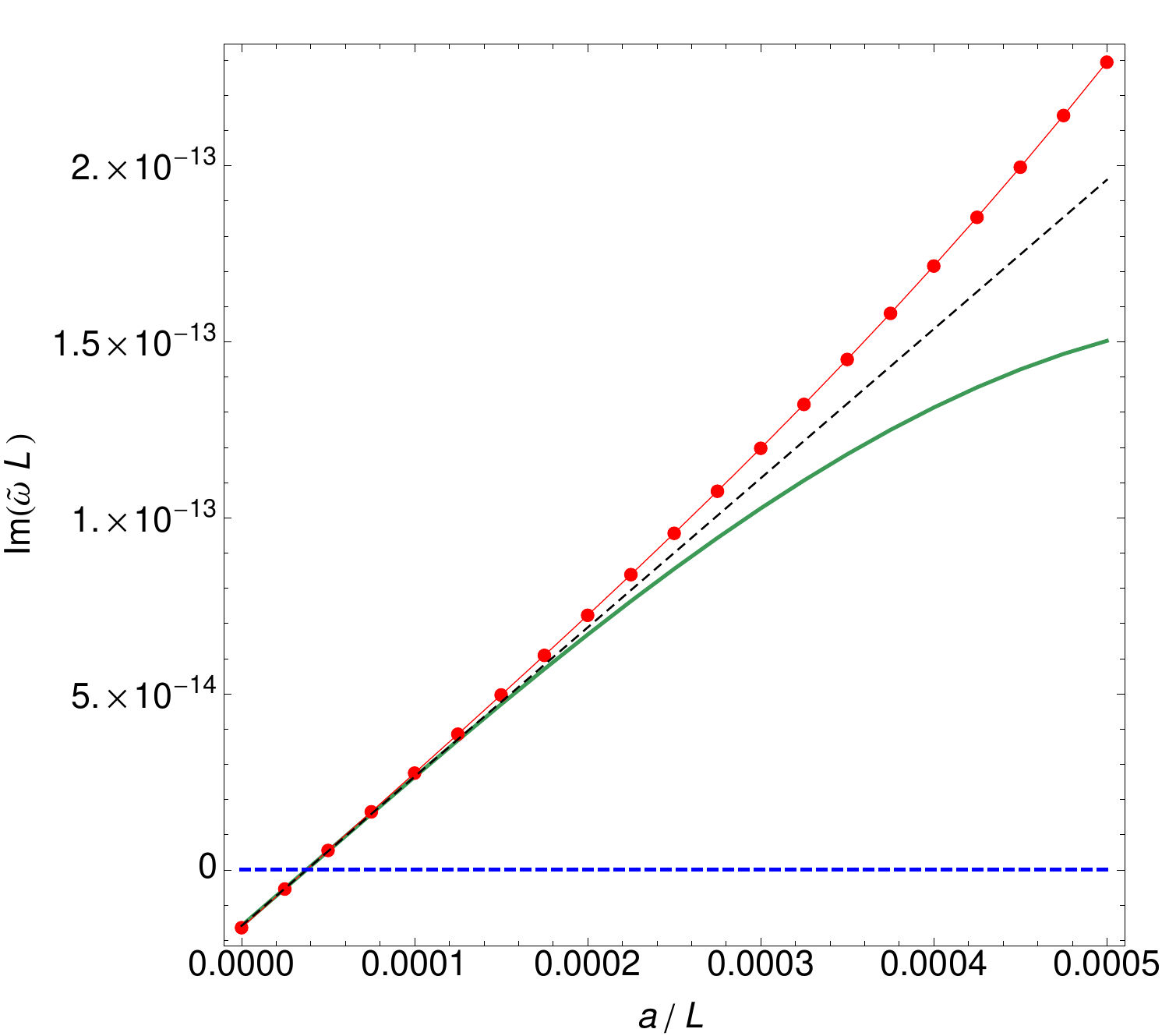}
\hspace{0.5cm}
\includegraphics[width=.47\textwidth]{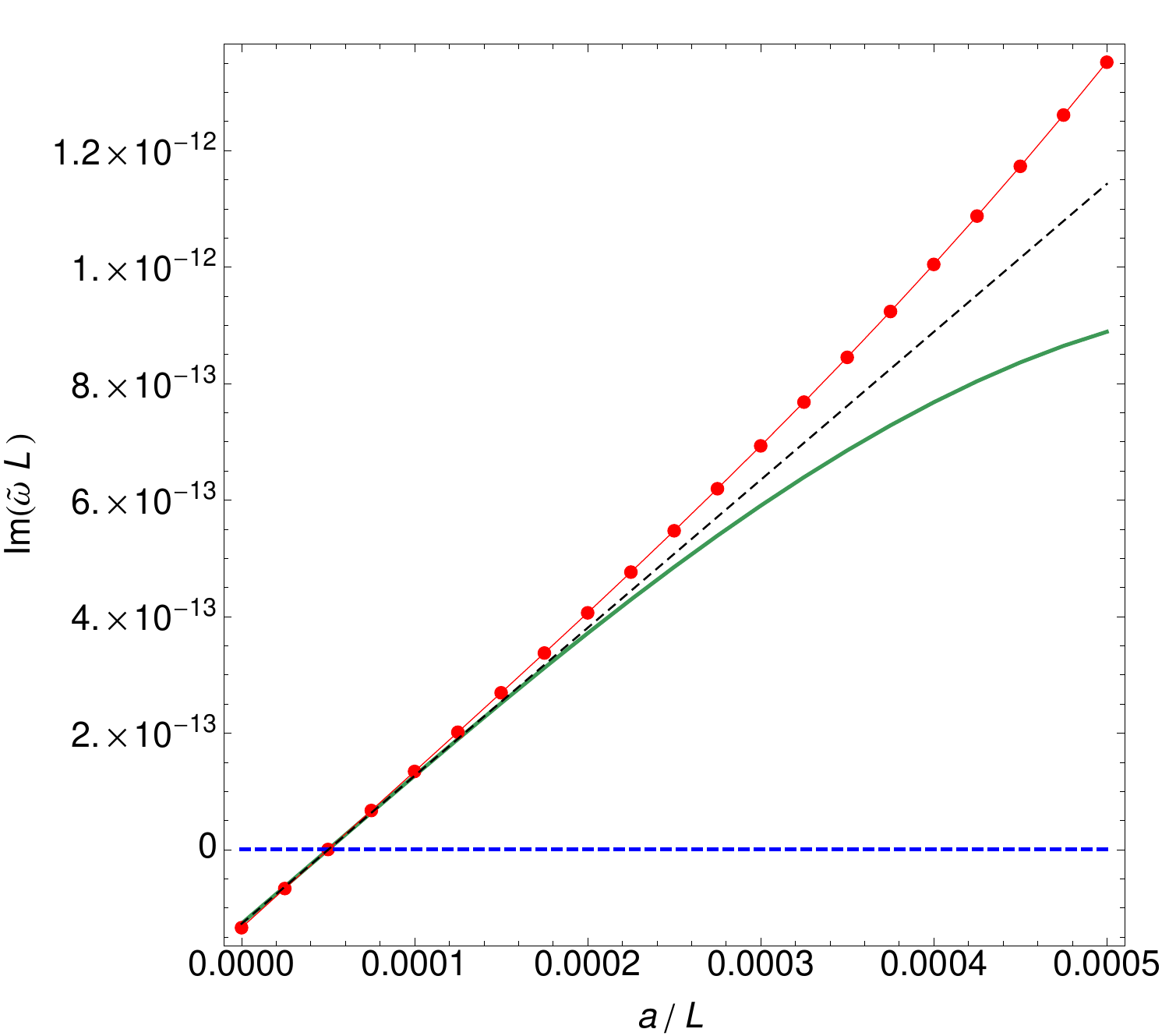}
\caption{
Imaginary part of the QNM frequency as a function of the rotation parameter $a/L$, for fixed horizon radius $r_+/L=0.005$,  for scalar ({\it Right Panel}) and vector modes  ({\it Left Panel}). This is for $\ell=2$ modes with no radial overtone.  These are an example of the QNM spectrum in the regime $a/L<r_+/L \ll 1$ where the analytical matching analysis is valid and its approximated results can be used to both test our numerical code (valid in any regime), and estimate more precisely the regime of validity of the analytical approximation. The red dots are the exact results from our numerical code. The green curve is the numerical solution of the matching transcendental equation \eqref{FreqMatchingl2}, while the dashed black curve is the approximated analytical solution \eqref{SnormalM} or \eqref{VnormalM} of \eqref{FreqMatchingl2}.
In both figures there is a critical rotation where ${\rm Im}(\tilde{\omega}  L)=0$ and ${\rm Re}(\tilde{\omega})-m\Omega_H\simeq 0$ to within $0.01\%$. For lower rotations the QNMs are damped and with ${\rm Re}(\tilde{\omega})-m\Omega_H>0$, while for higher rotations we have unstable superradiant modes  with ${\rm Re}(\tilde{\omega})-m\Omega_H<0$.}\label{Fig:matchNum}
\end{figure}  

In Appendix \ref{sec:AnalyticalQNMAppendix} we find that the matched asymptotic expansion analysis indicates 
that the frequency spectrum is quantized by the condition \eqref{FreqMatching}, for a generic mode with quantum numbers $\ell$ and $m$.
This frequency quantization condition simplifies considerably when we choose a particular harmonic $\ell$.
For instance, for the lowest harmonic $\ell=2$, the condition \eqref{FreqMatching} reads 
\begin{eqnarray}\label{FreqMatchingl2}
&& \hspace{-3.2cm} 
i (-1)^{L \tilde{\omega} +1} L^{-5}\left(r_+-\frac{a}{r_+^2}\right)^5 L \tilde{\omega}  \left(L^2 \tilde{\omega} ^2-1\right)\left(L^2 \tilde{\omega} ^2-4\right) \Gamma (5-2 i \varpi ) \nonumber\\
&&   \hspace{3cm} +5400 \left[\varepsilon_j+(-1)^{L \tilde{\omega} }\right] \Gamma (-2 i \varpi )=0\,,
\end{eqnarray}
where the superradiant factor $\varpi$ is defined in \eqref{superradiance factor}, and  $\varepsilon_j=1$ describes scalar modes with the BC  \eqref{TheBCsSa0} while  $\varepsilon_j=-1$ represents  vector modes with the BC  \eqref{TheBCsVa0}.
We can find the frequency that solves this transcendental equation numerically using a standard root-finder routine (for instance {\it Mathematica}'s
built-in {\it FindRoot} routine). Alternatively we can also provide an approximate analytic solution, still in the limit of $a/L\ll r_+/L \ll 1$, assuming that the frequency has a double expansion in the rotation and in the horizon radius, $ \tilde{\omega}(a,r_+) L= \sum _{j=0}^{n}\left( \frac{a}{L}\right)^j \sum _{i=0}^p \tilde{\omega}_{j,i}\left( \frac{r_+}{L}\right)^{i}$, and solving progressively \eqref{FreqMatchingl2} in a series expansion in $a/L$ and $r_+/L$. Here, $\tilde{\omega}_{0,0}$ is the global AdS frequency (see footnote \ref{foot}). Namely, the fundamental (no radial overtone) $\ell=2$ scalar and vector normal mode frequencies are $\tilde{\omega}_{0,0}^{({\rm s})} =3/L$ and $\tilde{\omega}_{0,0}^{({\rm v})} =4/L$, respectively. In the regime \eqref{AnalyticApprox} we work in this subsection, the correction to the real part of the frequency is very small (compared with $\tilde{\omega}_{0,0}$) and \eqref{FreqMatchingl2} fixes the the imaginary part of the frequency for fundamental $\ell=2$ modes to be 
\begin{eqnarray}
&&  \hspace{-1.7cm} 1) \quad \hbox{Scalar modes:} \quad  {\rm Im}(\tilde{\omega}  L)\simeq \frac{16 }{15 \pi }\left[-\frac{3 r_+^6}{L^6}+\frac{ m \,a \,r_+^4}{L^5}\left(1+15 (5 \gamma -7) \,\frac{r_+^2}{L^2}\right)\right] +\cdots \,, \label{SnormalM}
\\ 
&& \hspace{-1.7cm}  2) \quad \hbox{Vector modes:} \quad  {\rm Im}(\tilde{\omega}  L)\simeq \frac{96}{15 \pi }\left[-\frac{4 r_+^6}{L^6}+\frac{m\, a\, r_+^4}{L^5}\left(1+\frac{80 (5 \gamma -7)}{3}\,\frac{r_+^2}{L^2}\right)\right] +\cdots \,,\label{VnormalM}   
\end{eqnarray}
where $\gamma\simeq 0.577216$ is the Euler-Mascheroni constant. For both scalar and vector modes the imaginary part of the frequency starts negative for $a=0$, consistent with the fact that QNMs of Schwarzschild-AdS are always damped. However, as $a/L$ increases, ${\rm Im}(\tilde{\omega}  L)$ increases. A good check of our analytical matching analysis is that we find that at the critical rotation where the crossover occurs, i.e. ${\rm Im}(\tilde{\omega}  L)=0$, one has ${\rm Re}(\tilde{\omega})-m\Omega_H\simeq 0$ to within $0.01\%$. For smaller rotations one has  ${\rm Re}(\tilde{\omega})-m\Omega_H>0$ and for higher rotations one has  ${\rm Re}(\tilde{\omega})-m\Omega_H<0$ and ${\rm Im}(\tilde{\omega}  L)>0$. Therefore, the instability which is triggered at large rotation rates has a superradiant origin since the superradiant factor becomes negative, $\varpi<0$ precisely when the QNMs go from damped to unstable.
These analytical matching results provide also a good testbed check to our numerics. Indeed we find that our analytical and numerical results have a very good agreement in the regime of validity of the matching analysis. This is demonstrated  in Fig. \ref{Fig:matchNum} where we plot our numerical and analytical results for the fundamental $\ell=2$ scalar and vector modes. As a rough reference we can take this to be $r_+/L<5\times 10^{-3}$ and $a/L<10^{-4}$. (A similar analysis that lead to the results \eqref{FreqMatchingl2}-\eqref{VnormalM} can be repeated for any other harmonic starting from \eqref{FreqMatching}).

\subsection{Properties of superradiant unstable modes and QNMs \label{sec:ResultSuperQNMs}}

We are now ready to present the properties of  the superradiant unstable modes and QNMs for generic solutions in the parameter space. We use the numerical methods described in Section \ref{sec:num} to find the solution of the coupled ODE  angular \eqref{MasterAng} and radial \eqref{MasterRad} equations that describe the most general linear perturbation of a Kerr-AdS BH. We first present the gravitational scalar perturbations that obey the BCs \eqref{TheBCsS}, and then the gravitational vector perturbations that obey the BCs \eqref{TheBCsV}.

Consider a Kerr-AdS BH parametrized by particular values of the gauge invariant parameters  $\{ R_+/L,\Omega_h L \}$ described in the end of Section \ref{sec:KerrAdSbh}. A generic perturbation can have a frequency with negative, positive, or vanishing imaginary part.  Quasinormal modes are damped, Im$(\omega)<0$, whereas unstable modes grow exponentially in time, Im$(\omega)>0$. Thus, a particularly important set of modes, if present, are the marginal modes that define the stability boundary in a phase diagram. The marginal mode (or onset mode) curve is defined to be the locus of points in the parameter space $(R_+/L, \Omega_h L)$ for which a mode with Im$(\omega) = 0$ exists. There will be a marginal mode curve for each distinct pair of wave numbers $\{\ell,m\}$ resulting in an instability. To understand the nature of this instability it is useful to look into another useful characterization of linear perturbations. It comes from considering the difference between the real part of the frequency and $m\Omega_h$, which determines the sign of the the energy and angular momentum fluxes the perturbation carries through the future horizon; see Appendix \ref{sec:Fluxes}\footnote{Note that reflecting boundary conditions at the conformal boundary enforces the vanishing of the flux there; see Appendix \ref{sec:Fluxes}}. Modes with Re$(\omega) > m \Omega_h$ carry positive flux through the horizon, whereas modes with Re$(\omega) < m \Omega_h$ carry negative flux across the horizon, and are called superradiant. Vanishing flux at the horizon requires Re$(\omega) = m \Omega_h$. We find that  Re$(\omega) = m \Omega_h$ whenever  Im$(\omega) = 0$ and that  Re$(\omega) < m \Omega_h$ when Im$(\omega)<0$. Therefore, unstable modes in Kerr-AdS are always associated to the superradiant instability. 

\begin{figure}[t]
\centerline{
\includegraphics[width=1.02\textwidth]{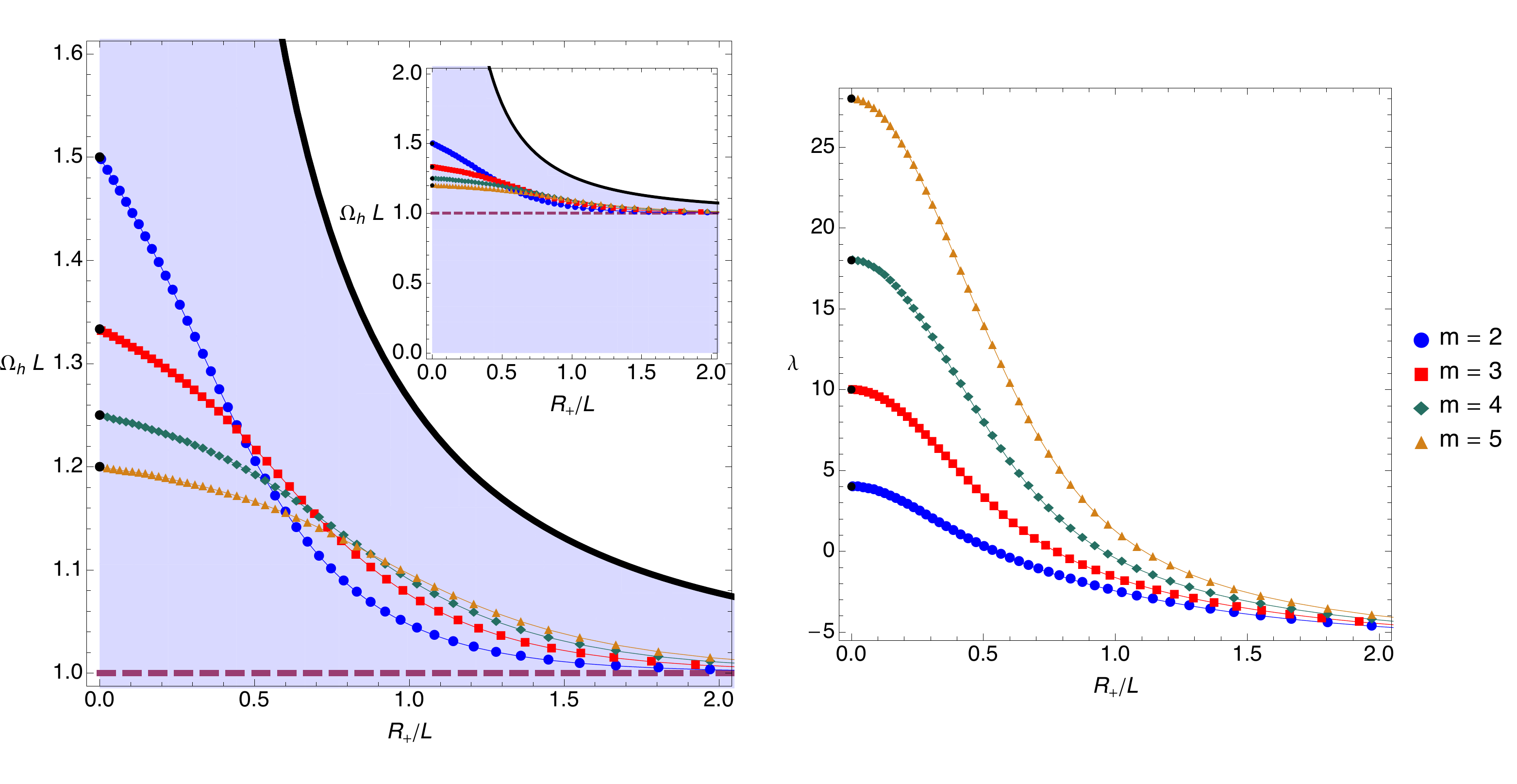}
}
\caption{The onset of superradiance for the first $\ell=m=2,3,4,5$ {\it scalar} modes of the Kerr-AdS BH. The {\it left panel} shows the OC in the phase diagram described by the gauge invariant  parameters $(R_+/L,\Omega_h L)$ (the inset plot zooms out the main plot to show an enlarged view of the parameter space).  Regular Kerr-AdS BHs exist in the blue shaded area all the way up to the black curve where extremality is attained.   In the {\it right panel} we show the value of the angular eigenvalue $\lambda$ as a function of the areal radius $R_+/L$ as we move along the OC. In both plots, the larger black points on the left with $R_+/L=0$ are fixed by the properties \eqref{wAdSscalar} of scalar normal modes of global AdS.
\label{Fig:KAdSscalarOnset}}
\end{figure}

As important illustrative examples, in the {\it left panel} of Fig. \ref{Fig:KAdSscalarOnset} we identify the superradiant onset curves (OC) for $\ell=m$ scalar modes (with vanishing radial overtone) in the phase diagram of Kerr-AdS BHs. The axes are given by the gauge invariant horizon radius $ R_+/L$ and the horizon angular velocity $\Omega_h L$ (for the frame that does not rotate at infinity), as previously introduced in Fig. \ref{Fig:domain}. Regular Kerr-AdS BHs exist in the blue shaded area, starting at $\Omega_h L=0$ and all the way up towards the black curve where extremality is attained. We identify the OC for the scalar modes with $\ell=m=2,3,4,5$. BHs that are above a particular $\ell=m$ OC are superradiantly unstable to modes with those particular values of $\ell=m$, while BHs below a particular OC are stable to the associated modes. For completeness, in the {\it right panel} of Fig. \ref{Fig:KAdSscalarOnset} we plot the angular eigenvalue $\lambda$ along the superradiant OC. Since Im$(\omega)=0$ along this OC, it follows from the mathematical structure of the coupled equations that we must also have Im$(\lambda)=0$.

The OCs have some properties that merit a detailed discussion. First, in both plots of Fig. \ref{Fig:KAdSscalarOnset} the large black points on the left at $R_+/L =0 $ are computed analytically and 
serve as additional checks for the numerical code. They describe the scalar normal mode frequencies and the associated angular eigenvalues of global AdS given by \cite{Dias:2011ss,Dias:2012tq},   
\begin{equation}\label{wAdSscalar}
L\,\omega_{\rm s}^{AdS}=1+\ell+2p \,, \qquad \lambda=\ell(\ell+1)-2\,,
\end{equation}
where $p=0,1,2,\cdots$ is the radial overtone (number of radial nodes). 
In more detail, to get the black points in the  {\it left panel} of Fig. \ref{Fig:KAdSscalarOnset} we use the superradiant onset condition to find $\Omega_h{\bigl |}_{R_+=0}=\omega_{s}^{AdS} / m $ and we set $p=0, \ell=m$, i.e.
\begin{equation}\label{OmAdSscalar}
L\, \Omega_h{\bigl |}_{R_+=0}=1+\frac{1}{m} \,,
\end{equation}
Note that given a $\{\ell,m\}$ pair there is an OC for each radial overtone $p$, but $p>0$ curves always lie above the $p=0$ curve, and therefore $p=0$ modes are the first to go unstable as the rotation is increased. For this reason only the $p=0$ curves are plotted.

The OCs always have $\Omega_h L >1$, monotonically approaching $\Omega_h L \to1$ (from above) asymptotically as $R_+/L \to \infty $, where all the scalar superradiant OCs pile up. This means that only  $\Omega_h L >1$ BHs can be unstable to superradiance, a property that was first proven in \cite{Hawking:1999dp}. 

Finally, note that for small BHs (say with $R_+/L\lesssim 0.45$) as $\ell=m$ increases the corresponding superradiant OC lowers. This means that, e.g.  we can have small BHs (those in the triangle-like region between the $\ell=m=2$ and $\ell=m=3$ curves) that are stable to $\ell=m=2$ modes but unstable to all other $\ell=m\geq 3$ modes, or e.g. BHs that are stable to  $\ell=m=2$ and  $\ell=m=3$ but always unstable to all other $\ell=m\geq 4$ modes. However, as the areal radius grows we find that the OCs start crossing each other. For example, the $\ell=m=2$ curve crosses the $\ell=m=3$ curve at  $R_+/L\sim0.45$ and for higher radius it crosses the $\ell=m=4$ and then the  $\ell=m=5$ curve. So, e.g. at $R_+/L=1$ the $\ell=m=2$ OC is below the three OCs $\ell=m=3,4,5$. This means that at this radius we can have Kerr-BHs that are unstable to $\ell=m=2$ modes but {\it not} to $\ell=m=3,4,5$ modes.

At first sight, this is of course exciting as it seems to indicate that there is a region of parameter space where Kerr-BHs are unstable to $\ell=m=2$ modes but stable to {\it any} other superradiant modes, with obvious consequences for the endpoint of the superradiant instability. However, this is not the case. Indeed, first notice that as $\ell=m \to \infty$  the corresponding OC still starts precisely at the point defined by \eqref{OmAdSscalar}. Thus, as $\ell=m$ grows large, its threshold modes are described by an OC that progressively approaches the line $\Omega_h L=1$,  becoming  horizontal in the limit $\ell=m\to +\infty$. Therefore as the BH rotation is increased, the first modes that become superradiantly unstable are the $m\to \infty$ modes. The conclusion that  $m\to +\infty$  modes are the ``first" to become unstable was first presented in the equal angular momenta Myers-Perry BHs in \cite{Kunduri:2006qa}. Furthermore, as we shall discuss later, all vector modes will be superradiantly unstable.

As stated previously, in the {\it left panel} of Fig. \ref{Fig:KAdSscalarOnset}, BHs that are above a particular $\ell=m$ OC are superradiant unstable to those particular $\ell=m$ modes. That is, their perturbations have frequencies with  Im$(\omega)>0$ and Re$ (\omega)<m\Omega$. On the other hand, BHs below a particular OC are damped and thus stable  (when perturbed these BHs return to equilibrium via the emission of QNMs with Im$(\omega)<0$ and Re$ (\omega)>m\Omega$). 

\begin{figure}[ht]
\centering
\includegraphics[width=.47\textwidth]{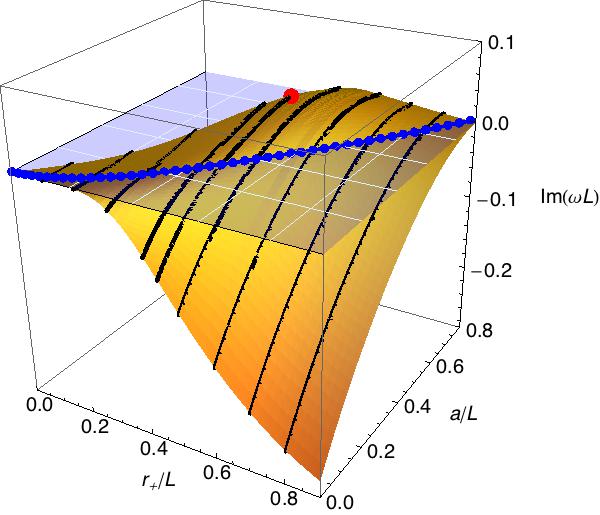}
\hspace{0.5cm}
\includegraphics[width=.47\textwidth]{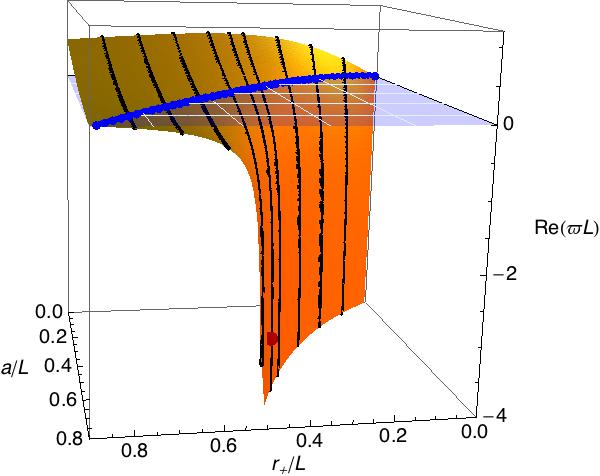}
\caption{Superradiant modes and QNMs for the $\ell=m=2$ scalar harmonic. The {\it left panel} plots the imaginary part Im$(\omega)$ of the frequencies while the {\it right panel} shows the real part of the superradiant factor, i.e. $(\text{Re}(\omega L)-m\Omega_h L)/(4\pi T_h)$, as a function of the horizon radius $r_+/L$ and rotation $a/L$ parameters. The blue curve is the superradiant OC with Im$(\omega)=0$ and  Re$(\varpi)=0$. The large red point signals the Kerr-AdS BH that is most unstable to scalar superradiance  described by \eqref{scalar:MaxInstab}..
The black curves have constant radius $r_+/L=0.1;\, 0.2;\, 0.3;\, 0.4;\, 0.445;\, 0.5;\, 0.6;\, 0.7;\, 0.8$. These plots are discussed in more detail in the text. }
\label{Fig:ScalarFreq}
\end{figure}

\begin{figure}[ht]
\centering
\includegraphics[width=.47\textwidth]{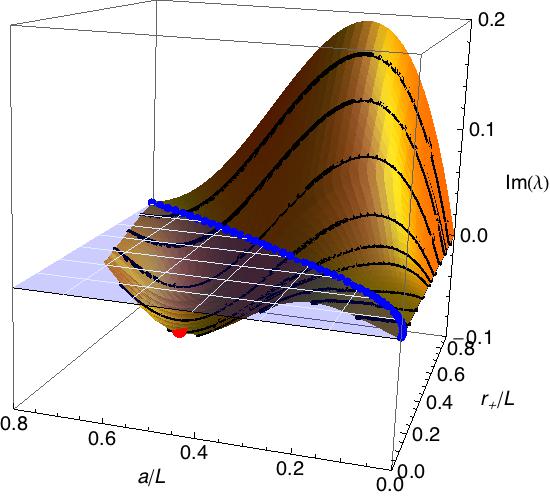}
\hspace{0.5cm}
\includegraphics[width=.47\textwidth]{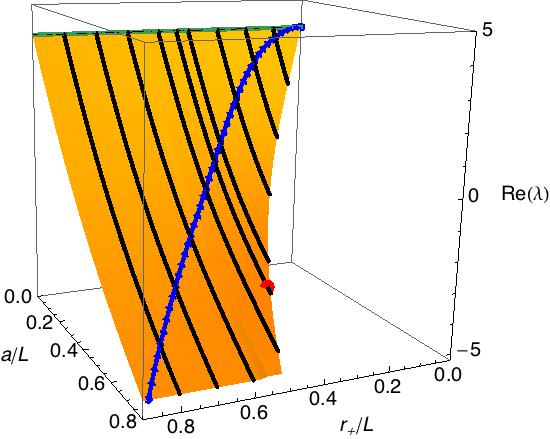}
\caption{Imaginary ({\it left panel}) and  real ({\it right panel}) part of the angular eigenvalues  of the superradiant modes and QNMs of the $\ell=m=2$ {\it scalar} harmonic whose frequencies are shown in Fig. \ref{Fig:ScalarFreq}. The color coding of the lines/points is the same as Fig. \ref{Fig:ScalarFreq}.}
\label{Fig:ScalarAng}
\end{figure}

Having studied the OCs for scalar modes with $\ell=m$, we now turn to consider one particular mode throughout a region of the parameter space to gain more insight into the stability properties of these black holes. A natural mode to consider is the $\ell=m=2$ one, as this is the mode with the largest value of the growth rate Im$(\omega)$ found in our study. The imaginary and real parts of the $\ell=m=2$ scalar mode frequencies are plotted in Fig. \ref{Fig:ScalarFreq}, and the imaginary and real part of the associated angular eigenvalues is shown in Fig. \ref{Fig:ScalarAng}. These quantities are plotted as a function of the dimensionless horizon radius $r_+/L$ and rotation $a/L$ and they define a 2-dimensional surface. To extract more efficiently the relevant physics, we plot in the {\it right panel} of Fig. \ref{Fig:ScalarFreq} is the real part of the superradiant factor Re$(\varpi)=\left(\text{Re}(\omega)-m\Omega_h\right)/(4\pi T_h)$, as introduced in \eqref{superradiance factor}. In all these plots the blue curve is the  $\ell=m=2$ OC  already identified in the phase diagram of Fig. \ref{Fig:KAdSscalarOnset}. To guide the eye (when appropriate) we draw an auxiliary plane with a grid that intersects the physical 2-dimensional surface along the OC and that has Re$(\varpi)=0$, Im$(\omega)=0$, and Im$(\lambda)=0$. We also plot some black curves at constant radius $r_+/L$.

In the {\it left panel} of Fig. \ref{Fig:ScalarFreq}, modes that are above the auxiliary plane grid are superradiant unstable modes.  In the {\it right panel} of Fig. \ref{Fig:ScalarFreq} and in the  {\it left panel} of Fig. \ref{Fig:ScalarAng} they correspond to the surface region below the auxiliary plane grid. Finally, in the  {\it right panel} of Fig. \ref{Fig:ScalarAng} these unstable modes are described by the surface region ``below" the blue line. In the four plots, the superradiant unstable surface region is a 2-dimensional surface bounded by the superradiant OC (blue line) and by the extremality curve (where the black curves at constant radius end).\footnote{Note that in the {\it right panel} of Fig. \ref{Fig:ScalarFreq} the shown surface would extend for smaller negative values of $Re(\varpi)$ but we stop it at $Re(\varpi)=-4$ for better visualization.} In all these plots, the surface region that starts at the blue OC that is complementary to the unstable region describes the QNMs of the Kerr-AdS BH. 

An important feature of the gravitational scalar superradiant instability concerns the order of magnitude of its timescale $\tau\sim 1/{\rm Im}(\omega)$. Inspecting the data we find that the maximum growth rate of the instability  is reached in a neighborhood of the point $\{r_+/L,a/L \}_{\rm max} \simeq \{ 0.445\pm 0.020, 0.589\pm 0.020 \}$ where the frequency is given by $\omega L \sim 1.397+ 0.032\, i$. So, the maximum growth rate for the scalar superradiant instability and the gauge invariant properties of the BH where it is attained are
\begin{equation}\label{scalar:MaxInstab}
\hbox{Scalar:}\quad
 \{ R_+/L,\Omega_h L \}\sim \{ 0.914,1.295 \}\,,\quad {\rm Im} (\omega L) \sim 0.032\,, \quad {\rm Re} (\varpi L) \sim -3.247\,.
\end{equation}
This maximum is denoted with a large red dot  in the plots of  Fig. \ref{Fig:ScalarFreq} and  Fig. \ref{Fig:ScalarAng}.
Note that this maximum occurs close to extremality but not at it. In particular, if we plot the instability growth rate  as a function of the rotation parameter $a/L$ at fixed radius (e.g. $r_+/L=0.445$), we find that, typically, starting from the onset the instability timescale first increases,  reaches a maximum for $a/L$ close to extremality, and then  decreases as we approach the $T_h=0$ Kerr-AdS BH. 
 
\begin{figure}[t]
\centerline{
\includegraphics[width=1.02\textwidth]{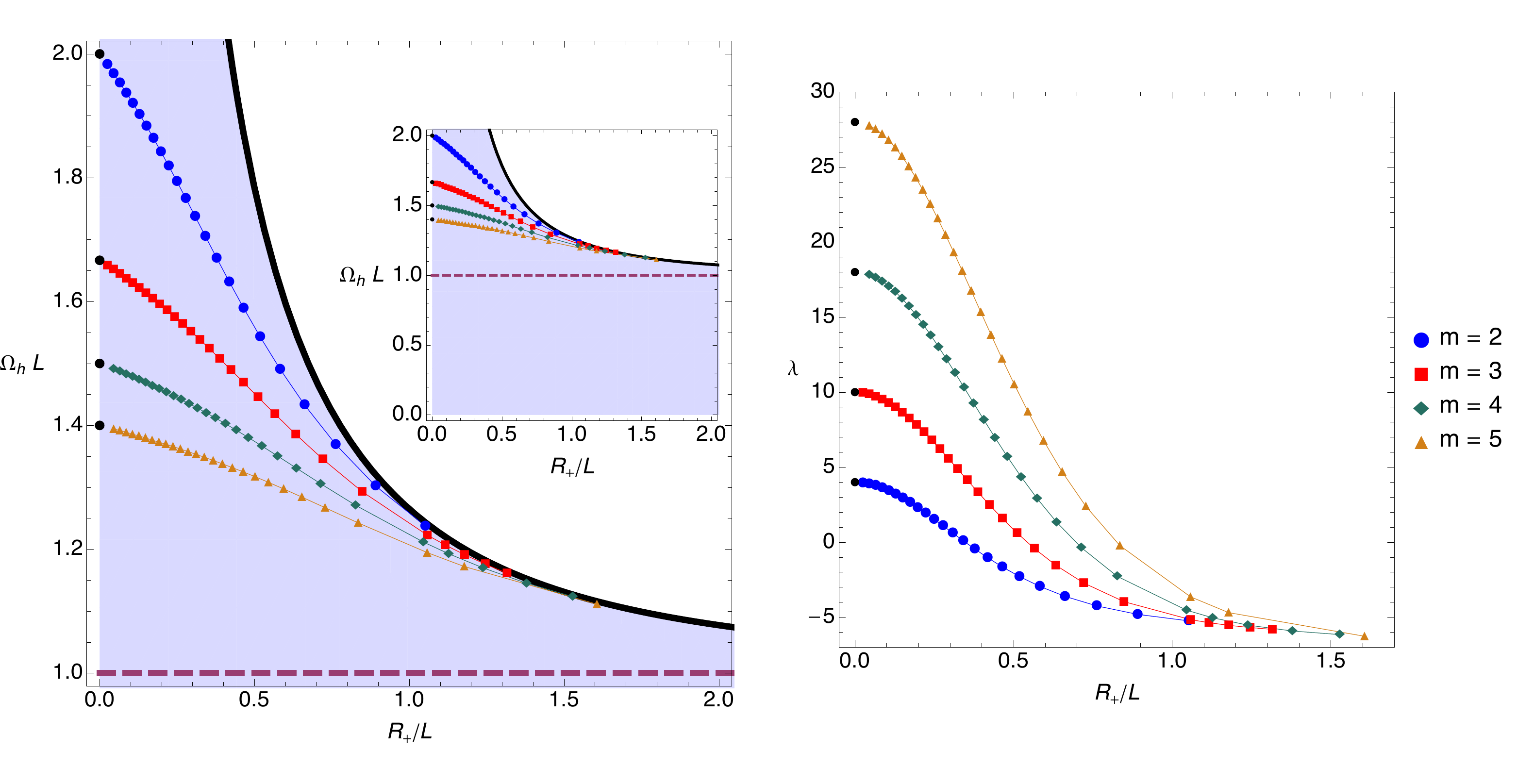}
}
\caption{Onset of superradiance for the first $\ell=m=2,3,4,5$ {\it vector} modes of the Kerr-AdS BH. The {\it left panel} shows the OC in the $(R_+/L,\Omega_h L)$ phase diagram (the inset plot zooms out the main plot).  The {\it right panel} shows how the angular eigenvalue $\lambda$ varies with $R_+/L$ along the OC. In both plots, the larger black points on the left with $R_+/L=0$ are fixed by the properties \eqref{wAdSvector} of vector normal modes of global AdS.
\label{Fig:KAdSvectorOnset}}
\end{figure}

Consider now the gravitational vector modes which obey the BCs \eqref{TheBCsV}. The {\it left panel} of Fig. \ref{Fig:KAdSvectorOnset} displays the phase diagram of Kerr-AdS BHs with the OCs for the $\ell=m=2,3,4,5$ vector modes displayed (again, only curves with vanishing radial overtone are shown). As in the scalar case, BHs that are above a particular $\ell=m$ vector OC are superradiantly unstable to modes with those particular values of $\ell=m$, while BHs below a particular OC are stable to the associated modes. In the {\it right panel} of Fig. \ref{Fig:KAdSvectorOnset} we plot the angular eigenvalue $\lambda$ along the  OC for vector modes.

\begin{figure}[ht]
\centering
\includegraphics[width=.47\textwidth]{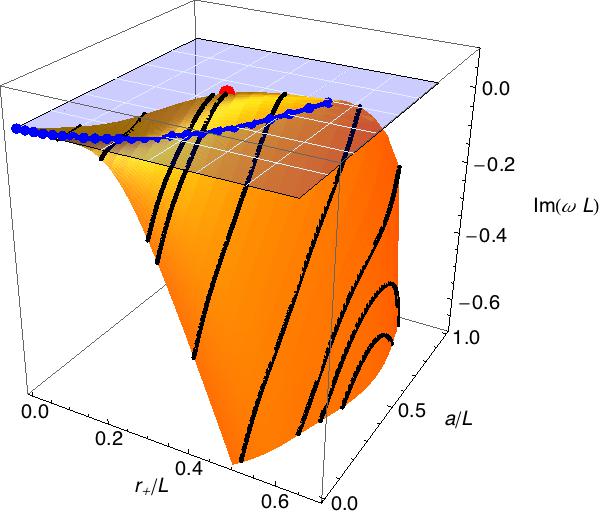}
\hspace{0.5cm}
\includegraphics[width=.47\textwidth]{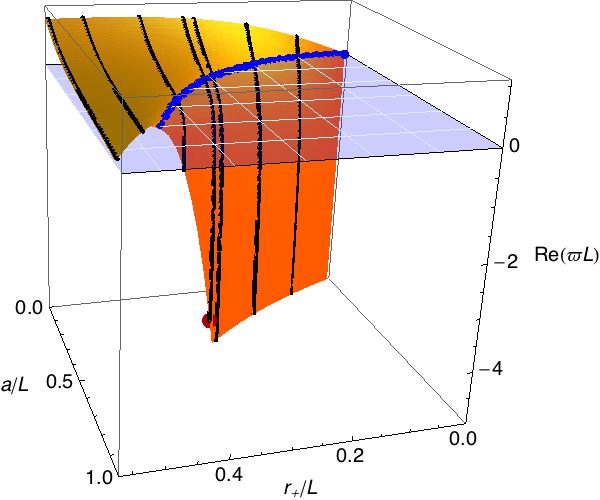}
\caption{Superradiant modes and QNMs for the $\ell=m=2$ {\it vector} harmonic. The {\it left panel} plots the imaginary part $Im(\omega)$ of the frequencies while the {\it right panel} shows the real part of the superradiant factor, i.e. $(Re(\omega L)-m\Omega_h L)/(4\pi T_h)$, as a function of the horizon radius $r_+/L$ and rotation $a/L$ parameters. The blue curve is the superradiant OC with $Im(\omega)=0$ and  $Re(\varpi)=0$. The large red point signals the Kerr-AdS BH that is most unstable to vector superradiance described by \eqref{vector:MaxInstab}.
The black curves have constant radius $r_+/L=0.1;\, 0.2;\, 0.3;\, 0.325;\, 0.4;\, 0.5;\, 0.565;\, 0.585;\, 0.6$ (the later two only in the {\it left panel}). These plot are discussed in more detail in the text. }
\label{Fig:VectorFreq}
\end{figure}  

The large black points at $R_+/L =0 $, in both plots of Fig. \ref{Fig:KAdSvectorOnset}, describe the vector normal modes of global AdS, namely \cite{Dias:2011ss,Dias:2012tq},   
\begin{equation}\label{wAdSvector}
L\,\omega_{\rm v}^{AdS}= 2+\ell+2p \:\:(p=0,1,2,\cdots)\,, \qquad \lambda=\ell(\ell+1)-2\,. \\
\end{equation}
Together with the superradiant onset condition (with $p=0$ and $\ell=m$) these normal modes give the black points of Fig. \ref{Fig:KAdSvectorOnset},
\begin{equation}\label{OmAdSvector}
L \Omega_h{\bigl |}_{R_+=0}=1+\frac{2}{m} \,,
\end{equation}

\begin{figure}[ht]
\centering
\includegraphics[width=.47\textwidth]{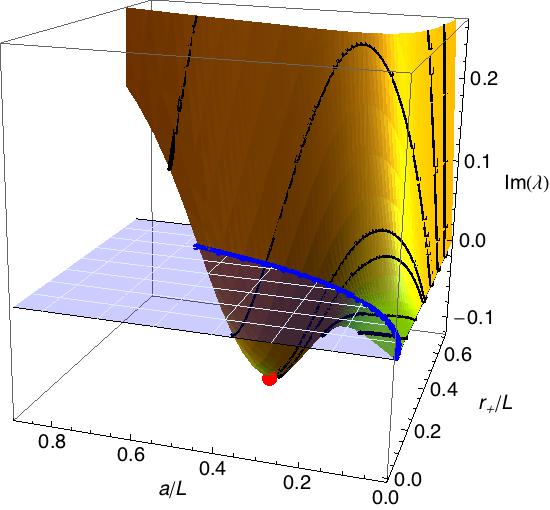}
\hspace{0.5cm}
\includegraphics[width=.47\textwidth]{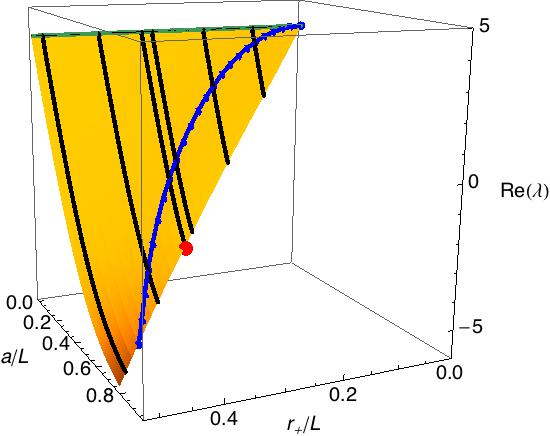}
\caption{Imaginary ({\it left panel}) and  real ({\it right panel}) part of the angular eigenvalues  of the superradiant modes and QNMs of the $\ell=m=2$ {\it vector} harmonic whose frequencies are shown in Fig. \ref{Fig:VectorFreq}. The color coding of the lines/points is the same as Fig. \ref{Fig:VectorFreq}.}
\label{Fig:VectorAng}
\end{figure}  

As in the scalar case, the vector OCs always have $\Omega_h L >1$ but contrary to the scalar case, these curves always end at extremality and the OCs for different $\ell=m$ never cross each other. In particular, this means that a BH that is unstable to $\ell=m=2$ modes must also be  unstable to all $\ell=m\geq 3$ modes. As $\ell=m$ grows, the curves hit extremality at a higher areal radius $R_+/L$ and they approach the  $\Omega L =1$ line. Modes with $m\to +\infty$ reach extremality only in the limit $R_+/L\to +\infty$.

To discuss details of the  superradiant and quasinormal modes of the vector sector, we focus again our attention in the $\ell=m=2$  case. The superradiant and QNM properties can be read from the plots of  Fig. \ref{Fig:VectorFreq} (imaginary and real part of the frequencies) and in  Fig. \ref{Fig:VectorAng} (imaginary and real part of the angular eigenvalues). We use a similar color coding and visualization angle as the ones used in the scalar case. Therefore, in all these plots the blue curve is the OC already studied in Fig. \ref{Fig:KAdSvectorOnset}; again the auxiliary plane with a grid intersects the physical 2-dimensional surface along the OC and helps visualizing the separation between unstable superradiant modes (Im$(\omega)>0$ and Re$ (\omega)<m\Omega$) and damped QNMs (Im$(\omega)<0$ and Re$ (\omega)>m\Omega$); and we plot some black curves at constant radius $r_+/L$. It follows that in the {\it left panel} of   Fig. \ref{Fig:VectorFreq} the unstable modes are in the upper region between the blue OC and extremality, while in {\it right panel} they are in the lower region (that we do not show it in all its extension).
The upper region of the {\it left panel} of Fig. \ref{Fig:VectorAng} shows the imaginary part of the eigenvalues of the QNMs (we do not show the upper surface in its full extension but its completion should be clear from the continuation of the interrupted black curves with constant $r_+/L=0.5$ and  $r_+/L=0.6$).

In the plots of Fig. \ref{Fig:VectorFreq} and Fig. \ref{Fig:VectorAng} the large red point signals the region where the gravitational vector superradiant instability reaches its maximum strength.
This occurs for a Kerr-AdS BH with $\{r_+/L,a/L \}_{\rm max} \simeq \{ 0.325\pm 0.020, 0.386\pm 0.020 \}$ where the frequency is given by $\omega L \sim 2.667+ 0.058\, i$. Stated in other words, the maximum growth rate  for the vector superradiant instability and the gauge invariant properties of the BH where it is achieved are
\begin{equation}\label{vector:MaxInstab}
\hbox{Vector:}\quad
 \{ R_+/L,\Omega_h L \}\sim \{ 0.530,1.687 \}\,,\quad {\rm Im} (\omega L) \sim 0.058\,, \quad {\rm Re} (\varpi L) \sim -4.451\,.
\end{equation}
In general, e.g. moving along a constant $r_+/L$, we find that the maximum of the vector superradiant instability is achieved much closer to extremality than in the scalar case. This property is probably related to the fact that the vector OC ends at extremality, as opposed to the scalar OC.

Comparing the properties of the maximum unstable cases \eqref{scalar:MaxInstab} and \eqref{vector:MaxInstab}, we see that the instability growth rate of the scalar and vector sectors is of the same order, with the maximum growth rate in the vector sector  being approximately twice stronger than in the scalar sector. Moreover, the most unstable case in the vector case occurs for a Kerr-AdS BH that is smaller (i.e. with smaller gauge invariant areal radius $R_+/L$) but rotates faster than the Kerr-AdS BH where the scalar instability is highest. 

Finally, note that the strength of the scalar or vector gravitational instabilities can can be orders of magnitude higher than the strength of the same superradiant instability sourced by a scalar field perturbation~\cite{Cardoso:2004hs,Uchikata:2009zz}.

\subsection{Large AdS limit and comparison with special QNMs in asymptotically flat cases \label{sec:CompareAsymptFlat}}
\begin{figure}
\centering
\includegraphics[width=0.97\textwidth]{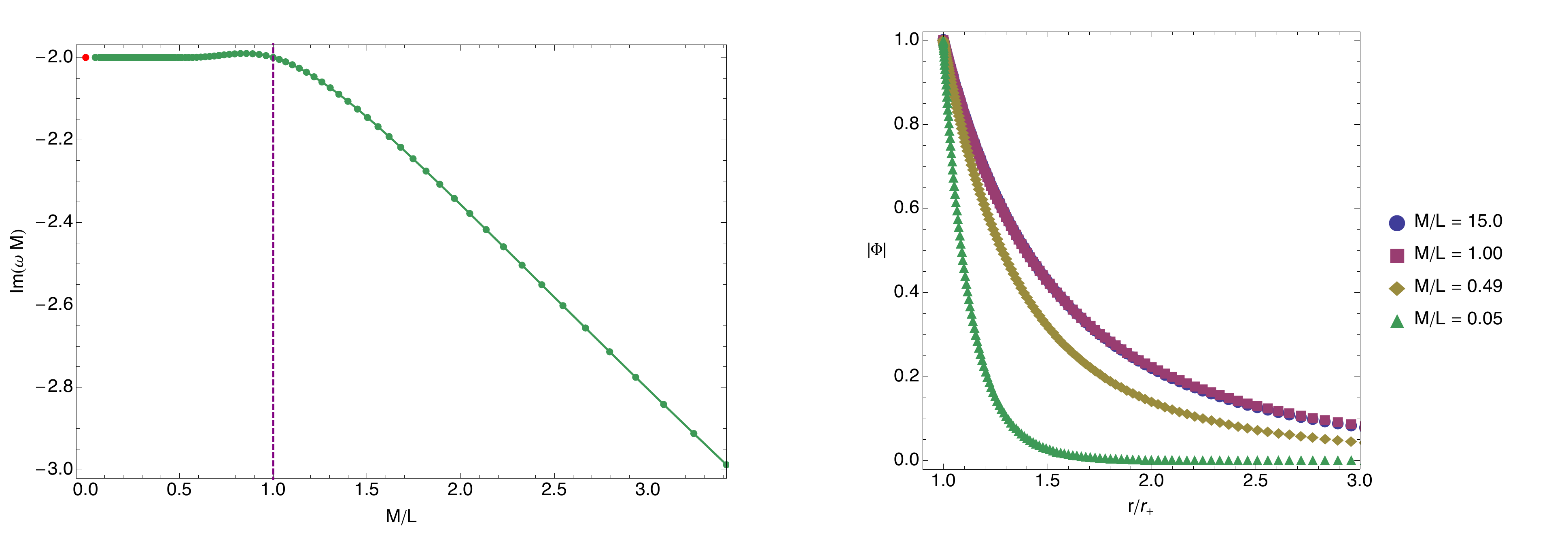}
\caption{\emph{Left panel}: Imaginary part of the ``shear mode'' as the cosmological constant is changed. The vertical purple line indicates where the Hawking-Page transition takes place. \emph{Right panel}: absolute value of the vectorial Kodama Ishibashi variable in ingoing Eddington-Finkelstein coordinates.}
\label{fig:modeshear}
\end{figure}

As we will discuss in section VI, the slowly decaying QNMs in Kerr-AdS play a key role
in the fluid/gravity correspondence. These modes have a particularly appealing interpretation in terms
of a relativistic hydrodynamic problem naturally induced at the AdS boundary. This correspondence
also indicates that rich and complex hydrodynamic phenomena have counterparts in the gravitational theory, as recently demonstrated in~\cite{Carrasco:2012nf,Adams:2013vsa,Green:2013zba}. 
Such a remarkable, and previously unexpected, phenomena displayed by gravity in the AdS context raises
the question of what analogues to hydrodynamic behavior arise in general scenarios. 
Studying such question is beyond the scope of this work (for recent works related to the gravity/hydro connection in AF settings 
see e.g.~\cite{Emparan:2009at,Freidel:2013jfa});
however, as we here are concerned with QNMs we can explore the connection of hydrodynamical modes in AdS with relevant ones in AF spacetimes. 
To this end, we  examine in particular the purely-imaginary QNM mode (often called ``shear mode'') in the limit $r_+/L \rightarrow 0$
for the non-spinning case, see left panel of Fig.~\ref{fig:modeshear}.
In this limit one makes contact with its possible asymptotically flat counterpart describing QNMs of a Schwarzschild black hole. Interestingly, we find the result obtained coincides with the ``algebraically special'' QNM mode. Furthermore, we can look at the profile of this mode, as we change the cosmological constant. It turns out it is very localized around the horizon (becoming more and more localized as we lower the cosmological constant), perhaps indicating that the dynamics involved here does not feel the boundary in any special way, see right panel of Fig.~\ref{fig:modeshear}. At this stage we stress this does not necessarily imply complex hydrodynamic phenomena has a gravitational analogue in AF cases as has been shown to be the case in the AdS case. Nevertheless this is certainly a tantalizing observation deserving further exploration.

\section{Superradiance and black holes with a single Killing field \label{sec:singleKVFBH}}

In the previous sections we confirmed that  Kerr-AdS BHs with $\Omega_h L>1$ are unstable to superradiance. An interesting observation is that at the onset of the superradiant instability there is an exact  zero mode with $\omega=m \Omega_h$ and $\hbox{Im}\,\omega=0$. This zero mode is special because it is invariant under the horizon-generating Killing field $K = \partial_T + \Omega_h \partial_{\Phi}$. Consequently it is regular on both the past ($\mathcal{H}^-$)  and future  ($\mathcal{H}^+$) horizons (generic perturbations can be made regular on the future or past horizons, but not both). In these conditions and for a given $m$,  \cite{Kunduri:2006qa} proposed that, in a phase diagram of stationary solutions, the OC of the instability should signal a bifurcation or merger of the Kerr-AdS BH with a new family of BH solutions that are stable to superradiant modes with the given $m$ and that preserve the same isometry of the superradiant onset mode (see also the nice discussion in \cite{Li:2013pra}).  That is, these new BHs have a single Killing vector field (KVF); the helical Killing field $K = \partial_T + \Omega_h \partial_{\Phi}$. In the context of superradiance of a {\it scalar field}, BHs with a similar helical single KVF that merge with the Kerr-AdS family have scalar hair orbiting around the central core. Examples of such hairy BHs were explicitly constructed perturbatively and non-linearly in \cite{Dias:2011at} \footnote{Recently, a single KVF was constructed analytically in $D=3$ Einstein-AdS theory \cite{Li:2013pra}. (In this case superradiance is absent.)}. Given this explicit proof of existence in the scalar field case, it is natural to expect that a similar new family of single KVF BHs with ``lumpy gravitational hair" merge with the Kerr-AdS BH at the OC of {\it gravitational} superradiance. The existence of such purely gravitational single KVF BHs was first proposed in  \cite{Kunduri:2006qa} and contact between these BHs and geons was made in \cite{Dias:2011ss}. In this section we will give the explicit construction (omitted in \cite{Dias:2011ss}) that leads to the leading order thermodynamics and properties of these BHs. Perhaps the most important consequence of this study is that Kerr-AdS BHs are not the only stationary BHs of Einstein-AdS gravity \cite{Dias:2011at,Dias:2011ss}.\footnote{The use of the word ``stationary" in this context requires a comment.  A solution is  static if $\partial_t$ is a KVF and the solution has the $t \to -t$ symmetry. Strickly speaking, a solution is said to be stationary if $\partial_t$ is still a KVF but the $t \to -t$ symmetry is no longer present. In addition,  $\partial_t$ must be timelike everywhere along the asymptotic boundary of the spacetime.  The single KVF BHs discussed here and in \cite{Dias:2011at,Dias:2011ss} certainly do not have $\partial_t$ as a KVF. Instead, they have a helical KVF. Moreover, this KVF is not timelike everywhere at spatial infinity; indeed it is timelike in the neighbourhood of the poles but spacelike near the equator of the sphere. Nevertheless these single KVF solutions are periodic. Now, a periodic solution can be considered to fit in the intuitive notion we have of stationarity.  For this reason we follow \cite{Dias:2011at,Dias:2011ss} who proposed extending the original definition of stationarity to accommodate these novel periodic BHs as members of the stationary class of solutions.}

We can discuss some of the main properties of the single KVF BHs \cite{Dias:2011at,Dias:2011ss} in terms of general arguments.
Recall again the main properties of superradiance in global AdS. A mode $e^{-i\omega T+im\Phi}$ can increase its amplitude by scattering off a rotating BH with angular velocity $\Omega_h$ satisfying $\omega<m\Omega_h$. In asymptotically global AdS spacetimes, the outgoing wave is reflected back onto the BH and scatters again further increasing its amplitude. This multiple amplification/reflection leads to an instability. The process decreases $\Omega_h$ and eventually results in a BH with ``lumpy hair" rotating around it. Such a BH is invariant under just a single Killing field which co-rotates with the hair, $K=\partial_T+\Omega_h \partial_\Phi$. 
Thus, the BH is stationary (periodic) but not time symmetric nor axisymmetric. However, it does not violate the rigidity theorems \cite{Hawking:1971vc,Hollands:2006rj,Moncrief:2008mr}. Indeed, these theorems assume the existence of a Killing vector, typically $\partial_T$, that is not normal to the horizon, and prove that a second Killing field   $\partial_\Phi$ must then be present. Such a BH is thus time symmetric and axisymmetric. The single KVF BHs evade the primary assumption of the rigidity theorem because in this case $K=\partial_T+\Omega_h \partial_\Phi$ is normal to the Killing horizon.

As stated previously, single KVF BHs and horizonless boson star solutions of this type with scalar hair have been constructed perturbatively as well as numerically at the full nonlinear level in \cite{Dias:2011at}. Alternatively, the leading order description of these BHs can also be found using a thermodynamic analysis \cite{Basu:2010uz,Bhattacharyya:2010yg,Dias:2011tj,Dias:2011at} similar to the one done below. The full nonlinear result confirms that this thermodynamic construction gives accurate leading order results \footnote{A similar thermodynamic model was introduced and proved to be correct, when compared with the exact non-linear results, also in the charged superradiant systems discussed in \cite{Basu:2010uz,Bhattacharyya:2010yg,Dias:2011tj}}. For small charges the single KVF hairy BHs exist in a region of the phase diagram that is bounded by the OC of scalar superradiance and by the boson star curve. 

In the purely gravitational sector of Einstein-AdS theory that we discuss here, the gravitational analogue of the horizonless boson stars are the geons constructed in \cite{Dias:2011ss}. Using the aforementioned thermodynamic model we will conclude that single KVF BHs exist in a region of the phase diagram that is bounded by the OC of gravitational superradiance and by the geon curve.\footnote{The hairy BHs of \cite{Dias:2011at} could be constructed non-linearly because they depend non-trivially only on the radial direction while the gravitational single KVF BHs we discuss here have an additional non-trivial dependence on the polar angle. It is challenging to solve the associated coupled system of PDEs and we leave its construction for future work.}  

We are ready to start the leading order thermodynamic construction of the single KVF BHs. We first review the geon and Kerr-AdS solutions, then we construct the single KVF BHs by placing a small Kerr-AdS BH on the top of a geon.

Geons are classical lumps of gravitational energy,  with harmonic time dependence $e^{-i \omega T+im\Phi}$, in which the centrifugal force balances the system against gravitational collapse \cite{Dias:2011ss}.  They are horizon-free, nonsingular, asymptotically globally AdS,  and can be viewed as  gravitational analogs of boson stars.  Each geon is specified by  $\ell$, which gives the number of zeros of the solution along the polar direction, and azimuthal quantum number $m$. It  is a one-parameter family of solutions parametrized e.g. by its frequency. At linear order, a geon is a small perturbation around the global AdS background and its possible frequencies are given by the AdS normal modes, namely \eqref{wAdSscalar} in the scalar sector, and \eqref{wAdSvector} in the vector sector. The energy and angular momentum of the geon are related by $E_{g}=\frac{\omega}{m} \,J_{g}+ O(J_{g}^2)$; they have zero entropy $S_{g}=0$ and undefined temperature,  and they obey the first law of thermodynamics, $dE_{g}=\frac{\omega}{m} \,dJ_{g}$.  \footnote{Back-reacting to higher order each of the individual normal modes of global AdS we approach the full nonlinear geon, but we do not need this knowledge for our argument \cite{Dias:2011ss}.}

Consider now the Kerr-AdS BH. For small $E$ and $J$ (i.e small  $r_+/L$ expansion), the leading and next-to-leading order thermodynamics of this solution is 
\begin{eqnarray}\label{KerrCharges}
&& \hspace{-1cm}E_{K}\simeq \frac{ r_+}{2} \left(1+\frac{r_+^2}{L^2} \left(1+\Omega _h^2L^2\right)\right)+O\left(\frac{r_+^4}{L^4}\right),\qquad
J_{K}\simeq\frac{1}{2}r_+^3\Omega _h+O\left(\frac{r_+^4}{L^4}\right),\nonumber\\
&& \hspace{-1cm} S \simeq \pi  r_+^2\left(1+ \Omega _h^2 r_+^2\right)+O\left(\frac{r_+^5}{L^5}\right),\qquad
 T_{h}\simeq \frac{1}{4 \pi  r_+}\left(1+\left(3-2 \Omega _h^2 L^2\right) \frac{r_+^2}{L^2}\right)+O\left(\frac{r_+^2}{L^2}\right),
\end{eqnarray}
which obeys the thermodynamic first law,  $dE_{K}=\Omega_h \,dJ_{K}+T_h \,dS$, up to next-to-leading order.

We can now construct perturbatively the single KVF BH of the theory  by placing a small Kerr-AdS BH at the core of the geon. The associated single KVF of the solution is inherited from the geon component of the system. To argue for the existence of this solution and to find its thermodynamic properties we can use a simple thermodynamic model where the leading order thermodynamics of the single KVF  BH is modeled by a non-interacting mixture of a Kerr-AdS BH and a geon
Absence of interaction between the two components of the system means that the charges $E$, $J$ of the final BH are simply the sum of the charges of its individual constituents: $E=E_{K}+E_{g}$, $J=J_{K}+J_{g}$.

In this mixture, the Kerr-AdS component controls the entropy and the temperature of the final BH (since by definition the geon has no entropy and has undefined temperature). The single KVF BH chooses the partition of its charges between the geon and the Kerr-AdS components in such a way that the total entropy $S$  of the system is extremized. Indeed, maximizing  $S=S_{K}(E-E_{g},J-J_{g})$ with respect to $J_{g}$ and using the first laws for the geon and for the Kerr-AdS, we find that the partition is such that the angular velocities of the two components are the same, $\Omega_h=\frac{\omega}{m}$, i.e. the two phases are in thermodynamic equilibrium. Actually, there is a much simpler way to derive this result. Since the geon has only one Killing field, $K=\partial_T+(\omega/m)\partial_\Phi$,  and we place a Kerr-AdS BH with a Killing horizon at its centre, the geon's Killing field must coincide with the horizon generator of the single KVF  BH.

The non-interacting and equilibrium conditions together with the leading order thermodynamics of the Kerr-AdS BH and of the geon yields that the final distribution of the charges among the system's constituents and the entropy and temperature of the single KVF BH are, respectively, 
\begin{eqnarray}
&& {\bigl \{} J_{g}\,, E_{g}{\bigr \}}=\left\{J\,,\frac{\omega}{m} J\right\}\,,\quad  {\bigl \{} J_{K}\,,E_{K}{\bigr \}}=\left\{0\,,E- \frac{\omega}{m} \, J\right\}\,,\nonumber\\
&&
   S= 4 \pi  \left(E-\frac{\omega}{m}\, J \right)^2 \,,\quad T_h= \frac{1}{8\pi }\left(E-\frac{\omega}{m}\,J\right)^{-1} \,.
\label{eq:BHleadingThermo} 
\end{eqnarray}
So, at leading order, the geon component carries all the rotation of the system and the Kerr-AdS component stores all the entropy. By construction, these relations obey the first law of thermodynamics $dE=T_h dS+\Omega_h dJ$, up to order $\mathcal{O}\left( M,J\right)$ with $\Omega_h=\omega/m$ and $\omega$ given by \eqref{wAdSscalar} in the scalar sector, or by  \eqref{wAdSvector} in the vector sector.

\begin{figure}[t]
\centerline{
\includegraphics[width=0.55\textwidth]{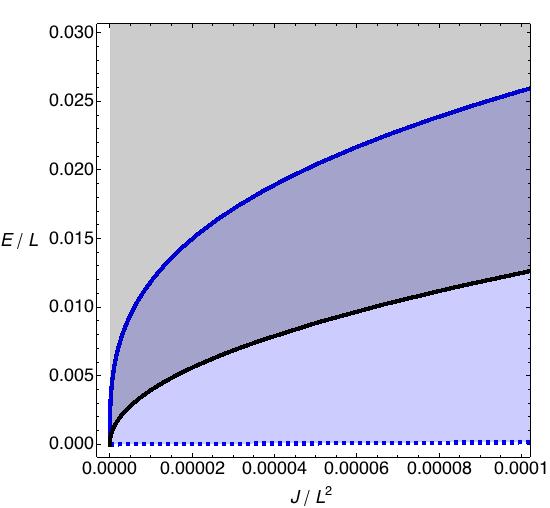}
}
\caption{Phase diagram of global AdS stationary solutions of the $d=4$ Einstein-AdS theory, for  small $E/L$ and $J/L^2$. Non-extremal Kerr-AdS BHs exist only above the extremal black line (grey region). The blue curve above the extremal line sets the onset of the gravitational superradiant instability to $\ell=m=2$ scalar modes (already represented e.g. in Fig. \ref{Fig:KAdSscalarOnset}). Kerr-AdS BHs below these curves are unstable to the associated $\ell=m=2$ superradiant scalar modes.  The dashed curve in the bottom represents the scalar $\ell=m=2$ geon described by $E=\frac{\omega}{m}\,J$ with $\omega=\omega_{\rm s}=(1+\ell)$ and $\ell=m=2$ (and $p=0$).  Single Killing field BHs with $m=\ell=2$ exist between the superradiant OC and the geon line (blue and blue/gray regions).
In the blue/gray shaded region between the black and the upper blue line, Kerr-AdS and single KVF BHs coexist, i.e. we have non-uniqueness.}
\label{fig:phasediag}
\end{figure}

Using this simple thermodynamic model we can further predict the region in phase space where single KVF BHs should exist. 
A single KVF BH merges with the Kerr-AdS family at a curve that describes the onset of the $m$-mode superradiant instability. This occurs at an angular velocity that saturates the superradiant condition, $\omega \leq m \Omega_h$, where $\{\omega,m\}$ are the frequency and azimuthal number of the linearized geon component of the single KVF BH. (It suffices to consider the linearized geon since the  gravitational  hair is very weak near the onset of the instability.) At the superradiant merger, the Kerr-AdS and single KVF  BH thermodynamics coincide. Thus, we can use the Kerr-AdS BH thermodynamics \eqref{KerrCharges} with  $\Omega_h = \omega/m$ to determine the charges of the final system.  In a phase diagram $\{E,\,J\}$ $-$ see Fig. \ref{fig:phasediag} $-$  this determines the upper bound curve of the region where single KVF BHs exist:
\begin{equation}
 E {\bigl |}_{merger}\simeq \frac{ r_+}{2}+\frac{r_+^3}{2L^2} \left(1+\frac{\omega^2L^2}{m^2}\right) ,\qquad  J{\bigl |}_{merger}\simeq\frac{1}{2}\,\frac{\omega}{m}\,r_+^3 \,.
 \end{equation}
Moving down from this curve, the Kerr-AdS contribution weakens and the leading order thermodynamics of the system is increasingly dominated by the geon component. In the limit where $r_+\to 0$, the lower bound curve of single KVF phase is expected to be the geon curve. This discussion is best illustrated in Fig. \ref{fig:phasediag}, where we represent the phase diagram  associated to the $\ell=m=2$ solutions of the scalar sector with frequency $\omega=\omega_{\rm s}$ given by \eqref{wAdSscalar}. 

Note that there is a region in the phase diagram (the blue/gray shaded region in Fig. \ref{fig:phasediag}) where the Kerr-AdS and single KVF BH families coexist, i.e. the present system provides the first example of non-uniqueness in Einstein gravity in four dimensions. The two families of BHs can have the same mass and angular momentum but different entropy.

As emphasized previously, in the scalar field superradiant system of \cite{Dias:2011at}, and in the charged superradiant system of  \cite{Basu:2010uz,Dias:2011tj}, the available full non-linear results confirm that the thermodynamic model we use also here gives the correct leading order thermodynamic properties of the system. We leave for the future the explicit non-linear construction of the single KVF BHs. 

We postpone for Sec. \ref{sec:Conc} the discussion of the stability properties of the single KVF BHs and the role they might have in the time evolution and endpoint of the superradiant instability of the Kerr-AdS BH.

\section{Hydrodynamic thermalization timescales in the AdS$_4$/CFT$_3$ duality \label{sec:Hydro}}

In the context of the gauge/gravity duality, a black hole is dual to a thermal state in the holographic quantum field theory (QFT). Moreover, QNMs are fundamental entities in this correspondence since the QNM frequencies in the bulk black hole describe the thermalization or relaxation timescales in the dual QFT. This map was first proposed in  \cite{Horowitz:1999jd,Danielsson:1999fa}
and later it was understood and established that the QNM spectrum of a given field perturbation coincides with the poles of retarded correlation functions of the gauge theory operator that is dual to the perturbation at hand \cite{Birmingham:2001pj,Son:2002sd,Kovtun:2005ev}. This was done in the framework of linear response theory appropriate for describing linearized fluctuations of any wavelength about AdS backgrounds as long as the perturbation amplitude is small. A particularly relevant family of perturbations are the lowest QNMs, i.e. those with small frequency  whose wavelength is large compared to the thermal scale of the field theory. The relaxation timescales of these modes have a hydrodynamic description and can be computed studying perturbations of the Navier-Stokes equation that describes the hydrodynamic regime of the holographic QFT \cite{Kovtun:2005ev,Policastro:2002se,Friess:2006kw,Michalogiorgakis:2006jc}.  These hydrodynamic modes are also captured by the fluid/gravity correspondence which is a formal one-to-one map between Einstein's equations in AdS and non-linear hydrodynamic equations \cite{Bhattacharyya:2008jc,Hubeny:2010ry}. It follows from a perturbation theory analysis where the small expansion parameter is the ratio of the mean free path of the theory (i.e. the thermal scale) over the typical variation wavelength of the fluid variables and gravitational field. With respect to linear response theory it has the advantage that it captures also non-linear physics but it is restricted to long wavelength physics. The two regimes therefore complement each other and 
intersect in a corner of the phase space corresponding to linearized long wavelength perturbations \cite{Hubeny:2010ry}. These are precisely the hydrodynamic QNMs that we want to study in this section.

A particular example of a gauge/gravity duality is the AdS$_4$/CFT$_3$ correspondence, whereby supergravity on the Kerr-AdS$\times S^7$ background is dual to a thermal conformal field theory (CFT) on the holographic boundary of the global AdS geometry. In this case the Kerr-AdS black hole is dual to a thermal state with a rotational chemical potential in the CFT$_3$ that is formulated on a sphere.

In this section  we aim to compare the  long wavelength gravitational QNMs of Kerr-AdS with the hydrodynamic relaxation timescales of the dual CFT$_3$. First, in Section \ref{sec:Hydro0} we compute the hydrodynamic modes both perturbatively and numerically and later, in Section \ref{sec:CompareQNMsHydro}, we compare them with with the long wavelength gravitational QNMs. The excellent match that we find  provides a further confirmation of the holographic interpretation of the QNM spectrum, of the shear viscosity to the entropy density bound, $\eta/s=1/(4\pi)$, and ultimately of the correspondence itself. Not less importantly, it provides the first non-trivial confirmation that the Robin boundary conditions for the Teukolsky gauge-invariant variable derived in the companion paper \cite{Dias:2013sdc} are indeed the ones that we must impose if we want the perturbations to preserve the asymptotic global AdS structure of the background. Indeed, had we chosen different BCs, e.g Dirichlet or Neumann BCs, and the QNM spectrum would not match the hydrodynamic timescales.

\subsection{Hydrodynamic thermalization timescales\label{sec:Hydro0}}
 
 The conformal boundary of the Kerr-AdS geometry is the static Einstein  universe $R_t \times S^2$ with line element that this time we write as
\begin{equation}\label{holoG}
ds^2_{\partial}=h_{bc}\,dx^bdx^c=-dT^2+L^2 d\Omega_{S^2}\,,\qquad  d\Omega_{S^2}=\frac{dX^2}{1-X^2}+\left(1-X^2\right) d\Phi^2\,.
\end{equation}
where $X$ is related to the standard polar angle on the sphere introduced in \eqref{metricAdS} by $X=\cos\Theta$. The CFT$_3$ is described by an holographic stress tensor $\langle T_{bc} \rangle$ which can be found using, e.g. the formulation of Haro, Skenderis and Solodukhin \cite{Haro:2000xn}.

We first introduce the Fefferman-Graham coordinate frame $\{T,z,X,\Phi \}$ whereby the Kerr-AdS geometry can be recast in an asymptotic expansion around the holographic boundary $z=0$ ($r=\infty$) as
\begin{equation}\label{FGexpansion}
ds^2 = \frac{L^2}{z^2}\left[dz^2+ds^2_{\partial}+\frac{z^2}{L^2}\, h_2+\frac{z^3}{L^3}\, h_3+\mathcal{O}(z^6)\right]\,,
\end{equation}
with $ds^2_{\partial}$ defined in \eqref{holoG}. The coordinate transformation that takes Kerr-AdS in the Chambers-Moss frame into the FG frame is obtained as an expansion in $z$, with the successive terms of the expansion being  fixed by requiring that $g_{zz}=L^2/z^2$ and $g_{zb}=0$ ($b=T,X,\Phi$) at all orders.
Up to the order relevant for our analysis, this FG coordinate transformation is explicitly given by
\begin{eqnarray}\label{FGcoordtransf}
&& t=\Xi \, T\,, \qquad \phi =\Phi -\frac{a}{L^2}\, T \,,\nonumber \\
&& r=\sqrt{L^2-a^2 \left(1-X^2\right)} \left(\frac{L}{z}+\frac{z}{L}\,\frac{a^4 \left(1-X^4\right)-L^4}{4 \left[L^2-a^2 \left(1-X^2\right)\right]^2}+ \frac{z^2}{L^2}\,\frac{\left(r_+^2+a^2\right) \left(r_+^2+L^2\right)}{6 r_+\left[L^2-a^2 \left(1-X^2\right)\right)^{3/2}}\right) \!\!+\!\mathcal{O}\!\left(\frac{z^3}{L^3}\right)\!,\nonumber \\
&& 
\chi =\frac{a \,L \,X}{\sqrt{L^2-a^2 \left(1-X^2\right)}}\left(1+\frac{z^2}{L^2}\,\frac{a^2 \left(L^2-a^2\right) \left(1-X^2\right)}{2 \left[L^2-a^2 \left(1-X^2\right)\right]^2}\right)+\mathcal{O}\left(\frac{z^4}{L^4}\right).
\end{eqnarray}
The leading terms in these expansions are fixed by our choice of conformal frame, namely we want the normalization where $g_{TT}=-1$ and the sphere has radius $L^2$ in the boundary metric $ds^2_{\partial}$.
On the other hand the azimuthal coordinate transformation guarantees that the conformal frame does not rotate. 

The holographic stress tensor can be read from the $h_3$ contribution of the expansion \eqref{FGexpansion} via \cite{Haro:2000xn}
\begin{equation}
\langle T_{bc} \rangle = \frac{3h_3}{16\pi G_4}\,,
\end{equation}
where $b,c$  run over the boundary metric coordinates $\{T,X,\Phi\}$. This stress tensor has the form of a perfect fluid with energy density $\rho$, pressure $p$, and fluid velocity $u$ given by
\begin{eqnarray}\label{perfF}
&& \langle T_{bc} \rangle_{(0)} =(\rho +p)u_b u_c+p \,h_{bc} \,, \nonumber\\
&& \rho{(0)} =2 p_{(0)},\qquad p_{(0)}=\frac{ \left(r_+^2+a^2\right) \left(r_+^2+L^2\right) }{3 r_+ \,L\, \gamma(X)^{-3} }\,,\qquad u_{(0)}=\gamma(X) {\bigl (} \partial_T -\Omega_\infty \partial_\Phi {\bigr )}, \\
&&  \hbox{where} \quad \Omega_\infty=\frac{a}{L^2}, \quad 
\gamma(X)=\left[1-\frac{a^2}{L^2} \left(1-X^2\right) \right]^{-1/2}\nonumber
\end{eqnarray}
are  the angular velocity $\Omega_\infty $ of the fluid, and the ratio $\gamma^{-1}=\frac{T}{\cal T}$ between the fluid temperature $T$ and the local temperature $\cal{T}$ (this gives the redshift factor relating measurements done in the laboratory and comoving frames), and $\partial_T,\partial_\Phi$ are the Killing vectors corresponding to the isometries of the boundary background \eqref{holoG},. Further,  $u^2=-1$ and the equation of state $\rho = 2 p$ follows from the fact that the holographic QFT and its fluid are conformal which  implies that the stress tensor is traceless. The stress tensor is conserved with respect to \eqref{holoG}, $\nabla_b\langle T^{bc}\rangle=0$, since there are no sources (e.g., scalar or Maxwell) in our system. Our bulk background is stationary and therefore the boundary fluid is also in stationary equilibrium fluid configuration with rigid roto-translational motion. 
Our choice for the fluid velocity definition is such that it obeys the Landau gauge condition
\begin{equation}\label{Landau}
u_b\langle T^{bc} \rangle=\rho u^c\,.
\end{equation}
This condition guarantees that the stress tensor components longitudinal to the velocity give the local energy density, in the local rest frame of a fluid element \cite{Romatschke:2009kr}.

A generic perturbation of the stationary fluid configuration will drive the system away from equilibrium and dissipation must be included to study the evolution of the system. This dissipative contribution to the total holographic stress tensor is encoded in the term $\langle \Pi_{bc} \rangle$ (this follows from a gradient expansion of Einstein equations around AdS in the regime where the thermodynamic variation lengthscales are much larger than the thermal scale of the stationary background  \cite{Bhattacharyya:2008jc,Hubeny:2010ry}),
\begin{eqnarray}\label{totTab}
&&  \langle T_{bc} \rangle =\langle T_{bc} \rangle_{(0)} +\langle \Pi_{bc} \rangle
\qquad \langle \Pi_{bc} \rangle=- 2\eta\sigma_{bc}\\
&& \hbox{where} \quad 
  \sigma^{bc} = \frac{1}{2} \left(P^{bd} \nabla_d u^c+ P^{cd} \nabla_d u^b\right)
  - \frac{1}{2} \vartheta P^{bc}\,,\qquad \vartheta = \nabla_c u^c\,,\qquad
P^{bc}\equiv h^{bc}+u^b u^c\,, \nonumber
\end{eqnarray}
are the shear viscosity tensor $\sigma^{bc} $, the fluid expansion $\vartheta$, and the  is the projector $P^{bc}$ onto the hypersurface orthogonal to $u$. The quantity $\eta$ is the shear viscosity. Since the fluid is conformal, its stress energy tensor must be traceless  (i.e. the conformal anomaly is proportional to $T_c^{\phantom{c}c}$ and vanishes\footnote{A CFT is invariant under Weyl transformations $h_{bc}\to h_{bc} e^{-2 \lambda(x)}$ which requires that its stress tensor is traceless. In a curved background the Weyl anomaly breaks in general the conformal symmetry and yields $T\propto R^2$, but this breaking occurs only at fourth order in a gradient expansion and the bulk viscosity appears at first order \cite{Romatschke:2009kr}.}). Consequently the fluid must have vanishing bulk viscosity. Also, the Landau frame condition \eqref{Landau} implies $u_a^{(0)}\Pi^{bc}=0$ which discards a possible heat diffusion contribution to the first order dissipative stress tensor (i.e. in this frame all the dissipative contributions are orthogonal to the velocity field) \cite{Romatschke:2009kr}.

We recall that a precise statement for the validity of the hydrodynamic regime of dual system can be made as follows. The mean free path of a theory is typically given by the ratio of the shear viscosity to the energy density, $\ell_{\rm mfp}\sim \frac{\eta}{\rho}$. We are working with a conformal theory so the associated fluid equation of state is $\rho=2p$ and the viscosity to entropy bound is saturated, $\eta=s/(4\pi)$ \cite{Policastro:2001yc}. For any fluid we also have the Euler-Gibbs relation $\rho+p={\cal T} s$, where the local temperature is related to the fluid temperature (dual to the black hole temperature $T=T_h$) by the Lorentz factor. Therefore we can write  $\ell_{\rm mfp}\sim \frac{\eta}{\rho}\sim \frac{3}{2}\frac{\eta}{\rho+P}\sim  \frac{3}{2} \frac{\eta}{{\cal T} s} \sim \frac{3}{8\pi}\frac{1}{\cal T}\sim \frac{3}{8\pi}\frac{\gamma^{-1}}{T_h}$. The hydrodynamic approximation is valid for when the thermodynamic quantities of the fluid and of its perturbations vary on lengthscales  that are much larger than $\ell_{\rm mfp}$, namely
\begin{equation}\label{validityHydro}
 \frac{r_+}{L}\gg 1\,,\qquad\hbox{and}\qquad \frac{a}{L}\ll 1\,,
\end{equation}  
where to get the first relation we used the fact that $0<\gamma^{-1}\leq 1$ and that the temperature scales as $T_h\sim r_+$ for large radius black holes $-$ see \eqref{OmegaT} $-$ while the second relation follows from the fact that the background pressure $p_{(0)}$ is not a constant and its gradient scales with the rotation parameter in AdS units.

According to the holographic dictionary, the fluid temperature is identified with the Hawking temperature $T_h$ of the black hole and it follows from the previous discussion that the angular velocity of fluid $\Omega_\infty=a/L^2$  is precisely the shift in the azimuthal coordinate such that the (non-dynamical) background on which the fluid flows is static. On the other hand, the viscosity is given in terms of the horizon radius of the bulk black hole as 
\begin{equation}\label{eta}
\eta=\frac{1}{3}\,L r_+^2\,.
\end{equation}
This is a universal relation for any fluid that is holographically dual to a black hole of Einstein-AdS$_4$ theory.
It follows from the celebrated viscosity to entropy density ratio of the theory namely $\eta=s/(4\pi)$ \cite{Policastro:2001yc}. This is a constitutive relation that is independent of the rotation of the fluid since it follows from measuring quantities in the rest frame of the fluid. Namely we can write the entropy density as $s=S/V=S\rho_{(0)}/E=2p_{(0)}S/E$ which yields \eqref{eta} after using the relations for the static black hole entropy and energy, $S=\pi r_+^2$ and $E=r_+\left(1+r_+^2/L^2\right)/2$, and taking the hydrodynamic limit $r_+/L\to \infty$.

The hydrodynamic equations of motion for the perturbed fluid, that will ultimately quantize the relaxation timescales of the system, follow from  the conservation of the total stress tensor,
\begin{equation}\label{consTab} 
\nabla_{b}\left( \langle T^{bc} \rangle_{(0)} +\langle \Pi^{bc} \rangle \right) =0\,. 
\end{equation}
These equations can be written as a set of two family of equations, namely the relativistic continuity and Navier-Stokes equations, \footnote{The continuity equation follows from projecting \eqref{totTab} along the fluid velocity. Plugging it into   \eqref{totTab} then yields  the Navier-Stokes equation, which is the projection of  \eqref{totTab} in the hypersurface orthogonal to the velocity.}
\begin{eqnarray}\label{fluidEOM} 
&& \hspace{-1.5cm}u^{c}\nabla_{c}\rho + (\rho+ p) \vartheta=
2\eta\sigma^{bc}\nabla_b u_c\,,\nonumber \\
&& \hspace{-1.5cm} (\rho+ p) u^b \nabla_b u^c = -P^{bc} \nabla_b  p
 +2\eta \left(  \nabla_b\sigma^{bc}-u^c\sigma^{bd}\nabla_b u_d\right).   
\end{eqnarray}

To study the perturbations of these fluid equations, we use the fact that $\partial_T$ and $\partial_\Phi$ are isometries of the background to write the most general perturbations for a conformal fluid as a sum of the following Fourier modes
\begin{eqnarray}\label{pertQ}
&&  \rho=2 p,\qquad p=p_{(0)}+ e^{-i \omega  T}e^{i m \Phi }\delta p(X)
\,, \nonumber\\
&& u=u_{(0)}+ \, e^{-i \omega  T}e^{i m \Phi } \delta u_c(X)\,dx^c.
\end{eqnarray}
The velocity normalization $u^2=-1$ requires  $u_{(0)}\cdot \delta u =0$ i.e
\begin{equation}\label{pertQ2}
 \delta u_T=-\frac{a}{L^2}\,\delta u_{\Phi }
\end{equation}
Plugging these fluctuations in the linearized version of the hydrodynamic equations \eqref{fluidEOM} we 
get the equations of motion (EoM) that the fluid perturbations $\{ \delta P, \delta u_X, \delta u_\Phi\}$ have to obey.
We solve these equations exactly using numerical methods like those we use to solve the gravitational equations. In addition, to get extra physical insight and check the numerics, we also find a perturbative explicit analytical expression for the fluid quantities of interest.   

To solve the linearized hydrodynamic equations using a perturbative method  \cite{Basu:2010uz,Bhattacharyya:2010yg,Dias:2011at,Dias:2011tj}, we assume a double expansion in the shear viscosity and in the rotation, both for  the fluid perturbations introduced in \eqref{pertQ}, $\{Q^{(f)}(X)\}=\{Q^{(1)},Q^{(2)},Q^{(3)}\}\equiv \{ \delta P/L, \delta u_X, \delta u_\Phi\}$, and for the perturbation frequency $\omega$:  
\begin{eqnarray}\label{fluidExpansion}
&&  Q^{(f)}(\eta,a;X) = \sum_{j=0}^{1}  Q^{(f)}_{j}(a;X)  \left( \frac{\eta}{L^3} \right)^j \qquad \hbox{and} \quad Q^{(f)}_{j}(a;X)= \sum_{i=0}^p Q^{(f)}_{j,i}(X) \left( \frac{a}{L}\right)^i, \nonumber\\
&& \omega(\eta,a)= \sum _{j=0}^{1}\omega_{j}(a)\,\left( \frac{\eta}{L^3}\right)^j,\qquad \hbox{and} \quad \omega_{j}(a)=\sum _{i=0}^p \omega_{j,i}\,\left( \frac{a}{L}\right)^{i},
\end{eqnarray}
and solve progressively \eqref{consTab} or \eqref{fluidEOM} in a series expansion in $\eta/L^3$ and $a/L$. For our purpose it will be enough to go up to third order  ($p=3$)  in the rotation expansion.  

Inspecting the EoM at leading order ${\mathcal O}\left(\eta^0,a^0\right)$, we immediately conclude that we have to split our analysis into two family of modes, namely the scalar and vector modes. The latter have $\omega_{0,0}=0$ and perturb the fluid velocity but not the pressure, while the former have $\omega_{0,0}\neq 0$ and perturb all fluid variables. At this  order rotation is absent and the hydrodynamic modes have an expansion in terms of the scalar ${\bf S}_i$ and vector  ${\bf V}_i$  Kodama-Ishibashi harmonics (which are both related to the associated Legendre polynomials  \cite{Kodama:2003jz,Michalogiorgakis:2006jc,Dias:2013sdc}). This is in agreement with the fact that the gravitational QNMs split also into two families as dictated by the two possible global AdS boundary conditions \eqref{TheBCs}-\eqref{TheBCsV}. As rotation and/or viscosity are turned on these two families naturally continue to follow different paths.

Our main goal is to find the characteristic damped oscillation frequencies of the fluid.  We leave the details  of  our computation to Appendix \ref{sec:HydroAppendix} and give here only its relevant outcome, namely the hydrodynamic CFT thermalization frequencies that can propagate in the CFT$_3$.
The frequencies of the hydrodynamic scalar modes are:\footnote{In  \eqref{CFTfreqScalar} and \eqref{CFTfreqVector} we discard terms of order ${\cal O}\left( \eta L^2/r_+^5 \right)$.}
\begin{eqnarray}\label{CFTfreqScalar}
&&  \omega L{\bigl |}_{\bf s}={\biggl [} \frac{ \sqrt{\ell (\ell +1)}}{\sqrt{2}}+ \frac{a}{L}\,\frac{m(\ell +2) (\ell -1)}{2 \ell  (\ell +1)} -\frac{a^2}{L^2}\,\frac{(\ell +2) (\ell -1)}{4 \sqrt{2}(2 \ell -1) (2 \ell +3) [\ell  (\ell +1)]^{5/2}}  \nonumber\\
&&\hspace{1.7cm}
\times  {\biggl (}2 (\ell -3) (\ell +4)\ell ^2 (\ell +1)^2+3 m^2 \left(6+\ell +\ell ^2\right) \left(1+2 \ell +2 \ell ^2\right) {\biggr )} \nonumber\\
&& \hspace{1.5cm}+\frac{a^3}{L^3}\,
\frac{m}{2 \ell ^4 (\ell +1)^4 (2 \ell -1) (2 \ell +3)}{\biggl( }\ell ^2 (\ell +1)^2 \left(\ell ^6+3 \ell ^5+6 \ell ^4+7 \ell ^3-53 \ell ^2-56 \ell -48\right)\nonumber\\
&&\hspace{2.4cm}
+m^2 \left(\ell ^8+4 \ell ^7-6 \ell ^6-32 \ell ^5+37 \ell ^4+132 \ell ^3+224 \ell ^2+152 \ell +48\right){\biggr) }
+{\mathcal O}\left(\frac{a^4}{L^4} \right) {\biggr ]} \nonumber\\
&& \hspace{1.2cm}
+\,i\,\frac{\eta }{r_+^3}\, (\ell -1)(\ell +2){\biggl [}  -\frac{1}{2}+\frac{a}{L}\,\frac{m \left(2+3 \ell +3 \ell ^2\right)}{2 \sqrt{2}  [\ell  (\ell+1 )]^{3/2}}   \nonumber\\
&&\hspace{2.3cm}
+\frac{a^2}{L^2}\,\frac{ 1}{2 \ell ^3 (\ell +1)^3 (2 \ell -1) (2 \ell +3)} {\biggl (} \ell ^2 (\ell +1)^2  \left(8+\ell +\ell ^2\right) \left(3+2 \ell +2 \ell ^2\right) \nonumber\\
&&\hspace{3.9cm}
-2 m^2  {\bigl [}\,12+\ell  (\ell +1) \left(32+14 \ell +15 \ell ^2+2 \ell ^3+\ell ^4\right){\bigr ]}{\biggr )} \nonumber\\
&& \hspace{2.3cm} + \frac{a^3}{L^3}\,\frac{m}{8 \sqrt{2}  \ell ^{9/2} (\ell +1)^{9/2} (2 \ell -1) (2 \ell +3) (\ell -1)(\ell +2)}\nonumber\\
&&
\hspace{2.8cm}  \times {\biggl \{}
m^2 {\biggl (}-34 \ell ^{10}-170 \ell ^9-387 \ell ^8-528 \ell ^7+164 \ell ^6 +1626 \ell ^5+7009 \ell ^4 \nonumber\\
&& \hspace{4cm}  +11032 \ell ^3+11608 \ell ^2+6400 \ell +1680{\biggr )}
 -2 \ell ^2 (\ell +1)^2 {\biggl (}15 \ell ^8+60 \ell ^7
\nonumber\\
&&\hspace{3.3 cm} +8 \ell ^6-186 \ell ^5+113 \ell ^4+606 \ell ^3+1832 \ell ^2+1488 \ell +864 {\biggr)} {\biggr \}}
+{\mathcal O}\left(\frac{a^4}{L^4}\right){\biggr ]},\nonumber\\
&&
\end{eqnarray}
while the frequencies of the hydrodynamic vector modes are: 
\begin{eqnarray}\label{CFTfreqVector}
&&\hspace{-0.3cm}  \omega L{\bigl |}_{\bf v}= \frac{m\,a}{L}\,{\biggl [}\frac{(\ell +2) (\ell -1)}{\ell  (\ell +1)} +4\frac{a^2}{L^2}\,{\biggl (}  \frac{12+\ell  (\ell +1) \left(2+\ell +\ell ^2\right)}{\ell ^2 (\ell +1)^2 (2 \ell -1) (2 \ell +3)}\nonumber\\
&&\hspace{5.2cm}-\frac{ m^2 \left[12+\ell  (\ell +1) \left(26-2 \ell +\ell ^2+6 \ell ^3+3 \ell ^4\right)\right]}{\ell ^4 (\ell +1)^4 (2 \ell -1) (2 \ell +3)} {\biggr )} +{\mathcal O}\left(\frac{a^3}{L^3}\right) {\biggr ]} \nonumber\\
&& \hspace{1cm}+\,i\,\frac{\eta }{r_+^3}\, (\ell -1)(\ell +2) {\biggl [} 
-1
+\frac{a^2}{L^2}{\biggl (}  \frac{24+\ell  (\ell +1) \left(4 \ell ^2+4 \ell -5\right)}{\ell  (\ell +1) (2 \ell -1) (2 \ell +3)}  \nonumber\\
&& \hspace{6cm}+\frac{8 m^2  \left[\ell  (\ell +1) \left(\ell ^4+2 \ell ^3+\ell ^2-5\right)-3\right]}{\ell ^3 (\ell +1)^3 (2 \ell -1) (2 \ell +3)}{\biggr )} +{\mathcal O}\left(\frac{a^3}{L^3}\right) 
{\biggr ]}. \nonumber\\
&&
\end{eqnarray}
 In these expansions (and associated Figs. \ref{Fig:hydroscalars}, \ref{Fig:hydrovectors} below) we assume the relation \eqref{eta} for the viscosity. When the rotation vanishes, \eqref{CFTfreqScalar} and \eqref{CFTfreqVector} reduce to the hydrodynamic frequencies first computed in \cite{Michalogiorgakis:2006jc}.

\begin{figure}[ht]
\centering
\includegraphics[width=.99\textwidth]{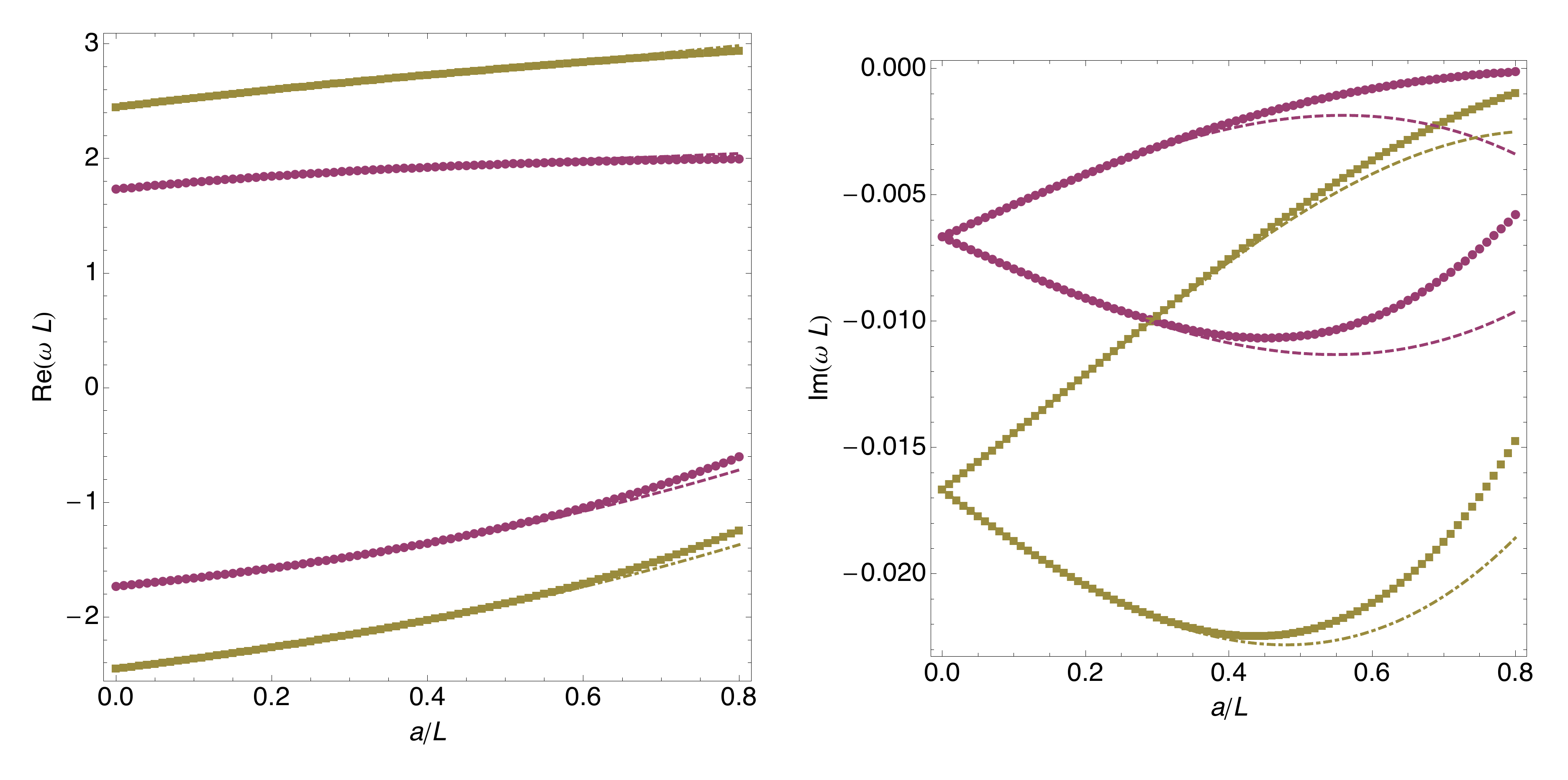}
\caption{
Real and Imaginary part of the frequency for the scalar hydrodynamic modes as a function of the adimensional rotation parameter. The disks (squares) are the exact numerical solutions of the hydrodynamic equations for the $\ell=m=2$ ($\ell=m=3$)  harmonics. On the other hand the dashed line ($\ell=2$ ) and the dashed-dotted line ($\ell=3$) are the curves dictated by the perturbative analytical expression  \eqref{CFTfreqScalar}. In the right panel the upper (lower) branches of each harmonic pair describe the imaginary part of the modes with positive (negative) real part. 
}\label{Fig:hydroscalars}
\end{figure}  

\begin{figure}[ht]
\centering
\includegraphics[width=.99\textwidth]{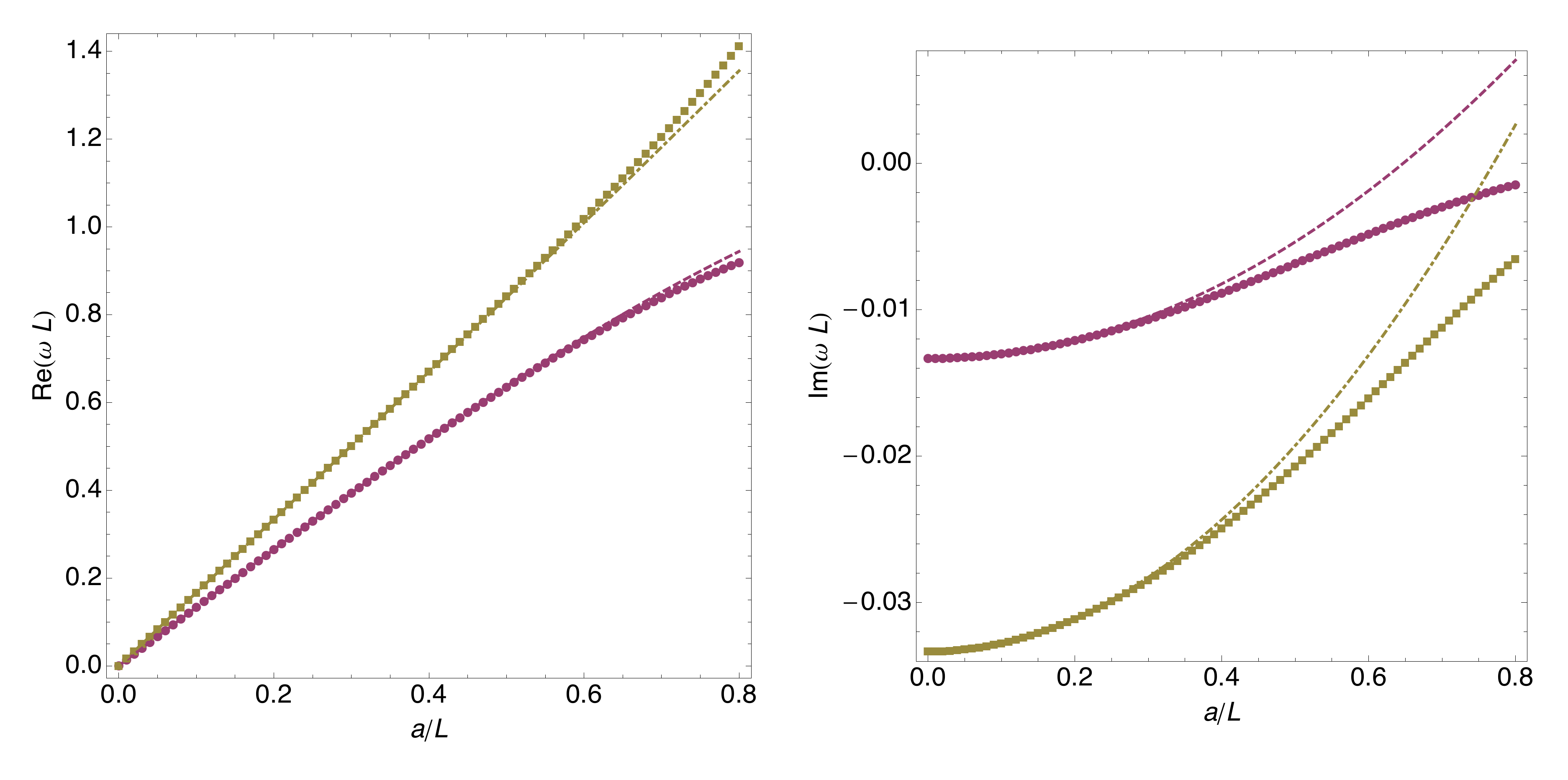}
\caption{
Real and Imaginary part of the frequency for the vector hydrodynamic modes as a function of the adimensional rotation parameter. The disks (squares) describe the exact numerical solution of the hydrodynamic equations for the $\ell=m=2$ ($\ell=m=3$)  harmonics. On the other hand the dashed line ($\ell=2$ ) and the dashed-dotted line ($\ell=3$) are the curves predicted by the perturbative analytical expression  \eqref{CFTfreqVector}.
}\label{Fig:hydrovectors}
\end{figure}  

As illustrative examples, Figs. \ref{Fig:hydroscalars} and \ref{Fig:hydrovectors} show the regime of validity of the perturbative expressions 
\eqref{CFTfreqScalar} and \eqref{CFTfreqVector} by comparing them against the exact numerical solutions of the linearized hydrodynamic equations \eqref{fluidEOM} for the $\ell=m=2$ and $\ell=m=3$ harmonics in both the scalar and vector sectors. As is evident from the figures, the match is excellent in the small rotation regime as expected.

Fig. \ref{Fig:hydroscalars} describes the hydrodynamic scalar modes. For each harmonic $\ell$ there is a pair of solutions, one with positive and the other with negative real part of the frequency. At zero rotation and only in this case,  the background has the $t-\phi$ symmetry and thus the two solutions are physically the same: they form a pair $\{\omega,-\omega^*\}$ related by complex conjugation. Rotation breaks this degeneracy. Fig. \ref{Fig:hydrovectors} describes the hydrodynamic vector modes. These are characterized by having vanishing frequency real part when the rotation vanishes, so there is only one family of solutions for each harmonic.

\subsection{Long wavelength QNMs and hydrodynamic modes \label{sec:CompareQNMsHydro}}

In the last subsection we computed analytically and numerically the  hydrodynamic relaxation timescales.
In this section we compare these timescales with the long wavelength gravitational QNMs.

\begin{figure}[t]
\centerline{
\includegraphics[width=0.52\textwidth]{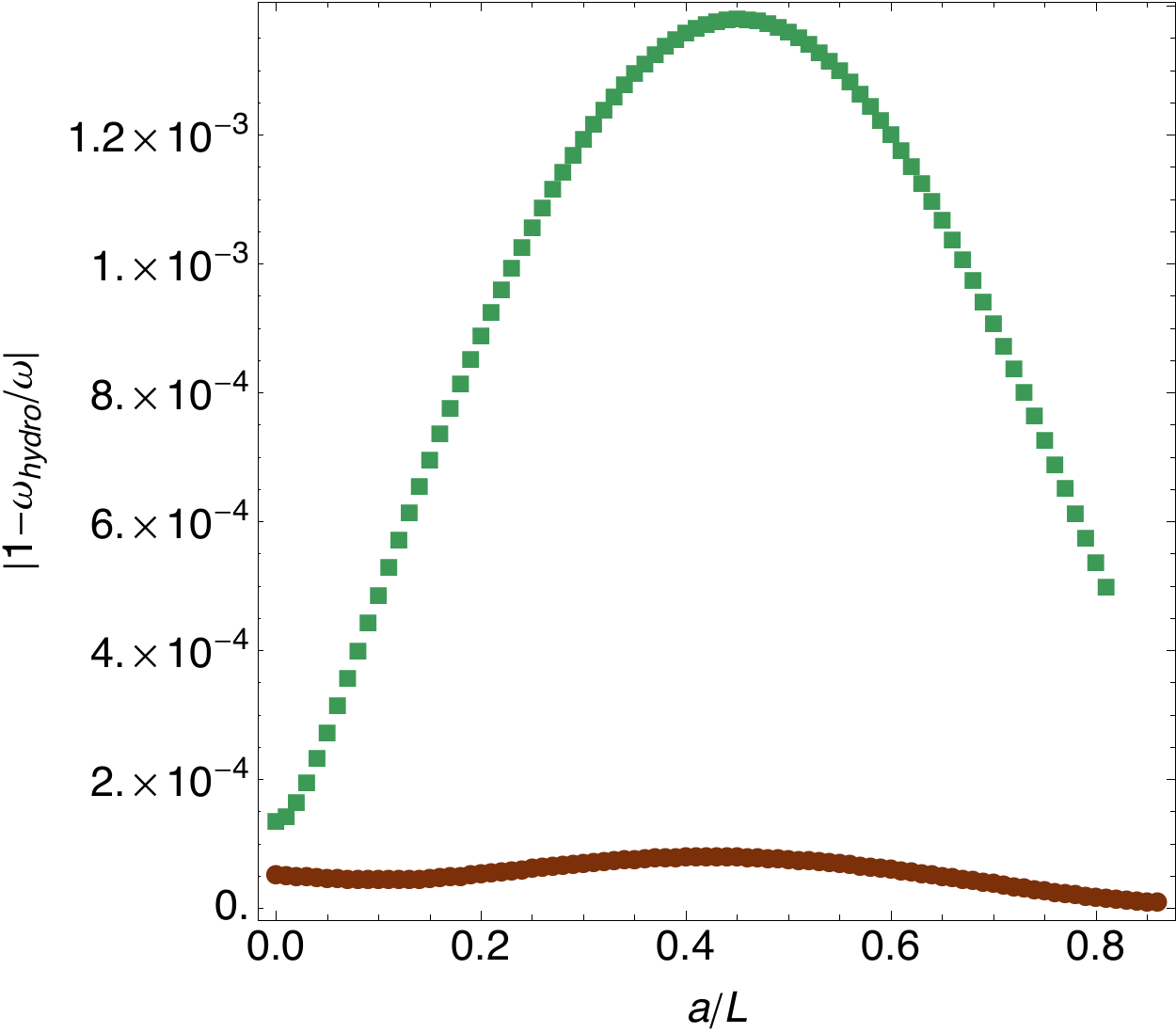}
}
\caption{Comparison between the long wavelength gravitational QNMs and the hydrodynamic modes (for the later we use the exact numerical results) for $r_+/L = 100$. The brown (disks) curve describes the scalar modes while the green (squares) curve is for the vector modes.
}\label{Fig:hydroComparison}
\end{figure}

To perform the comparison we recall that the hydrodynamic and gravitational modes are expected to match in the regime of parameters \eqref{validityHydro}, namely $r_+/L\gg 1$ and $a/L\ll 1$.
We thus consider a Kerr-AdS black hole with radius parameter $r_+/L=100$ to do the comparison.
A measure of the deviation between the numerical hydrodynamic frequencies (call them $\omega_{\rm hydro}$) and the numerical gravitational QNM frequencies (call them $\omega$) is given by $\left | 1-\omega_{\rm hydro}/\omega \right |$. 
In Fig. \ref{Fig:hydroComparison} we plot this deviation measure as a function of the rotation parameter $a/L$ ($a/L<1$ for regular black holes) for a Kerr-AdS BH with $r_+/L=100\gg 1$. The brown curve (disks) is for scalar modes, while the green curve (squares) is for vector modes. We see that the match between the hydrodynamic and long wavelength QNM frequencies is very good even when the rotation grows large and thus moves away from the hydrodynamic validity regime $a/L\ll 1$: for scalar (vector) modes the maximum deviation is below $10^{-4}$ ($2\times 10^{-3}$).  

This perfect match when the rotating chemical potential is present is a further confirmation of the holographic interpretation of the QNM spectrum, of the shear viscosity to the entropy density bound, $\eta/s=1/(4\pi)$, and ultimately of the AdS/CFT correspondence itself. 

\section{QNMs and superradiance in 5 dimensions \label{sec:5d}}
In this section we extend the study of thermalization, quasinormal modes, and superradiance to five dimensions. There are many motivations for doing so. Firstly, there is currently a general interest in studying gravity  in higher dimensions, for a modern review see \cite{Horowitz:2012nnc}. It is interesting to ask how the solutions to the Einstein equations and their properties vary with $D$. As the dimension increases, the types of black hole solutions increase dramatically. Some examples of the non-standard black holes possible in higher dimensions are black rings, black Saturns, and black branes. Many of these solutions challenge the intuition gained from studying four dimensions by exhibiting non-uniqueness and a variety of interesting instabilities, some of which likely lead to topology-changing transitions once quantum effects are included. In addition to this very general motivation, we shall see that superradiance in five dimensions is qualitatively very similar to four dimensional Kerr-AdS case. Thus we expect that the intuition we gain from studying superradiance in four and five dimensions will be useful for thinking about other dimensions. Additionally, because of the properties of the particular class of black holes we chose to study, certain aspects of the problem will turn out to be more tractable than the Kerr-AdS case.

A second motivation for studying five dimensional asymptotically AdS black holes is that they play an important role in understanding strongly coupled field theories in four dimensions via gauge/gravity duality. In particular, the most well-developed example of this duality is Type IIB string theory on AdS$_5 \times S^5$ spacetimes, which is dual to $\mathcal{N} = 4$ Supersymmetric Yang-Mills. The physics of five dimensional AdS black holes can thus lead to an improved understanding of this specific duality, and more generally about the physics of four dimensional field theories at finite temperature. As in the Kerr-AdS case, large black holes will be particularly interesting in this regard as they will be dual to field theories admitting a hydrodynamic description.

\subsection{Myers-Perry$-$AdS black holes with equal angular momenta}

The generalization of the Kerr metric to higher dimensions was found by Myers and Perry \cite{Myers:1986un}. It was then further generalized to include negative cosmological constant for the case of five dimensions in \cite{Hawking:1998kw}, and then to arbitrary dimensions by \cite{Gibbons:2004js,Gibbons:2004uw}. Although our numerical results are for five dimensions only, in the presentation that follows we will keep the dimension general whenever possible. We therefore refer to these black holes as Myers-Perry$-$AdS (MP-AdS) black holes.

In higher dimensions there are more planes for an object to rotate in than in four dimensions, and therefore these black holes are described by $n = \lfloor (D-1)/2 \rfloor$ angular momenta parameters. For generic choices of the angular momenta, the symmetry of the MP-AdS black hole is $\mathbb{R} \times U(1)^n$. The first factor is due to time translations, and the $n$ $U(1)$'s are due to the $n$ independent planes of rotation. For certain choices of the angular momenta, this symmetry can be increased. For odd dimensions, a particularly dramatic enhancement occurs when all angular momenta are equal. In this case, the isometry becomes $\mathbb{R}\times U(1) \times SU(N+1)$, where $D=2N+3$. For this case, the line element becomes cohomogeneity-1, which is to say that it depends non-trivially only on the radial coordinate. This feature makes the study of linear stability particularly tractable for these black holes, as the linearized perturbation equations can be easily separated and reduced to ODE's. Therefore, in what follows, we shall restrict ourselves to odd dimensions and the equal angular momenta sector of the full parameter space. 

Here we introduce the MP-AdS black holes for odd dimension $D=2N+3$ and with all angular momenta equal. The line element is
\be ds^2 = -f(r)^2 dt^2 + g(r)^2 dr^2 + h(r)^2(d\psi + A_a dx^a - \Omega(r) dt)^2 + r^2 \hat{g}_{ab}dx^a dx^b, \ee
with the metric functions defined as follows:
\be g(r)^2 = \Big(1+\frac{r^2}{L^2}-\frac{r_M^{2N}}{r^{2N}}+\frac{r_M^{2N}}{r^{2N}}\frac{a^2}{L^2}+\frac{r_M^{2N}a^2}{r^{2N+2}} \Big)^{-1}, \quad h(r)^2 = r^2 \Big(1+\frac{r_M^{2N} a^2}{r^{2N+2}}\Big), \ee
\be \Omega(r) = \frac{r_M^{2N}a}{r^{2N}h(r)^2}, \quad f(r) = \frac{r}{g(r)h(r)}. \ee
Here $\hat{g}_{ab}$ is the Fubini-Study metric on $\mathbb{CP}^N$. We will adopt the convention that lowercase latin indices run over $\mathbb{CP}^N$ coordinates, and that hatted tensors are associated with this space. The Fubini-Study metric is Einstein, with the following proportionality constant
\be \hat{R}_{ab} = 2(N+1)\hat{g}_{ab}. \ee
When the angular momenta are set to zero, this metric reduces to the usual Schwarzschild-AdS metric with the unit sphere written in terms of the Hopf fibration:
\be d\Omega_{2N+1}^2 = (d\psi + A_a dx^a)^2 + \hat{g}_{ab}dx^a dx^b, \ee
where here $A$ is related to the K\"{a}hler  form by $2J=dA$. Constant $t,r$ slices have the geometry of homogeneously squashed spheres, with the amount of squashing determined by $h(r)$.

The energy, angular momenta, and angular velocity of the horizon are \cite{Gibbons:2004ai}:
\be E = \frac{A_{2N+1}}{8\pi G}r_M^{2N} \Big(N + \frac{1}{2} + \frac{a^2}{2L^2}\Big), \qquad J = \frac{A_{2N+1}}{8\pi G} (N+1) r_M^{2N}  a \ee
\be \Omega_H = \frac{r_M^{2N} a}{r_+^{2N+2} + r_M^{2N} a^2}. \ee
The horizon-generating Killing field is $K = \partial_t + \Omega_H \partial_{\psi}$. The physics of perturbations of this black hole will depend crucially on $\Omega_H$. For $\Omega_H L < 1$, $K$ is timelike everywhere outside the horizon. If $\Omega_H L > 1$, then it becomes spacelike sufficiently far away from the horizon, and in particular is spacelike at the conformal boundary. For $\Omega_H L = 1,$ $K$ is exactly null at the conformal boundary.

The metric is described by three dimensional parameters, $(L,r_M,a)$. We will find it useful to use instead the parameters $(L,r_+,\Omega_H)$, where $r_+$ is the horizon radius, to describe the solution. 
Note that these parameters are defined independently of the parameters that appear in the Kerr-AdS case, and also note that the boundary metric is not rotating in these coordinates. 

\begin{figure}[h]
\centering
\includegraphics[width=.6\textwidth]{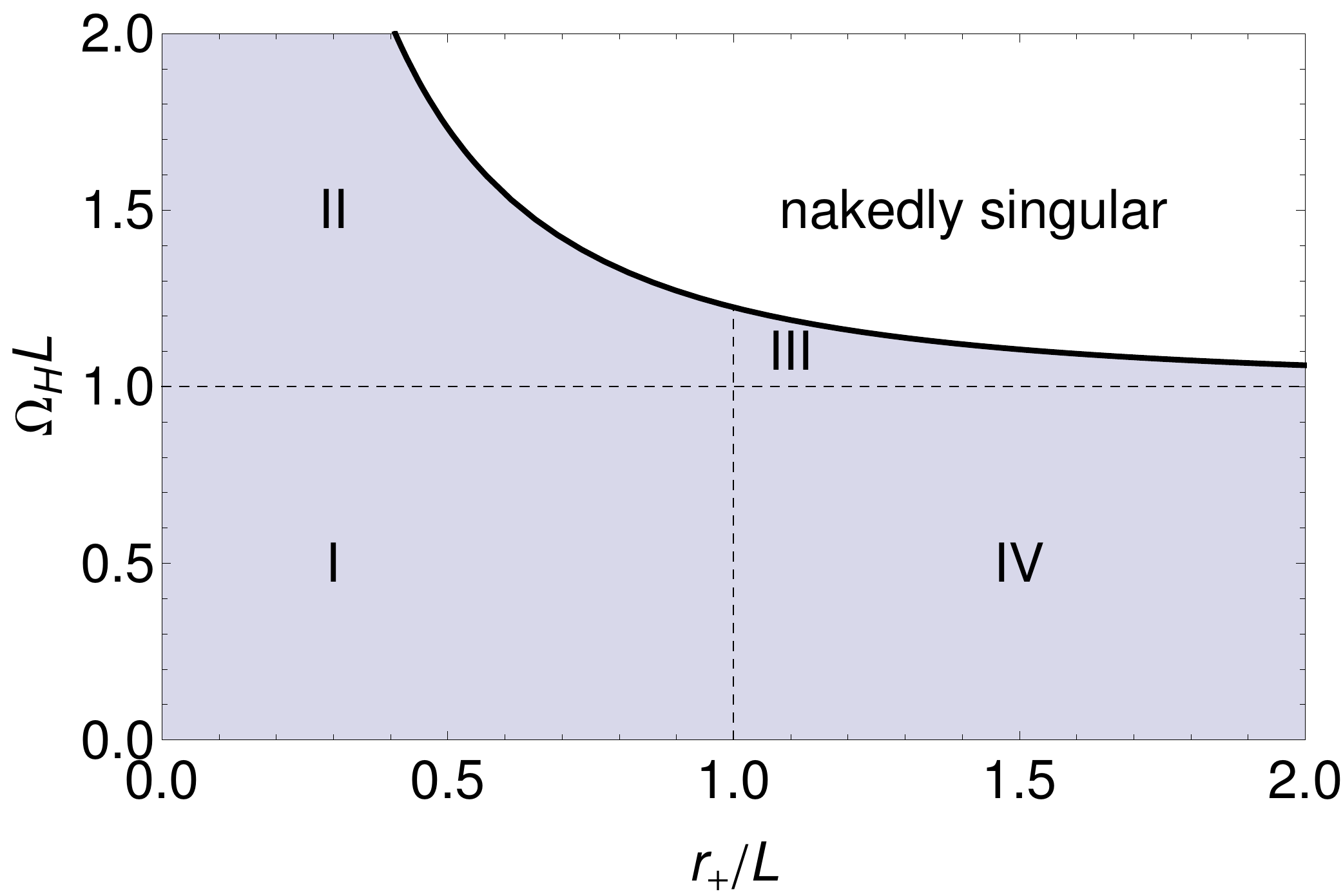}
\caption{The parameter space for $5D$ equal angular momenta black holes. Below the line $\Omega_H L = 1$ the black holes are expected to be stable, and above they can potentially be unstable. Black holes to the right of the line $r_+/L = 1$ are thermodynamically preferred in the grand canonical ensemble, whereas black holes to the left are not. Note that the domain extends infinitely in both directions.
\label{Fig:5Ddomain}}
\end{figure}

For these equal angular momenta black holes the angular velocity cannot be arbitrarily large and must obey the extremal bound
\be \Omega_H L \le \sqrt{1+ \frac{N}{N+1} \frac{L^2}{r_+^2}}. \ee
In Fig.~\ref{Fig:5Ddomain} we plot the domain of the parameter space for the case $D=5$. We also plot the lines $\Omega_H L = 1$ and $r_+/L = 1$. These divide the sub-extremal parameter space into four distinct regions. This division comes from the analysis of Hawking and Reall \cite{Hawking:1999dp}. The importance of the line $\Omega_H L = 1$ is that black holes with $\Omega_H L < 1$ are expected to be stable, whereas those with $\Omega_H L > 1$ are expected to be susceptible to instabilities. Additionally, the partition function of these black holes in a grand canonical ensemble becomes ill-defined for  $\Omega_H L \ge 1$, as the dual CFT will be rotating faster than the speed of light. The importance of $r_+/L = 1$ is that black holes larger than this are thermodynamically preferred over thermal, rotating AdS in the grand canonical ensemble, whereas smaller black holes are not. 

\subsection{Scalar-gravitational perturbations}
We now review the problem of linear perturbations of the above metric. The decomposition we employ was first utilized in \cite{Dias:2010eu}, where scalar-gravitational perturbations of asymptotically flat MP black holes with equal angular momenta were studied. Metric perturbations may be decomposed according to how they transform under the isometries of the $\mathbb{CP}^N$ base space. There are three sectors of perturbations to consider: scalar, vector, and tensor. Tensor and scalar field perturbations of the MP-AdS black holes were studied in \cite{Kunduri:2006qa}. A major simplifying feature of five dimensions is that vector and tensor perturbations do not exist, because the associated vector and tensor harmonics do not exist on $\mathbb{CP}^1$ \cite{Kunduri:2006qa, Durkee:2010ea}. Thus, we need only consider the scalar sector of perturbations in five dimensions. We now briefly review charged scalar harmonics on $\mathbb{CP}^N$ following \cite{Hoxha:2000jf, Dias:2010eu}. First, introduce a charged covariant derivative:
\be \DD_a \equiv \hat{\nabla}_a-i m A_a. \ee
That this is the natural derivative operator to consider can be seen from the dimensional reduction of the fibre coordinate in the Hopf fibration. Charged scalar harmonics (with charge $m$) are then those functions of the $\mathbb{CP}^N$ coordinates that satisfy
\be (\DD^2 + \lambda) \mathbb{Y} = 0. \ee
Here the eigenvalue is a function of two quantized parameters, $(\kappa,m)$:
\be \lambda = l(l+2N)-m^2, \quad l = 2\kappa + |m|, \ee
where $\kappa = 0,1,2...$, and $m \in \mathbb{Z}$. Charged scalar-derived vectors can be obtained by differentiating,
\be \mathbb{Y}_a = - \frac{1}{\sqrt{\lambda}} 
\DD_a \mathbb{Y}. \ee
These can be further decomposed into  Hermitian and anti-Hermitian parts
\be J_a^b \mathbb{Y}_b^{\pm} = \mp  i \mathbb{Y}^{\pm}_a. \ee
Lastly, the scalar-derived tensors are given by
\be
\mathbb{Y}_{ab}^{++} = \DD^+_{(a}\mathbb{Y}^{+}_{b)}, \qquad \mathbb{Y}_{ab}^{--} = \DD^-_{(a}\mathbb{Y}^{-}_{b)}, \qquad \mathbb{Y}^{+-}_{ab} = \DD^+_{(a}\mathbb{Y}^-_{b)} + \DD^-_{(a}\mathbb{Y}^+_{b)} - \frac{1}{2N}\hat{g}_{ab} \DD \cdot \mathbb{Y} . \ee

In order to implement the harmonic decomposition of the perturbation, it will be useful to introduce the 1-forms,
\be e^0 = f(r) dt, \quad e^1 = g(r) dr, \quad e^2 = h(r) (d\psi + A_a dx^a - \Omega(r)dt). \ee
The $\mathbb{CP}^N$ scalar sector of metric perturbations can then be written as
\begin{eqnarray}
h_{AB} &=& f_{AB} \mathbb{Y}, \\
h_{Aa} &=& r(f_A^+ \mathbb{Y}^+_a + f_A^- \mathbb{Y}^-_a ), \\
h_{ab} &=& -\frac{r^2}{\lambda^{1/2}}(H^{++} \mathbb{Y}^{++}_{ab} + H^{--} \mathbb{Y}^{--}_{ab} + H^{+-} \mathbb{Y}^{+-}_{ab}) + r^2 H_L \hat{g}_{ab}\mathbb{Y},
\end{eqnarray}
where upper case latin indices run over 0,1,2, and lower case latin letters run over the $\mathbb{CP}^N$ coordinates. Adopting the traceless transverse gauge,
\be h = g^{\mu\nu}h_{\mu\nu} = 0, \quad \nabla^{\mu} h_{\mu\nu} = 0, \ee
the linearized Einstein equations are then
\be \nabla^2 h_{\mu\nu} + 2 R_{\mu\rho\nu\sigma}h^{\rho\sigma} = 0. \ee
With the above parametrization, the $\mathbb{CP}^N$ dependence in the Einstein equations will separate, resulting in a system of coupled ODE's. These equations are rather lengthy, and depend non-trivially on the particular harmonic $(\kappa,m)$ under consideration, and we therefore omit their presentation here. The non-trivial way in which the equations depend on the $\mathbb{CP}^N$ harmonic in question is as follows: For certain values of $(\kappa,m)$, some of the harmonic tensors do not exist and therefore their coefficient functions are zero. As an example, for $N=1$  $(D=5)$, and $m>0$, $\mathbb{Y}^+_a = \mathbb{Y}^{++}_{ab} = \mathbb{Y}^{+-}_{ab} = 0$, and so therefore the functions $f_A^+, H^{++}, H^{+-}$ do not enter into the perturbation. 
%
\subsection{Boundary conditions}
We now turn to a discussion of the boundary conditions. Boundary conditions must be supplied at the horizon and at the conformal boundary. The appropriate boundary conditions for quasinormal modes are those that correspond to an ingoing perturbation on the future horizon $\mathcal{H}^+$, and a normalizable perturbation at the boundary. For the case of asymptotically flat equal angular momenta MP black holes, the requirement of being ingoing at the horizon was translated into boundary conditions in Ref. \cite{Dias:2010eu}. The same method applies here, and so we omit a detailed discussion of the horizon boundary conditions.

The boundary conditions at infinity are less straightforward. Here we describe a general method for finding the boundary conditions of a normalizable gravitational perturbation of asymptotically locally AdS spacetimes. Recall that in the Kerr-AdS case the perturbation equations were reduced to a single gauge invariant equation, the Teukolsky equation. There were two steps in the process of finding the boundary conditions at infinity. First, a Frobenius expansion analysis yielded the allowed fall-off's, and then the requiring that the perturbation be normalizable fixed a certain linear combination of the two solution branches.

We wish to generalize this method to a system of $n$ 2nd order ODE's for $n$ functions $f_i$. In will be convenient to convert to a new radial coordinate, $z= 1/r$. First we demonstrate that $z=0$ is a regular singular point of the equations. To do so, change to a new basis of functions $q_i = z^{\alpha_i} f_i$, and write the equations as 
\be A_{ij}(z) z^2  \partial_z^2 q_j +  B_{ij}(z) z \partial_z q_j + C_{ij}(z) q_j = 0. \ee
If some choice of the $\alpha_i$ and overall multiplicative factors can be made such that the coefficient matrices approach finite and non-zero constant matrices in the limit $z \rightarrow 0$, then we will have demonstrated that $z=0$ is a regular singular point of this system of equations. We do not know of an algorithmic way to determine the $\alpha$'s, but for the problem at hand we have demonstrated that the equations can be put into this form. Then, after changing coordinates to $\partial_t = z\partial_z$, the equations become:
\[ A_{ij}(0) \partial_t^2 q_j + (B_{ij}(0) - A_{ij}(0))\partial_t q_j + C_{ij}(0) q_j = 0. \]
This is a 2nd order system of ODE's with constant coefficients, which can be solved via the standard method of writing it as a system of coupled 1st order ODE's:
\be \dot{V_a} = M_{ab} V_b \ee
Where
\be V_a = \begin{bmatrix}
q_i \\ \partial_t q_i \end{bmatrix}, \qquad M_{ab} = \begin{bmatrix} 0 & \mathbb{I} \\ -A(0)^{-1}.C(0) & (\mathbb{I} - A(0)^{-1}.B(0)) \end{bmatrix}, \ee
and $a = 1,...,2n$. The generic solution (excluding possible logarithmic terms in z) is then
\[ V_a = \begin{bmatrix}
 q_i \\ \partial_t q_i \end{bmatrix} =  \sum_{b=1}^{2n} c_b \exp\Big(t \lambda_b \Big) v^{(b)}_a, \]
where the $\lambda_b, v^{(b)}_a$ are the eigenvalues and vectors of $M$. There are $2n$ coefficients $c_b$, which is expected: for $n$ 2nd order ODE's there should be $2n$ constants of integration. This is  the generalization of the first step in the Frobenius method, in which the two independent branches of a solution to an ODE around a regular singular point are determined. The next and last step is to use physical considerations to determine the $c_b$ that correspond to the type of perturbation being studied. For normalizable metric perturbations, the boundary metric is held fixed and the boundary stress tensor is varied. For the purpose of translating this condition into choices of the constants of integration $c_b$, it is useful to put the full metric (background plus perturbation) into Fefferman-Graham gauge. In this gauge the metric takes the form
\begin{eqnarray} ds^2 &=& \frac{L^2}{z^2}(dz^2 + g_{ab}(z,x)dx^a dx^b), \\
g_{ab}(z,x) &=& g_0(x) + ... + z^d g_d(x) + z^d \log(z^2) h_d(x) + ...\end{eqnarray}
Here $D = d + 1$ and the lower case Latin indices run over all but the radial coordinate. In order to make the perturbation normalizable we require that it only affects the terms $g_d$ and higher in the above expansion. This corresponds to  holding fixed the boundary metric and only allowing the metric perturbation to affect the expectation value of the stress tensor of the dual field theory. This requirement will fix $n$ of the constants $c_b$. This method generalizes the usual Frobenius method for finding normalizable fall-off's for fields in AdS.

As an explicit example, we display the fall-off's for $D=5$ and for the $(0,m)$ harmonic (for $m > 0$. For this mode, the non-zero perturbation functions are 
\be (f_{00}, f_{01}, f_{02}, f_{11}, f_{12}, f_{22}, f^-_0,  f^-_1, f^-_2, H^{--}, H_L), \ee 
and the gauge conditions can be used to algebraically solve for the functions $(f_{00}, f_0^-, f_1^-, f_2^-, H^{--})$. Then a new auxiliary set of functions $q_i(r)$ can be defined which are finite and non-zero at both boundaries via:
\begin{eqnarray} 
f_{01}(r) &=& \Big(1-\frac{r_+}{r}\Big)^{-i \alpha(\omega-m \Omega_H) -1} \Big(\frac{r_+}{r}\Big)^{5} q_1(r) \\
f_{02}(r) &=& \Big(1-\frac{r_+}{r}\Big)^{-i \alpha(\omega-m \Omega_H) -1/2} \Big(\frac{r_+}{r}\Big)^{4} q_2(r) \\
f_{11}(r) &=& \Big(1-\frac{r_+}{r}\Big)^{-i \alpha(\omega-m \Omega_H) -1} \Big(\frac{r_+}{r}\Big)^{6}  q_3(r) \\
f_{12}(r) &=& \Big(1-\frac{r_+}{r}\Big)^{-i \alpha(\omega-m \Omega_H) -1/2} \Big(\frac{r_+}{r}\Big)^{5}  q_4(r) \\
f_{22}(r) &=& \Big(1-\frac{r_+}{r}\Big)^{-i \alpha(\omega-m \Omega_H)} \Big(\frac{r_+}{r}\Big)^{4}  q_5(r) \\
H_L(r) &=& \Big(1-\frac{r_+}{r}\Big)^{-i \alpha(\omega-m \Omega_H)} \Big(\frac{r_+}{r}\Big)^{4}  q_6(r).
\end{eqnarray}
Here we have introduced the quantities
\be \alpha = \frac{h(r_+)}{r_+ \Delta'(r_+)}, \qquad \Delta(r) = g(r)^{-2}. \ee
Once these $q_i$ functions are known, one can easily find the boundary conditions by expanding the equations near the endpoints.

\subsection{Numerical results}
Here we present our numerical results. We calculated scalar-gravitational QNM frequencies for the five dimensional, equal angular momenta MP-AdS black hole. The possible perturbations are parametrized by two integers $(\kappa,m)$. We are particularly interested in the onset of superradiant instabilities, which was discussed earlier in Sec.~\ref{sec:ResultSuperQNMs}. We remind the reader that superradiant modes are characterized by the condition Re$(\omega) < m \Omega_H$, and linear instabilities by the condition that Im$(\omega) > 0$. For superradiant instabilities in AdS, these two conditions are satisfied simultaneously as the mode becomes unstable. In considering the onset of these  instabilities, we will find it useful to consider the function $\Omega_{\text{H,onset}}(r_+/L)$, which for a given $(\kappa,m)$, is the value of the rotation such that $\omega = m \Omega_{\text{H,onset}}$.

Before presenting our results on scalar perturbations, we briefly review the results for tensor perturbations \cite{Kunduri:2006qa}, which exist for $D$ odd and $D \ge 7$. There an infinite number of superradiant instabilities were found, all for $\Omega_H L > 1$. Lower $m$ modes become unstable for larger values of the rotation, and in the limit $m\rightarrow \infty$, the critical rotation approaches $\Omega_{\text{H,onset}} \rightarrow 1$. Also, $\Omega_{\text{H,onset}}$ depends only very weakly on the size of the black hole, and in particular the ordering of $\Omega_{\text{H,onset}}$'s for different $m$'s is independent of the size of the black hole. For a given $m$ the instability terminates once the black hole is taken to be sufficiently large. We will find that some scalar perturbations behave qualitatively differently than these tensor perturbations.

\begin{figure}[h]
\centering
\includegraphics[width=.7\textwidth]{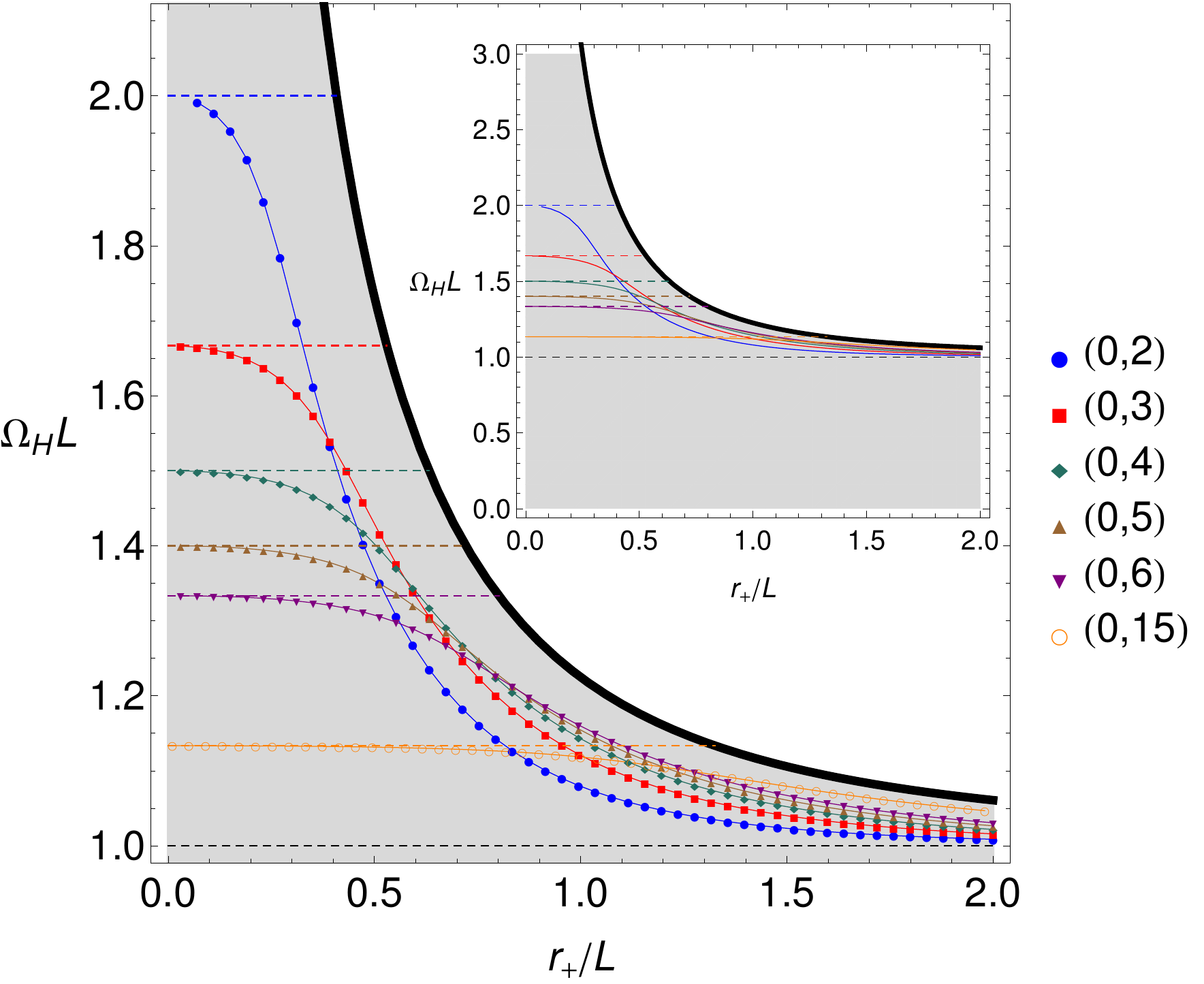}
\caption{The onset of superradiance for $(0,m)$ modes. The dashed horizonal lines are the analytic prediction for small black holes, discussed above. We extend the curves past $r_+/L = 0$ to emphasize the fact that the onset curves are all monotonically decreasing. Two major features of the plot are 1) the crossing of the onset curves, and 2) that as $m\rightarrow \infty$ the onset curve approaches $\Omega_H L = 1$. \emph{Inset}: a zoomed-out plot showing an enlarged view of the parameter space.
\label{Fig:5Dk0onset}}
\end{figure}

\begin{figure}[h]
\centering
\includegraphics[width=.7\textwidth]{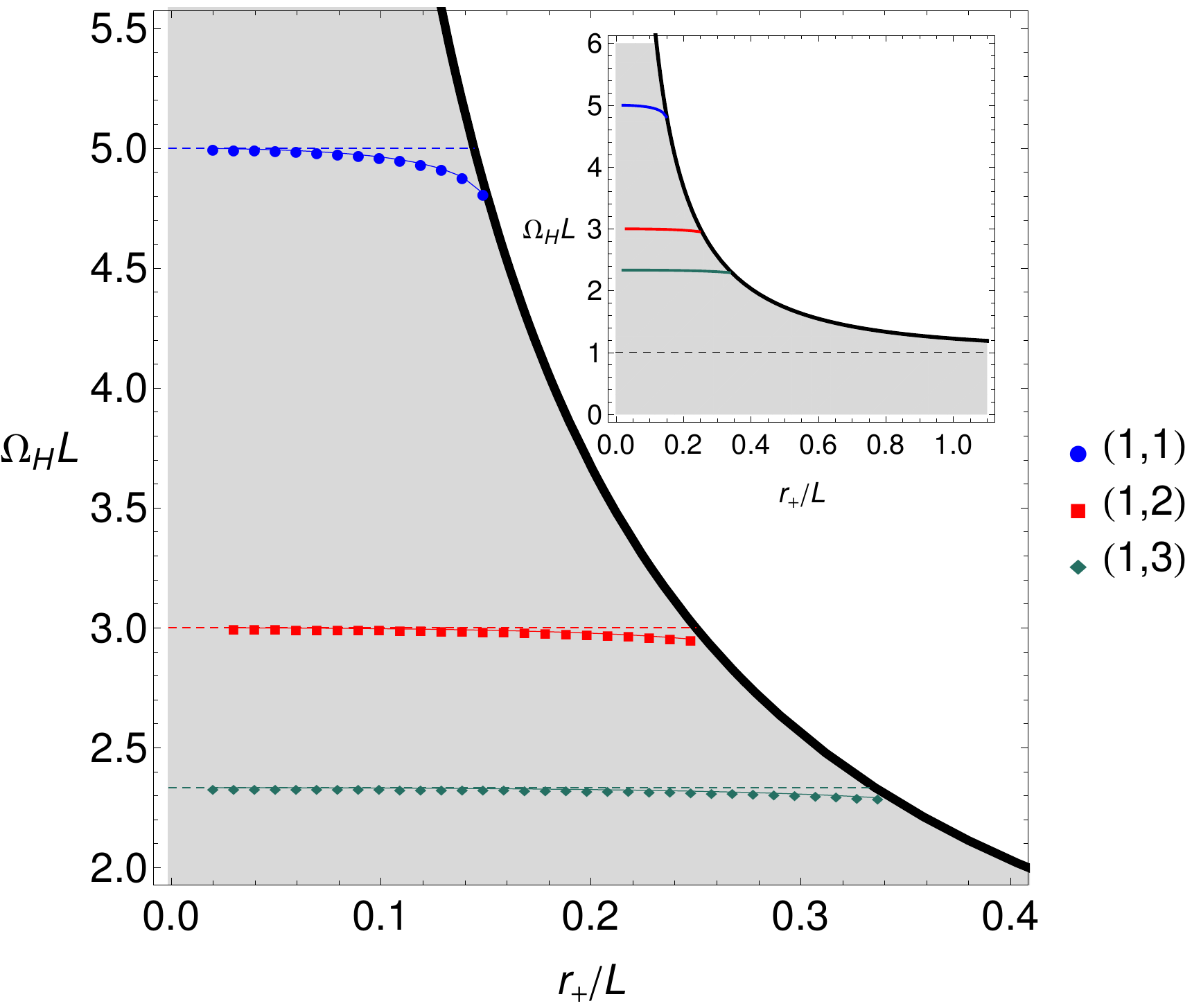}
\caption{The onset of superradiance for $(1,m)$ modes. The dashed horizonal lines are the analytic prediction for small black holes, which we extend pass $r_+/L = 0$ to emphasize the fact that the onset curves are monotonically decreasing. Notice that these curves do not cross each other and terminate at finite $\Omega_H$, in contrast to the $(0,m)$ modes. \emph{Inset}: a zoomed-out plot showing an enlarged view of the parameter space. Due to numerical limitations, the onset curves do not extend all the way to $r_+/L = 0$.
\label{Fig:5Dk1onset}}
\end{figure}

In Fig.~\ref{Fig:5Dk0onset} we plot the onset of the superradiant instability for $(0,m)$ modes, for a range of $m$. We calculated these threshold unstable modes using the Newton-Raphson method presented in Sec.~\ref{sec:num}. As a check on our results, it is useful to first consider small black holes. For small black holes, the onset of the instability can be very easily predicted by first setting $\omega = m \Omega_H$ for the onset of superradiance, and then also setting $\omega = \omega_{\text{AdS}}$, where $\omega_{\text{AdS}}$ is the \emph{normal} mode frequency of AdS. For $(0,m)$ modes, this results in
\be \Omega_{\text{H,onset}} L = 1 + \frac{2}{m}, \ee
Notice that for $m \rightarrow \infty$, $\Omega_{\text{H,onset}} \rightarrow 1$. 
As the size of the black hole is increased, the onset curves for different $m$ begin to cross. This behaviour was also observed for scalar perturbations of Kerr-AdS in Sec.~\ref{sec:ResultSuperQNMs}, and is qualitatively different from the behavior of the tensor instabilities discussed above. Although the onset curves cross as the size is increased, the fact that the $m \rightarrow \infty$ instability hugs the line $\Omega_H L = 1$ indicates that the ``$m=\infty$ mode will never be crossed'', i.e. arbitrarily large $m$-modes will be the first ones to go unstable as $\Omega_H L$ is increased, even for arbitrarily large black holes. This same phenomenon occurs for scalar perturbations of Kerr-AdS, and is discussed in more detail in Sec.~\ref{sec:ResultSuperQNMs}. Another difference between these scalar instabilities and the tensor ones is that, for a given $m$, the scalar onset curves extend for arbitrarily large black holes, whereas the tensor curves terminate. For very large black holes, we can examine the approach of the onset curve to it's limiting value of $\Omega_H L \rightarrow 1$. Interestingly, the approach is power-law, with the exponent independent of $m$:
\be \Omega_{\text{H,onset}} L \sim 1 + \text{const} \Big(\frac{L}{r_+}\Big)^4. \ee
In Fig.~\ref{Fig:5Dk1onset} we plot the onset of the superradiant instability for $(1,m)$ modes. These instabilities are evidently qualitatively different from the $(0,m)$ instabilities in that a) the onset curves do not cross, and b) for a given m, the instabilities do not persist for arbitrarily large black holes. In fact, these modes appear to be qualitatively very similar to the tensor modes on $\mathbb{CP}^N$ for $N \ge 2$.

\begin{figure}[h]
\centering
\includegraphics[width=.95 \textwidth]{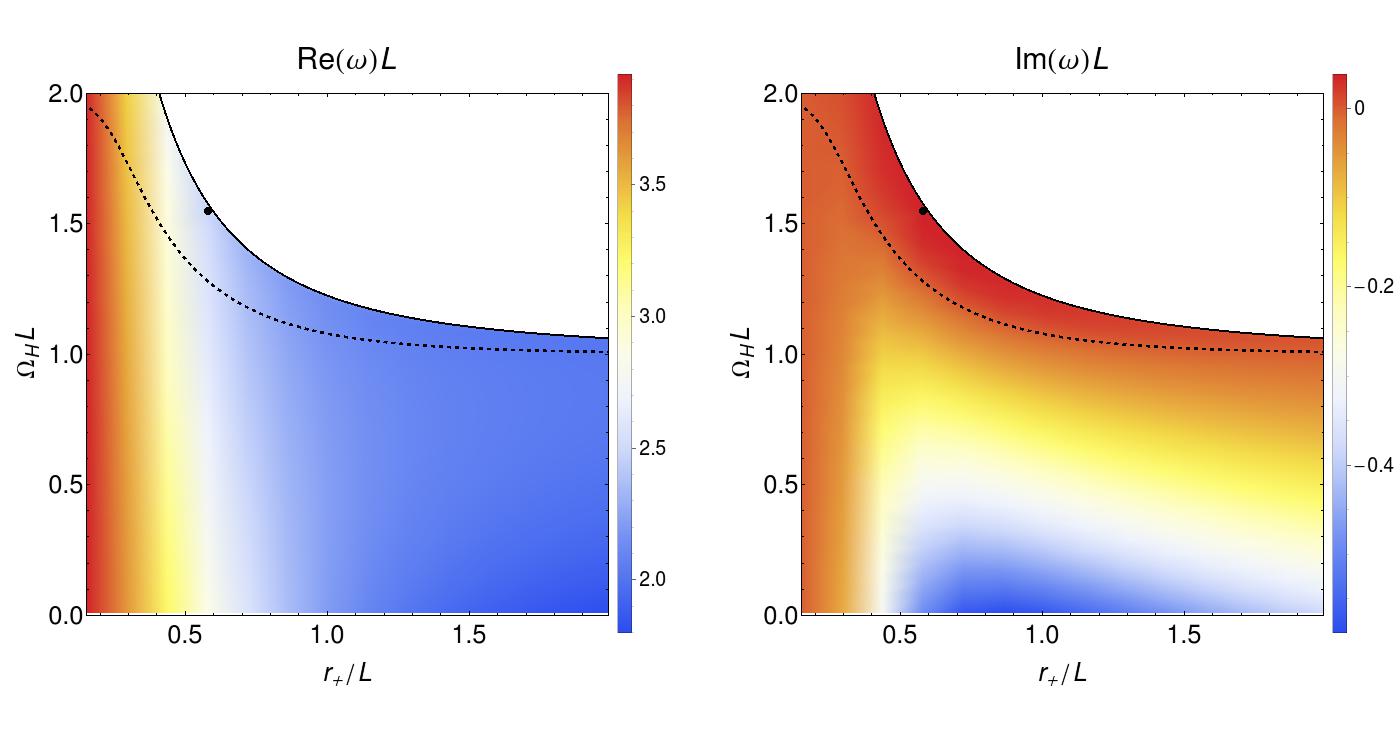}
\caption{Contour plots of $\text{Re}(\omega)$ (\emph{Left plot}), and $\text{Im}(\omega)$ (\emph{Right plot}). Above the solid black line the black hole is nakedly singular. The dashed lines correspond to the onset of the superradiant instability, $\omega = m \Omega_H$. The isolated black dots correspond to the data point with the largest positive value of Im$(\omega)$ found in our scan of the parameter space.
\label{Fig:5Dk0m2Contour}}
\end{figure}

In Fig.~\ref{Fig:5Dk0m2Contour} we plot  contour plots of the real and imaginary parts of $\omega$ for the $(0,2)$ mode. The black dashed line corresponds to the onset mode, $\omega = m \Omega_H$, and the black dot corresponds to the  data point with the largest positive value of Im$(\omega)$. This point is near the  extremality bound, and lies at the end of the grid used to scan the parameter space, and so it is likely that true maximum either lies along or very near the extremality curve. This point is given by
\be (r_+/L, \Omega_H L) = (0.579, 1.548), \qquad \omega L = 2.481 + 0.039 i .
\ee

Lastly, we remark on the importance of the onset modes, $\omega = m \Omega_H$, on the phase diagram of black hole solutions in five dimensions. In Sec.~\ref{sec:singleKVFBH} it was discussed how threshold unstable modes with $\omega = m \Omega_H$ signified new branches of single KVF black hole solutions. We expect much of this analysis to carry over to five dimensions. It would be interesting to study these putative solutions further. Also, as in the Kerr-AdS case, the endpoint of the superradiant instability remains a very interesting open question, especially given the crossing of the onset curves observed in the $(0,m)$ scalar sector of perturbations.

\subsection{Hydrodynamic thermalization timescales in the AdS$_5$/CFT$_4$ duality}

We now turn to study the hydrodynamic QNM's of the five dimensional MP-AdS black hole. As discussed in Sec.~\ref{sec:Hydro}, this serves as a powerful check on our numerics, the hydrodynamic approximation, and more generally, gauge/gravity duality itself. Compared to the Kerr-AdS case, the hydrodynamic approximation for the cohomogeneity-1 MP-AdS black holes is conceptually simpler and has a wider range of validity. Recall that in the Kerr-AdS case, the pressure was a function of the angular coordinate $\theta$, and this introduced another length scale into the approximation which limited it's domain of applicability (although the agreement turned out to be excellent even for large rotations). In contrast, the pressure for these five dimensional black holes is constant, and the approximation is valid \emph{for all} rotations $\Omega_H L < 1$. As is often the case, we will find that the hydrodynamic approximation agrees excellently with our numerical for  reasonably large black holes.

The general theory of hydrodynamic modes was reviewed in Sec.~\ref{sec:Hydro}, and so here we only remark on the differences that occur for these five dimensional equal angular momenta MP-AdS black holes. 
The boundary metric is now 
\be h_{\mu\nu}dx^{\mu}dx^{\nu} = - dt^2 + L^2\Big( (d\psi + A_a dx^a)^2 + \hat{g}_{ab}dx^a dx^b \Big). \ee
The energy density, pressure, and fluid velocity are related to the black hole parameters via
\be \rho = (D-2)p, \qquad p = \frac{r_M^{2N}}{2L} \Big(1 - \frac{a^2}{L^2}\Big), \qquad u_{\mu}dx^{\mu} = \Big(1-\frac{a^2}{L^2}\Big)^{-1/2} \Big(-dt + a (d\psi + A_a dx^a) \Big). \ee
We will also need to use the viscosity to entropy relation, $\eta = s/4\pi$, which in our conventions  yields $\eta = r_+^3/(2L)$.

\begin{figure}[h]
\centering
\includegraphics[width=1.1 \textwidth]{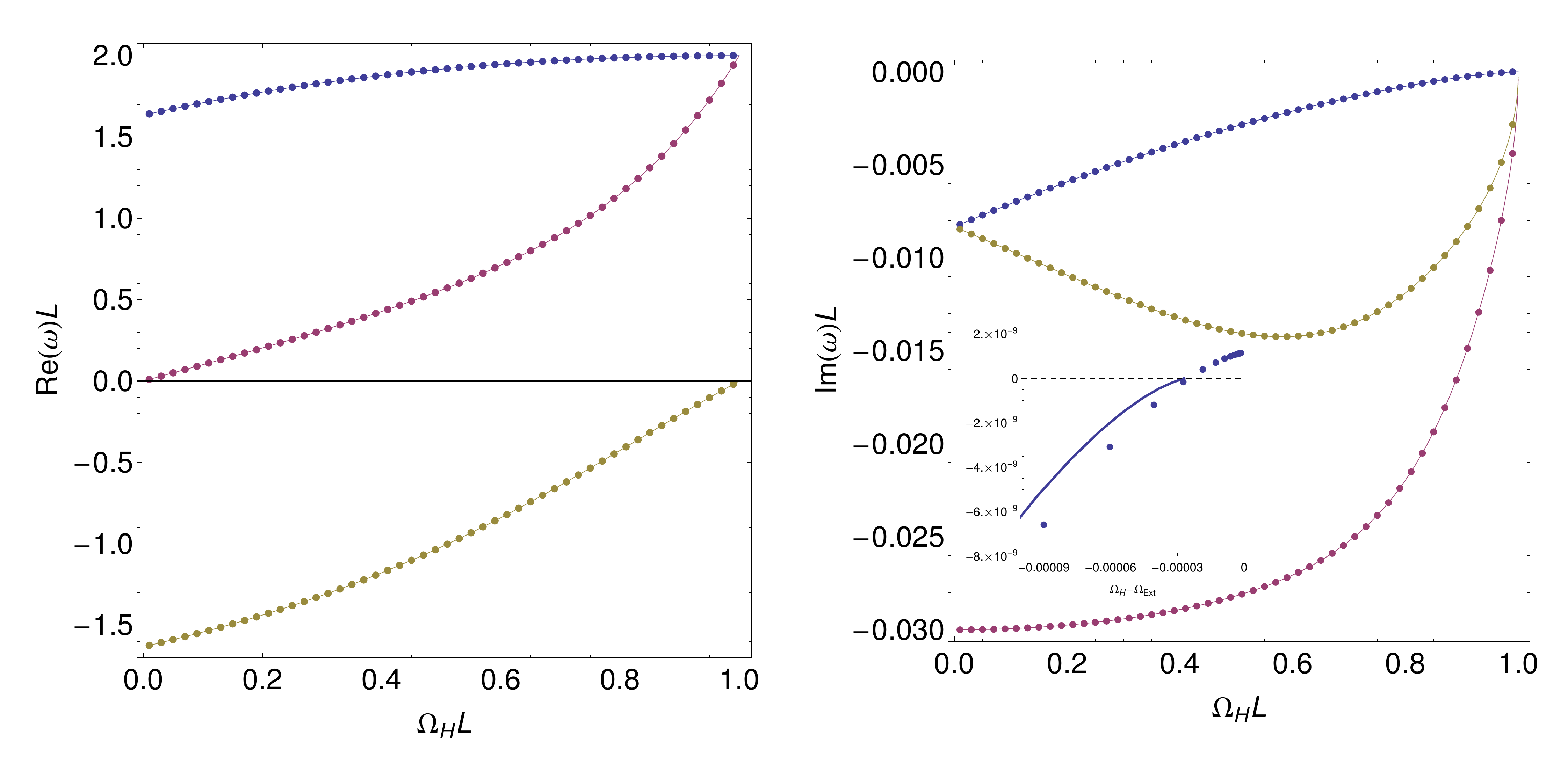}
\caption{Comparision of the leading order hydrodynamic approximation for $(\kappa,m) = (0,2)$ modes. Here $r_+/L = 100$. \emph{Right Inset}: For $\Omega_H L \ge 1$ the hydrodynamic approximation breaks down, and despite the otherwise excellent agreement, fails to predict the superradiant instability.
\label{Fig:5Dk0hydro}}
\end{figure}

\begin{figure}[h]
\centering
\includegraphics[width=1.1 \textwidth]{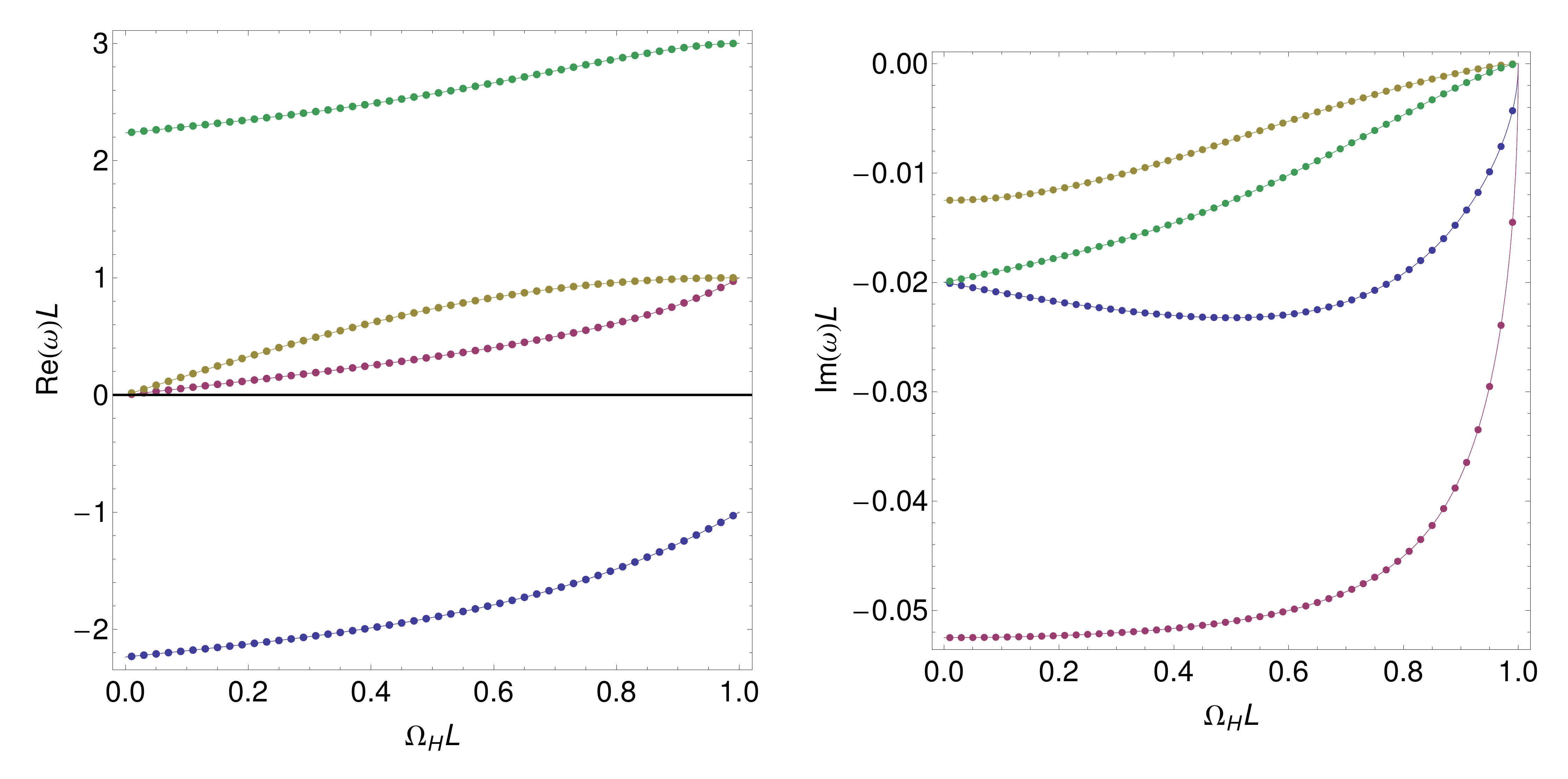}
\caption{Comparision of the leading order hydrodynamic approximation for $(\kappa,m) = (1,1)$ modes. Here $r_+/L = 100$. Once again, we find excellent agreement between our data and the hydrodynamic prediction. \label{Fig:5Dk1hydro}}
\end{figure}

Hydrodynamic modes are obtained through perturbations of the stress tensor that are traceless and divergenceless. The $\mathbb{CP}^N$ dependence of these equations can again be separated using the charged harmonics introduced above. Since our numerical data is for scalar perturbations, we will restrict our attention to hydrodynamic scalar modes. The decomposition of the fluid variables is:
\begin{eqnarray} 
\delta \rho &=& (D-2)\delta p, \qquad \delta p = \delta \hat{p} \mathbb{Y}e^{-i(\omega t - m \psi)}, \\
\delta u_{\mu}dx^{\mu} &=& \Big((\delta u_t dt + \delta u_{\psi} d\psi)\mathbb{Y} + (\delta u_+ \mathbb{Y}^+_a + \delta u_- \mathbb{Y}^-_a)dx^a \Big) e^{-i(\omega t - m \psi)}. 
\end{eqnarray}
The perturbed quantities $ \{ \delta \hat{p}, \delta u_t, \delta u_{\psi}, \delta u_+ \delta u_- \}$ and $\omega$ can then be solved for by using the conservation of the stress tensor. As the expressions are rather lengthy, and depend non-trivially on the harmonic under consideration, we omit their full presentation here. Given the difficulty of obtaining analytic predictions for black hole QNM's, we will however include the expressions in an  expansion about $\Omega_H = 0$. For $D=5$ and the $(0,m)$, harmonics (with $m\ge 2$), and to first order in both $L/r_+$, $\Omega_L = \Omega_H L$, the hydrodynamic modes are
\begin{eqnarray}
\omega_{\pm} 
&=& \left(\pm \frac{\sqrt{m (m+2)}}{\sqrt{3}}+\frac{2 \left(m^2+2 m-3\right) \Omega_L}{3
   (m+2)} + \mathcal{O} \left(\Omega_L^2\right)\right) + \\
&& + \frac{L}{r_+} \left(-\frac{1}{6} i \left(m^2+2m-3\right) \pm \frac{i \left(m^4+4 m^3+3 m^2-2 m-6\right) \Omega_L}{2 \sqrt{3m} (m+2)^{3/2}} + \mathcal{O}\left(\Omega_L^2\right)\right) + \mathcal{O} \left(L^2/r_+^2\right) \nonumber, \\
\omega_0 
&=& \left(\frac{m^2 \Omega_L}{m+2} + \mathcal{O}\left(\Omega_L^2\right)\right)+ \frac{L}{r_+}
   \left(-\frac{1}{4} i m (m+4) + \mathcal{O}\left(\Omega_L^2\right)\right) + \mathcal{O}\left(L^2/r_+^2\right).
\end{eqnarray}
Next, we compare our numerical data for large black holes with the hydrodynamic approximation. In Fig.'s ~\ref{Fig:5Dk0hydro} and ~\ref{Fig:5Dk1hydro}, we fix $r_+/L = 100$ and plot the analytic prediction of the hydro modes against our numerical data for two choices of $\mathbb{CP}^1$ harmonics: $(0,2)$ and $(1,1)$. We work to leading order in $L/r_+$ and find excellent agreement for all $\Omega_H L < 1$. The typical error of the hydro approximation is $10^{-4}$, which is exactly what we'd expect as the next term in the expansion comes in at $\mathcal{O}(L^2/r_+^2) \sim 10^{-4}$. Although we find excellent agreement between the hydrodynamic approximation and our numerical data, the approximation fails to capture the superradiant instabilities. As mentioned earlier, all superradiant instabilities necessarily have $\Omega_H L > 1$. For these black holes, the boundary is rotating faster than the speed of light, and the hydrodynamic approximation shouldn't be expected to be valid. So, while the hydrodynamic approximation has again proven to be an excellent approximation scheme, in this case it fails to capture the instabilities.

\section{Discussion and open problems \label{sec:Conc}}

We studied the most general linear perturbations of the Kerr-AdS BH and of the equal angular momentum Myers-Perry-AdS BH in $D=5$. We imposed asymptotic BCs that preserve the conformal metric \cite{Haro:2000xn,Dias:2013sdc}. These BCs also guarantee that the energy and angular momentum fluxes  across the asymptotic boundary vanish.
Using a novel numerical approach, which we believe might also be useful for other applications, we computed the QNM spectrum of these BHs and the growth rate of their instabilities. The only linear instability that we find in the $D=4$ and $D=5$ stationary BHs have a superradiant nature and they appear only in BHs with  $\Omega_h L>1$. We focused on these spacetimes because of their interest for $AdS_4/CFT_3$ and  $AdS_5/CFT_4$ dualities formulated on the static Einstein Universe, i.e. on the sphere.  
Higher dimensional stationary BHs with $D\geq 6$ were not considered here but they should have novel features that might be worth investigating. Indeed, it is established that $D\geq 6$ stationary BHs are also unstable to the ultraspinning instability, whose onset  was identified in \cite{Dias:2010gk} (this instability was first studied in asymptotically flat stationary BHs in $D\geq 6$ \cite{Emparan:2003sy,Dias:2009iu,Dias:2010maa,Dias:2010eu,Dias:2011jg}). However, it is still an open question whether another instability that is present in $D\geq 6$ vacuum stationary BHs, namely the bar-mode instability \cite{Emparan:2003sy,Shibata:2009ad,Shibata:2010wz,Hartnett:2013fba}, is present in AAdS rotating BHs.  

The onset  of the superradiant instability is an exact  zero mode that is invariant under the horizon-generating Killing field of the Kerr-AdS (MP-AdS) BH. On the shoulders of an idea originally proposed in \cite{Kunduri:2006qa}, we argued that, in a phase diagram of stationary solutions, the superradiant onset curve is a bifurcation line to a new family of BH solutions with a single Killing field that span a region that is further limited by the geon family constructed in  \cite{Dias:2011ss}. We have constructed perturbatively the leading order thermodynamics of these novel BHs using a simple thermodynamic model  \cite{Basu:2010uz,Bhattacharyya:2010yg,Dias:2011tj,Dias:2011at}. In the future, it is important to construct explicitly (numerically) these single KVF BHs and geons at full nonlinear level to confirm the ideas here discussed (in the context of scalar superradince, similar single KVF BHs and boson stars have already been constructed nonlinearly in \cite{Dias:2011at}. Their properties are in agreement with the thermodynamic model we use here, in the regime of small charges). It is worth emphasizing that these single KVF BHs are periodic but not time symmetric neither axisymmetric and their existence shows that the Kerr-AdS and MP-AdS BHs are not the only stationary BHs of Einstein-AdS theory (as discussed in detail before, their existence is not in conflict with the rigidity theorems). 

An interesting open question concerns the time evolution and endpoint of the superradiant instability. Before addressing this issue for rotating systems, it is useful to discuss first the situation for  global AdS Reissner-Nordstr$\ddot{o}$m BH (RN-AdS BH) with chemical potential $\mu$ that are unstable to charged superradiance. This is  the case if the RN-AdS BH is scattered by a charged scalar wave with frequency $\omega$ and charge $e$ that obeys $\omega\leq e\mu$. Here, the marginal mode with $\omega= e\mu$ signals a bifurcation curve, in a phase diagram of static solutions, to a new family of charged BHs with scalar hair that have been explicitly constructed (perturbatively and nonlinearly) in \cite{Basu:2010uz,Bhattacharyya:2010yg,Gentle:2011kv,Dias:2011tj}. When they coexist, the entropy of the hairy BHs is always higher than the entropy of the RN-AdS BH with same mass and charge, for fixed $e$. Moreover, the hairy BHs are not unstable to superradiance since for a given mass and charge (and fixed $e$)  the chemical potential of the hairy BH is smaller than the chemical potential of the RN-AdS BH; therefore the would-be superradiant modes of hairy BHs no longer fit inside the global AdS box. So far we have just discussed the phase diagram of solutions but said nothing about the time evolution of the original RN-AdS BH. The expectation, to be confirmed by a full time evolution, is that the endpoint of the charged superradiant instability in the RN-AdS BH is one of the hairy BHs constructed in \cite{Basu:2010uz,Bhattacharyya:2010yg,Dias:2011tj}. This follows from the fact that for a given mass and charge (and fixed scalar charge $e$) the hairy BH has higher entropy and lower chemical potential than the RN-AdS BH. Therefore a time evolution towards the hairy BH is compatible with the second law of thermodynamics and the endpoint would be stable (to superradiant modes with the given fixed $e$). Given the properties of this  charged system, and the obvious similarities with the rotating superradiant system, it is often assumed that we can use the charged system to extrapolate on evolution properties of the rotating system. However, we next argue that such an extrapolation for time evolution properties is not appropriate. To begin, notice a fundamental difference between the charged and rotating systems. The charge of the scalar field $e$, that enters the superradiant condition $\omega\leq e\mu$, is fixed. However, the azimuthal quantum number $m$,  that enters  the superradiant condition $\omega\leq m\Omega_h$, is not fixed since the nonlinearities of Einstein equation will excite other $m$ modes during a time evolution. This means that a given single KVF BH, constructed in association with a given $m$ mode, can be at most just a metastable state but never the endpoint of the superradiant instability. This is because the single KVF BH is stable to the particular $m$-mode but not to other $m$ superradiant modes that are inevitably excited in a time evolution. Therefore the endpoint of the superradiant instability in rotating BHs is not known at all and finding it is one of the most interesting open questions in BH perturbation physics. Not much can be said about it without performing the full time evolution but it is interesting to observe that, typically, stable BHs to a given $m$-mode are nevertheless unstable to higher $m$-modes. So one possibility is that the system will evolve to configurations with higher and higher $m$ structure. Another important observation is that only BHs with angular velocity $\Omega_h L<1$ are stable to superradiance, as first proved in \cite{Hawking:1999dp}. So a natural expectation for the endpoint of the superradiant instability would be a (single KVF) BH with $\Omega_h L<1$. Finding whether such a BH exists requires constructing the single KVF BHs at full nonlinear level. However, in the  similar scalar superradiant system of  \cite{Dias:2011at},  where much of the present discussion about the time evolution also applies, the single KVF BHs of the theory have been explicitly constructed nonlinearly but none of them has $\Omega_h L<1$. 

Within the gauge/gravity correspondence, black hole QNMs are dual to thermalization timescales in the dual CFT. An explicit check of this statement is possible in the regime where the CFT  admits a near equilibrium, long wavelength effective hydrodynamic description. We have explicitly checked that for BHs with radius much larger than the AdS length (and small rotations in the four dimensional case), the long wavelength gravitational QNMs of stationary BHs match  the hydrodynamic relaxation timescales of the dual CFT. This confirms that the holographic interpretation of the QNM spectrum extends to systems with a rotating chemical potential. It is also a further check of the validity of the shear viscosity to the entropy density bound, $\eta/s=1/(4\pi)$, and a non-trivial confirmation that the global AdS BCs derived in the companion paper \cite{Dias:2013sdc} preserve the conformal boundary.

The damped QNM modes of a BH have a well defined dual CFT interpretation, but not much is known about the holographic interpretation of the superradiant instability (for discussions in this direction see\cite{Hawking:1999dp}). From the gravity side it is clear that the superradiance instability has to do with a quenched cooling of the system, since increasing the angular momentum very rapidly, cools the system down. It would be interesting to connect this interpretation with a simple CFT model for such a phenomena, where one could perhaps understand the final state of the system. In addition, it would also be important to understand the novel holographic phases or states that are dual to the  single KVF BHs and geons that appear in the superradiant context. 
From a different perspective, in the bulk we have discussed BHs only at the classical level. However, when quantum effects are included, Hawking radiation is also present and it is entangled with spontaneous superradiant emission. These phenomena should have a microscopic or statistical description.  Within string theory, certain BHs can be described by a configuration of  $D$-branes. In this context, Hawking radiation can be microscopically understood as the emission of a closed string off the $D$-branes as a result of the collision of two open strings that are attached to the $D$-branes  (see \cite{Dias:2007nj} and references therein). Similarly, superradiant emission has a microscopic description in terms of collisions of fermionic (spinning) left and right moving string excitations, and the superradiant condition $\omega\leq m\Omega_h$ follows from the Fermi-Dirac statistics for the fermionic open strings  \cite{Dias:2007nj}. It would be interesting to extend this microscopic description to  BHs (like Kerr-AdS) that do not have a $D$-brane description.

The properties of spheroidal wavefunctions and eigenvalues are known analytically for some time in AF spacetimes~\cite{Berti:2005gp}.
Our (numerical) analysis leaves the corresponding analytical analysis of spheroidal harmonics in AAdS unexplored, but clearly a compelling topic.
Specially interesting is the extremal regime $a=L$, which might be amenable to a full analytic treatment, both in the angular eigenvalue and in the eigenfrequency.

In a similar vein, a detailed analysis of superradiance in extremal, AF and AAdS geometries is seemingly lacking. Quasinormal mode results for the extremal, AF Reissner-Nordstrom geometry uncovered an interesting symmetry between different perturbations~\cite{Onozawa:1995vu} which might propagate to superradiant amplification factors and to other geometries. Note that our analysis does not apply to the extremal Kerr-AdS BH because it has a double horizon and thus our BCs are not appropriate. However, as one approaches the extremal BH, superradiance emission persists as first observed for the Kerr BH in \cite{Page76}. An interesting observation is that in the AF case, when the Kerr BH is extremal and the perturbations have a frequency that saturates the superradiant bound, i.e. $\omega=m \Omega_h^{\rm ext}$, the radial Teukolsky equation has an exact solution in terms of hypergeometric functions \cite{Teukolsky:1974yv}. However, for the Kerr-AdS BH we can no longer solve the radial equation analytically, even in the above particular conditions \cite{Dias:2012pp}. 
Extremal Kerr(-AdS) BHs are also interesting because they have a near-horizon limit where a Kerr/CFT correspondence can be formulated (see e.g.  \cite{Dias:2009ex,Dias:2012pp} and references therein).
The study of gravitational perturbations in the Kerr(-AdS) near-horizon geometries was done in  \cite{Dias:2009ex,Dias:2012pp}. It is interesting to note that all frequencies in the near-horizon geometry correspond to the single frequency $\omega=m \Omega_h^{\rm ext}$ in the original full geometry. Probably this is the reason why no signature of superradiance is found in the near-horizon geometry. It might be useful to explore further this system since its radial solutions are analytical.

We conclude this discussion section with some important general remarks concerning perturbations of AAdS spacetimes. One might wonder whether Robin boundary conditions, such as the ones used throughout this paper, lead to a well defined initial value problem for fields propagating in arbitrary asymptotically AdS backgrounds. This has been shown to be the case for the propagation of a real scalar field in \cite{Warnick:2012fi,Holzegel:2012wt}, where no assumption about the stationarity of the background or separability of the wave equation was made. The proof given there can be readily extended to complex of multi-component fields, including the gravitational perturbations discussed here. In the absence of a linear instability, one might think that linear perturbations about Kerr-AdS will decay exponentially with time, in a manner dictated by the QNM spectrum. However, it turns out that this is not the generic case, and indeed, depending on the smoothness of the initial data, the decay might be a lot slower than that. A simple argument suggest logarithmic decay: modes with very large angular momentum have a very large timescale, their growth rate can be shown (using for instance the WKB approximation) to decay exponentially with increasing angular quantum number $\ell$, i.e. $\tau\equiv\mathrm{Im}(\omega)^{-1} \sim \exp(\alpha \ell)$, where $\alpha$ is independent of $\ell$. This suggests that very long timescales can be achieved if the initial data contains support in very large angular momentum quantum numbers, that is to say $\ell \sim \log \tau$. Since the initial data has to live in a Sobolev space of sufficiently high order, we conclude that $||\psi||\sim (\log \tau)^{-p}$, where $p$ is related to the Sobolev norm we are considering. Note that even real analytic data might not decay exponentially, this would correspond to taking the limit $p\to+\infty$, which would lead to $||\psi||\sim \tau^{-\beta}$, for some constant $\beta$. The logarithmic behavior has been rigorously shown to be sharp in \cite{Holzegel:2013kna,Holzegel:2011uu}. Perhaps more worrying, the long time behavior of generic perturbations about black hole in global AdS might not even be related with quasinormal modes at all! In \cite{Warnick:2013hba}, it was shown, by counterexample, that quasinormal modes do not form a complete basis and that some perturbations can never be described by their dynamics. However, we should stress that in the presence of a linearly unstable mode, such as a superradiant instability, the linear spectrum of perturbations does provide an accurate description of the dynamics at early times.

\section*{Acknowledgments}
It is a pleasure to thank D. Marolf and G. T. Horowitz for helpful discussions.
This work was supported in part by the NRHEP 295189 FP7-PEOPLE-2011-IRSES Grant,
the ANR grant 08-JCJC-0001-0, the ERC Starting Grant 240210 - String-QCD-BH, NSERC Discovery Grant (LL) and CIFAR (LL). 
J.E.S.'s work is partially supported by the John Templeton Foundation. V.C. acknowledges partial financial
support provided under the European Union's FP7 ERC Starting Grant ``The dynamics of black holes:
testing the limits of Einstein's theory'' grant agreement no. DyBHo--256667. G.S.H.'s work is partially supported by the National Science Foundation under Grant No. PHY12-05500.

This research was supported in part by Perimeter Institute for Theoretical Physics. 
Research at Perimeter Institute is supported by the Government of Canada through 
Industry Canada and by the Province of Ontario through the Ministry of Economic Development 
\& Innovation.

\begin{appendix}

\section{Fluxes across the horizon and asymptotic boundary  \label{sec:Fluxes}}

In this Appendix we explicitly show that the energy and angular momentum fluxes across the asymptotic  boundary vanish if we impose boundary conditions (BCs) that preserve the conformal metric. We also review why the flux across the horizon is proportional to the superradiant factor. 

The energy and angular momentum fluxes of gravitational perturbations are calculated using the
Landau-Lifshitz ``pseudotensor" whose definition we review next (see e.g. \cite{Dias:2009ex}). Consider metric perturbations
$h_{\mu\nu}$ around a background $\bar{g}_{\mu\nu}$ up to second order in the amplitude,
 \begin{equation}
 g_{\mu\nu}=\bar{g}_{\mu\nu}+h_{\mu\nu}=\bar{g}_{\mu\nu}+h^{(1)}_{\mu\nu}+h^{(2)}_{\mu\nu}+\mathcal{O}{(h^3)}.
\end{equation} 
The linearized Einstein equation reads
 \begin{equation}
G^{(1)}_{\mu\nu}[h^{(1)}]=0 \,. \label{Einstein1st}
\end{equation}
At second order, the Einstein equation relates terms linear in
$h^{(2)}$  to terms quadratic in $h^{(1)}$:
 \begin{equation}
G^{(1)}_{\mu\nu}[h^{(2)}] = -G^{(2)}_{\mu\nu}[h^{(1)}]\ \equiv 8\pi G {\cal T}_{\mu\nu}[h^{(1)}] \,, \label{Einstein2nd}
\end{equation}
where the RHS is quadratic in $h^{(1)}$. Written out explicitly, for
generic perturbations it reads (here, we use the notation
$h_{\mu\nu}\equiv h_{\mu\nu}^{(1)}$; if we choose the traceless-transverse gauge this is known as the Landau-Lifshitz ``pseudotensor"):
 \begin{eqnarray}
 8 \pi G_N {\cal T}_{\mu\nu}&=& -\frac{1}{2}{\biggl[}\frac{1}{2}\left( \nabla_\mu h_{\alpha\beta}\right) \nabla_\nu
 h^{\alpha\beta}
 + h^{\alpha\beta}\left( \nabla_\nu \nabla_\mu h_{\alpha\beta}+\nabla_\alpha \nabla_\beta h_{\mu\nu}
   - \nabla_\alpha \nabla_\mu h_{\nu\beta} -\nabla_\alpha \nabla_\nu
   h_{\mu\beta}\right) \nonumber\\
&& \hspace{0.7cm}+  \nabla_\alpha h^{\beta}_{\:\:\mu}\left(
\nabla^\alpha h_{\beta\nu}-\nabla_\beta h^{\alpha}_{\:\:\nu} \right) -
\nabla_\alpha h^{\alpha\beta}\left( \nabla_\mu h_{\beta\nu}+\nabla_\nu
h_{\mu\beta}-\nabla_\beta h_{\mu\nu} \right)\nonumber \\
&& \hspace{0.7cm}+\frac{1}{2} \nabla_\alpha h \left( \nabla_\mu h_{\beta\nu}+\nabla_\nu
h_{\mu\beta}-\nabla_\beta h_{\mu\nu} \right)  {\biggl]}\nonumber\\
&& +\frac{1}{4}\,\bar{g}_{\mu\nu}{\biggl[}\frac{1}{2}\left(
\nabla_\gamma h_{\alpha\beta}\right) \nabla^\gamma
 h^{\alpha\beta}
 + h^{\alpha\beta}\left( \nabla_\gamma \nabla^\gamma h_{\alpha\beta}
   - 2 \nabla_\alpha \nabla^\gamma h_{\gamma\beta} \right) - 2 \left(
\nabla_\alpha h^{\alpha\beta}\right) \nabla^\gamma
h_{\beta\gamma} \nonumber\\
&&\hspace{1.3cm} +  \nabla_\alpha h^{\beta\gamma}\left( \nabla^\alpha
h_{\beta\gamma}-\nabla_\beta h^{\alpha}_{\:\:\gamma} \right)  +\frac{1}{2}\nabla_\alpha h \left( 2\nabla_\beta h^{\alpha\beta} -\nabla^\alpha h \right){\biggl]}. \label{gravTLL}
\end{eqnarray}
We can now define the fluxes associated with the first order perturbation.

Let $\xi$ be one of the Killing vector fields $\xi=\partial_t$ or $\xi=-\partial_\phi$ of (Kerr-)AdS, that are conjugate to the energy ($E$) and angular momentum ($J$) of the solution, respectively. Conservation of the ``pseudotensor" ${\cal T}_{\mu\nu}$,
$\nabla_\mu {\cal T}^{\mu\nu}=0$, and the Killing equation, $\nabla_{(\mu}\xi_{\nu)}=0$, imply that the 1-form $\mathcal{J}_\mu=-T_{\mu\nu} \xi^\nu$  is conserved, $d\star \mathcal{J}=0$, where $\star$ is the Hodge dual. 
We can then define the energy or angular momentum flux across a hypersurface $\Sigma$ (like the horizon or the asymptotic boundary) as   
\begin{equation}
\Phi_\xi \equiv -\int_\Sigma \star \mathcal{J}
 = -\int_\Sigma \!\!dV_\Sigma \;{\cal T}_{\mu\nu}\xi^\mu n^\nu \label{def:flux}
 \end{equation}
where $n^\nu$ is the normal vector to $\Sigma$ and $dV_\Sigma$ is the induced
volume on $\Sigma$. 

Consider first the asymptotic boundary  $\Sigma=\Sigma_\infty$ which is the timelike hypersurface defined by $z=0$ (where $z$ is the FG radial coordinate). This has unit normal $n=z/L\, dz$. As discussed in association with the FG expansion \eqref{introFG}, AAdS backgrounds start differing from each other only at order ${\cal O} \left( g^{(d)}z^{d-2} \right)$. This in particular also implies that the most general perturbation of a global AdS background that preserves the asymptotic structure of the background has an asymptotic expansion around $z=0$ that starts at order ${\cal O} \left( z^{d-2} \right)$, i.e. $h_{z \mu}=0$ and $h_{ab}=e^{-i\omega t} e^{i m \phi} f(X) z^d+\cdots $ (the Fourier decomposition in $t,\phi$ follows from the fact that these directions are isometries of the background). Inserting this general perturbation in the  ``pseudotensor"  \eqref{gravTLL} and computing the fluxes  \eqref{def:flux} we find that they vanish because the integrand of the fluxes has a polynomial expansion that starts at ${\cal O} \left( z^d \right)$ (for both Killing fields):
\begin{equation}
\Phi_\xi{\bigl |}_\infty= -\int_{\Sigma_\infty} \star \mathcal{J}=0 \,. \label{fluxI}
 \end{equation}
 That is, perturbations that preserve the conformal metric (the static Einstein Universe) have vanishing energy and angular momentum fluxes at the asymptotic boundary. 

Take now the Killing horizon (null) hypersurface, $\Sigma=\Sigma_H$, defined by $r=r_+$.
To find the flux across the horizon we work with the ingoing Eddington-Finkelstein coordinates $\{v,r,\chi,\widetilde{\phi} \}$, introduced in \eqref{EF}, that extend the solution through the horizon. The horizon generator is by definition normal to the horizon, i.e. $n\equiv K=\partial_v+\Omega_h \partial \widetilde{\phi}$.  The metric perturbation $h_{\mu\nu}\equiv h_{\mu\nu}^{(1)}$ is constructed applying a differential operator to the Teukolsky variable $\delta \Psi _4$ (this is known as the Hertz map; see the companion paper \cite{Dias:2013sdc}). This yields  long expressions for the components of $h_{\mu\nu}$ that are not at all illuminating. The keypoint is that inserting them in  the ``pseudotensor"  \eqref{gravTLL} we find that the fluxes across the horizon are proportional to the superradiant factor ($C_\xi$ are positive constants if $\{\omega>0,m>0\}$),\footnote{This property is universal to scalar, electromagnetic and gravitational perturbations. A massless real scalar field perturbation obeying the Klein-Gordon equation is the simplest case that illustrates the origin of the superradiant factor. Indeed, inserting a scalar perturbation $\Psi=e^{-i\omega v} e^{i m \widetilde{\phi} }\Psi(r,\chi) +c.c.$ in its energy-momentum tensor ${\cal T}_{\mu\nu}=\partial_\mu\Psi\partial_\nu\Psi-(1/2)\left(\partial\Psi\right)^2$, and computing the flux vector across the horizon we find
$$-n^\mu \xi^\nu{\cal T}_{\mu\nu}=-n^\mu\partial_\mu\Psi  \xi^\nu\partial_\nu\Psi=- (\partial_v \Psi+\Omega_h \partial_{\widetilde{\phi}} \Psi )  \xi^\nu\partial_\nu\Psi=-(\omega-m\Omega_h) c_\xi \left[{\rm Re}(-i \Psi)\right ]^2,$$ where $c_{t}=\omega$ and $c_{\widetilde{\phi}} =m$. We used $n\equiv K=\partial_v+\Omega_h \partial_{\widetilde{\phi}}$ and $\xi\cdot K{\bigl |}_H=0$.}
\begin{equation}
\Phi_\xi{\bigl |}_H=- \int_{\Sigma_H} \star \mathcal{J}= -(\omega-m\Omega_h) C_\xi \,. \label{fluxH}
 \end{equation}
That is, these fluxes are negative (inwards the BH) if $\omega>m\Omega_h$ for which we perturbations are damped (QNMs);  positive (outwards the BH) if $\omega<m\Omega_h$ in which case this energy and angular momentum fluxes feed the superradiant instability growth; and finally they vanish when $\omega=m\Omega_h$, i.e. at the onset of superradiance.

\section{Details of the hydrodynamic QNM computation ($D=4$)   \label{sec:HydroAppendix}}

In this  appendix  we  give details of  the hydrodynamic computation that leads to the frequency quantization \eqref{CFTfreqScalar} and \eqref{CFTfreqVector}.

Our starting point is the double expansion \eqref{fluidExpansion} in the shear viscosity and in the rotation, both for  the fluid perturbations introduced in \eqref{pertQ}, $\{Q^{(f)}(X)\}=\{Q^{(1)},Q^{(2)},Q^{(3)}\}\equiv \{ \delta P/L, \delta u_X, \delta u_\Phi\}$, and for the perturbation frequency $\omega$.
These expansions are inserted in the hydrodynamic equations of motion \eqref{consTab} or \eqref{fluidEOM} that are then solved progressively in a series expansion in $\eta/L^3$ and $a/L$. For our purpose it will be enough to go up to first order in the viscosity ($n=1$) and up to second order in the rotation ($p=2$) expansions.  
There are two families of modes, namely the scalar and the vector modes. 

\subsection{Scalar modes  \label{sec:HydroScalars}}

Consider first the scalar modes. At leading order in the aforementioned expansion, the viscosity and rotational effects are absent, and we are interested in finding the quantities $ S_{0,0}^{(f)}$ and $\omega_{0,0}$ (for scalar modes we use the notation $ S_{j,i}^{(f)}\equiv Q_{j,i}^{(f)}$ ). In these conditions, the pressure perturbation is proportional to the Kodama-Ishibashi scalar harmonic $S(X,\Phi)\sim e^{i m \Phi}P_{\ell}^m(X) $, where $P_{\ell}^m(x)$ is the associated Legendre polynomial, while the velocity perturbation is proportional to the  vector derived scalar harmonics obtained by taking angular derivatives of the scalar harmonic $ \scalar_i \propto D_i \scalar$ (where
 $D_j$ is the covariant derivative associated to the unit radius metric on $S^2$).  We thus have $S_{0,0}^{(1)}\,e^{i m \Phi}=A_1\,e^{i m \Phi} P_{\ell }^m(X)$, $S_{0,1}^{(2)}\,e^{i m \Phi}=A_2\,e^{i m \Phi} P_{\ell }^m(X)'$ and $S_{0,1}^{(3)}\,e^{i m \Phi}=i \,m \,A_2\,e^{i m \Phi} P_{\ell }^m(X)$, for arbitrary amplitudes $A_k$. Inserting these expressions in the equations of motion (EoM) we fix the ratio $A_1/A_2$ and quantize the frequency $\omega_{0,0}$. This yields 
(we introduce the notation $z_+=\frac{r_+}{L}$)
\begin{eqnarray}\label{S:Order00}
&& S_{0,0}^{(1)}=i \,A_2 \,z_+ \left(1+z_+^2\right) \frac{ \sqrt{\ell (\ell +1)}}{\sqrt{2}} P_{\ell }^m(X) , \nonumber\\
&& S_{0,0}^{(2)}=A_2 \,P_{\ell }^m(X)' \,,\nonumber\\
&& 
S_{0,0}^{(3)}=i \,m \,A_2 \,P_{\ell }^m(X),
\end{eqnarray}
and $\omega_{0,0}$ that can be read from  \eqref{CFTfreqScalar}. This conclusion agrees with the static results  first  derived in  \cite{Michalogiorgakis:2006jc}. 

Still at leading order in the viscosity, we now consider the first order correction introduced by the rotation.
It follows from two of the EoM at this order that the perturbations $Q^{(2)}_{0,1}$ and $Q^{(3)}_{0,1}$ can be algebraically expressed as a function of  $S^{(2)}_{0,1}$ and/or its derivative. Plugging these relations in the third EoM we fix the frequency correction $\omega_{0,1}$ as written in \eqref{CFTfreqScalar} (this is done doing the procedure exemplified below for the $\omega_{0,2}$ conribution) and the differential equation for $S^{(1)}_{0,1}$ that is left is the familiar associated Legendre equation.  
Altogether, the perturbation eigenfunctions at order ${\cal O}\left(\eta^0,a^1\right)$ are then
\begin{eqnarray}\label{S:Order01}
&&  S^{(1)}_{0,1}=B_0\,P_{\ell }^m(X),\nonumber\\
&& S^{(2)}_{0,1}=\frac{A_2 m z_+ \left(1+z_+^2\right) \left(\ell ^2+5\ell +2\right)-2 i B_0 \ell  (\ell +1)}{\sqrt{2} z_+ \left(1+z_+^2\right)\left[ \ell  (\ell +1)\right]^{3/2}} \, \frac{(\ell +1)X}{ 1-X^2}P_{\ell }^m(X)    \nonumber\\
&& \hspace{1.2cm}-\frac{A_2 m z_+ \left(1+z_+^2\right) \left(\ell ^2+\ell +2\right)-2 i B_0 \ell  (\ell +1)}{\sqrt{2} z_+ \left(1+z_+^2\right)\left[ \ell  (\ell +1)\right]^{3/2}} \, \frac{(\ell +1-m) }{ 1-X^2}P_{\ell +1}^m(X), \nonumber\\
&& 
S^{(3)}_{0,1}=\frac{ i\,A_2 z_+ \left(1+z_+^2\right) \left(m^2 \left(\ell ^2+\ell +2\right)+\ell  (\ell +1)^2 \left(X^2 (\ell +4)-\ell \right)\right)-2 i B_0 m \ell  (\ell +1)}{\sqrt{2} z_+ \left(1+z_+^2\right)\left[ \ell  (\ell +1)\right]^{3/2}}\,P_{\ell }^m(X)  \nonumber\\
&& \hspace{1.2cm} +\frac{2 i \sqrt{2} A_2 \ell  (\ell +1) (m-\ell -1)}{\left[ \ell  (\ell +1)\right]^{3/2}}\,X P_{\ell +1}^m(X).
\end{eqnarray}
where $B_0$ is a new arbitrary amplitude that is introduced at this order.

We can improve our approximation by finding the correction up to second order in the rotation (at this point still at vanishing viscosity). This requires looking to the EoM at order ${\cal O}\left(\eta^0,a^2\right)$ that involve the unknown quantities $S_{0,2}^{(f)}$ and $\omega_{0,2}$. We use this case to exemplify in detail how we typically solve equations of our problem to get the perturbative frequency corrections. 
Two of the EoM at order ${\cal O}\left(\eta^0, a^2 \right)$ yield two algebraically equations for $S^{(2)}_{0,2}$ and $S^{(3)}_{0,2}$ in terms of $S^{(1)}_{0,2}$ and its derivative (in addition to Legendre polynomial contributions sourced by the lower order solutions). Inserting these algebraic relations in the third EoM we get a second order ODE for $S_{0,2}^{(1)}$. Explicitly, the equations discussed in this paragraph are:  
\begin{eqnarray}\label{S:Order02}
&& \hspace{-0.2cm} S_{0,2}^{(1)}(X)''-\frac{2 X}{1-X^2}S_{0,2}^{(1)}(X)'-\frac{m^2+\left(X^2-1\right) \ell  (\ell +1)}{\left(1-X^2\right)^2}S_{0,2}^{(1)}(X) 
\nonumber\\
&&\hspace{1.4cm}
+\frac{i A_2 z_+ \left(1+z_+^2\right) \left(5 \ell ^2+5 \ell +16\right)}{\sqrt{2} \sqrt{\ell  (\ell +1)}}X P_{\ell }^m(X)'
\nonumber\\
&&\hspace{1.4cm}
+\frac{i A_2 z_+ \left(1+z_+^2\right)}{2 \sqrt{2} (\ell  (\ell +1))^{3/2}} \frac{ P_{\ell }^m(X)}{1-X^2}
 {\biggl[}m^2 \left[12+\ell  (\ell +1) \left(\ell ^2+\ell +4\right)\right]  \nonumber\\
&&\hspace{2.4cm}
-\ell ^2 (\ell +1)^2 \left[20-\left(1-X^2\right) \left(\ell ^2+\ell +22\right)-4 \sqrt{2} \sqrt{\ell  (\ell +1)}L \omega _{0,2}\right] {\biggr ]}=0
\nonumber\\
&&\hspace{-0.2cm}S_{0,2}^{(2)} (X)=-\frac{i \sqrt{2}S_{0,2}^{(1)}(X)'}{z_+ \left(1+z_+^2\right) \sqrt{\ell  (\ell +1)}}+P_{\ell }^m(X)'\left[-\frac{i B_0 m \left(\ell ^2+\ell +2\right)}{z_+ \left(z_+^2+1\right) \ell ^2 (\ell +1)^2}+A_2 \left(\frac{m^2 \left(\ell ^2+\ell +2\right)^2}{2 \ell ^3 (\ell +1)^3}\right.\right.  \nonumber\\
&& \hspace{1.5cm} 
\left.\left.+\frac{2 \sqrt{2} X^2 z_+^2 \left(\ell ^2+\ell +4\right)-\sqrt{2} \left(2 z_+^2+1\right) \ell  (\ell +1)-2 \omega _{0,2}L z_+^2 \sqrt{\ell  (\ell +1)}}{\sqrt{2} z_+^2 \ell  (\ell +1)}\right)\right]
 \nonumber\\
&& \hspace{1.6cm} 
  +\left(\frac{A_2 \left[4 m^2 \left(\ell ^2+\ell +2\right)+ \ell ^2 (\ell +1)^2\left(1-X^2\right)\right]}{\ell  (\ell +1)}-\frac{4 i B_0 m}{z_+ \left(z_+^2+1\right)}\right)\frac{X P_{\ell }^m(X) }{ \ell  (\ell +1)\left(1-X^2\right)}\nonumber\\
&&\hspace{-0.2cm} S_{0,2}^{(3)} =\frac{\sqrt{2} m S_{0,2}^{(1)}(X)}{z_+ \left(1+z_+^2\right)\sqrt{\ell  (\ell +1)}}+P_{\ell }^m(X) \left(\frac{i A_2 m^3 \left(\ell ^2+\ell +2\right)^2}{2 \ell ^3 (\ell +1)^3}+\frac{2 i A_2 m X^2 \left(\ell ^2+\ell +4\right)}{\ell  (\ell +1)}\right. \nonumber\\
&& \hspace{1cm} 
\left.-\frac{i A_2 m \left(2 z_+^2+1\right)}{z_+^2}-\frac{i \sqrt{2} A_2 m L \omega _{0,2}}{\sqrt{\ell  (\ell +1)}}+\frac{B_0 \left[m^2 \left(\ell ^2+\ell +2\right)+\left(X^2-1\right) \ell ^2 (\ell +1)^2\right]}{z_+ \left(1+z_+^2\right) \ell ^2 (\ell +1)^2}\right) \nonumber\\
&& \hspace{1cm} 
+\frac{4 i X \left(1-X^2\right) P_{\ell }^m(X)'}{z_+ \left(1+z_+^2\right)\ell ^2 (\ell +1)^2}\left[A_2 m z_+ \left(1+z_+^2\right) \left(\ell ^2+\ell +2\right)-i B_0 \ell  (\ell +1)\right].
\end{eqnarray}
Note that the ODE for $S_{0,2}^{(1)}$ is of the form
$f_1S_{0,2}^{(1)}{}^{ \prime \prime }+f_2S_{0,2}^{(1)}{}^{ \prime }+f_3S_{0,2}^{(1)}+s_2P_{\ell }^m{}^{ \prime }+s_1P_{\ell }^m=0$, with $f_k$ and $s_k$ being functions of $X$ that can be read from the first equation in \eqref{S:Order02}. Contracting this equation with $\int dX P_{\ell }^m$ we can now use the properties of integration by parts. Namely, we can subtract the vanishing total divergence contribution  $\int dX \partial_X\! \left( P_{\ell }^m f_1 S_{0,2}^{(1)}{}^{ \prime }\right)$ to the previous equation and integrate by parts the $\int dX P_{\ell }^mf_1S_{0,2}^{(1)}{}^{ \prime \prime }$ term to rewrite the ``EoM" as $\int dX P_{\ell }^m\left[\hat{f}_2S_{0,2}^{(1)}{}^{ \prime }+f_3Q_{0,2}^{(1)}+\hat{s}_2 P_{\ell }^m{}^{ \prime }+s_1P_{\ell }^m\right]=0$, where we have redefined the coefficients $f_2\to\hat{f}_2$ and $s_2\to \hat{s}_2$ to absorb the new contributions arising from the integration by parts. We use again a similar approach, namely we subtract the total divergence term $\int dX \partial_X \! \left( P_{\ell }^m\hat{f}_2S_{0,2}^{(1)}\right)=0$ and use integration by parts to  get  $\int dX P_{\ell }^m \left(\tilde{s}_2P_{\ell }^m{}^{ \prime }+s_1P_{\ell }^m\right)=0$ where we made the redefinition $\hat{s}\to \tilde{s}_2$  and a would be $S_{0,2}^{(1)}$ contribution is absent since  $f_3-\hat{f}_2{}^{ \prime}=0$. Subtracting the total divergence $\int dX \partial _X \!\left[ \tilde{s}_2\left(P_{\ell }^m\right){}^2\right]$ and a third final integration by parts finally yields $\int dXP_{\ell }^m \hat{s}_1 P_{\ell }^m=0$ with $\hat{s}_1=\tilde{s}_2{}^{ \prime }+s_1$. Explicitly, this final condition is
\begin{eqnarray}\label{w02aux1}
&&  \hspace{-2cm} \frac{A_2 z_+ \left(1+z_+^2\right) }{2 \ell  (\ell +1)}{\biggl\{} \left[\ell  (\ell +1) \left(\ell ^2+\ell +7\right)-48\right]\int_{-1}^{1} dX\, X^2 \,P_{\ell }^m(X){}^2\nonumber\\
&&\hspace{0.8cm}  -{\biggl [} \ell  (\ell +1) \left[\ell  (\ell +1) \left(4 \sqrt{2} \sqrt{\ell  (\ell +1)} \omega _{0,2}+\ell ^2+\ell -3\right)-16\right]\nonumber\\
&&\hspace{1.3cm}
+m^2 \left[\ell  (\ell +1) \left(\ell ^2+\ell +4\right)+12\right] {\biggr ]} \frac{\int_{-1}^{1} dX\, P_{\ell }^m(X){}^2}{\ell  (\ell +1)}   {\biggr\}} =0\,.
\end{eqnarray}
To proceed we use the integrals
\begin{eqnarray}\label{w02aux2}
&&  \int P_{\ell }^m(X)P_{\ell }^m(X)dX= \frac{2}{(2\ell +1)}\frac{(\ell +m)!}{(\ell -m)!}\,  \nonumber\\
&&\int X^2 P_{\ell }^m(X)P_{\ell }^m(X)dX=\frac{2\left(2 \ell ^2+2\ell  -2 m^2-1\right)}{(2\ell -1)(2\ell +1)(2\ell +3)}\frac{(\ell +m)!}{(\ell -m)!}\,,
\end{eqnarray}
to  rewrite \eqref{w02aux1} as
\begin{eqnarray}
&& \hspace{-0.2cm}0=\frac{A_2 z_+ \left(1+z_+^2\right) (2 \ell -3)\text{!!} (\ell +m)!}{\ell ^2 (\ell +1)^2 (2 \ell +3)\text{!!} (\ell -m)!}
{\biggl \{}3 m^2 (\ell -1) (\ell +2) \left(\ell ^2+\ell +6\right) \left(2 \ell ^2+2 \ell +1\right)  \nonumber\\
&&\hspace{0.2cm}+2 \ell ^2 (\ell +1)^2 \left[2 \sqrt{2} \sqrt{\ell  (\ell +1)} (2 \ell -1) (2 \ell +3) \omega _{0,2}L+(\ell +1) \ell ^3+(\ell +1) \ell ^2-14 (\ell +1) \ell +24\right]\!\!{\biggr \}}. \nonumber
 \end{eqnarray}
This condition finally quantizes the frequency contribution $\omega_{0,2}$ as  
\begin{equation}\label{w02}
\hspace{-0.3cm}\omega_{0,2}L=-\frac{(\ell +2) (\ell -1) \left[2 (\ell -3) (\ell +4)\ell ^2 (\ell +1)^2+3 m^2 \left(6+\ell +\ell ^2\right) \left(1+2 \ell +2 \ell ^2\right)\right]}{4 \sqrt{2}(2 \ell -1) (2 \ell +3) [\ell  (\ell +1)]^{5/2}}\,.
 \end{equation}

To include the effects of dissipation we now consider the linear order contribution in the viscosity, while still doing also an expansion in the (adimensional) rotation parameter, i.e. we solve the  perturbative EoM at order ${\mathcal O}\left( \eta, a^0\right)$, ${\mathcal O}\left( \eta, a^1\right)$, ${\mathcal O}\left( \eta, a^2\right)$. The technical analysis proceeds in a way that is very similar to the procedure already outlined for the zero-order contribution in the viscosity so we now omit further details and just give the final results for the frequencies $\omega_{1,i}$ and for the perturbation eigenfunctions $S_{1,i}^{(f)}(X)$.
At order  ${\mathcal O}\left( \eta, a^0\right)$ the eigenfunctions are
\begin{eqnarray}\label{S:Order10}
&& S_{1,0}^{(1)}=\left[i \frac{1}{\sqrt{2}} \,K_2 \,z_+ \left(1+z_+^2\right)\sqrt{\ell  (\ell +1)}-\frac{1}{2}\, A_2\,(\ell +2)(\ell -1)\right]P_{\ell }^m(X) , \nonumber\\
&& S_{1,0}^{(2)}=K_2\, P_{\ell }^m(X)' ,\nonumber\\
&& 
S_{1,0}^{(3)}=i\, m\, K_2 \,P_{\ell }^m(X). 
\end{eqnarray}
where $K_2$ is a new arbitrary amplitude, and the frequency $\omega_{1,0}$ is written in  \eqref{CFTfreqScalar}. 

At order  ${\mathcal O}\left( \eta, a^1\right)$ the eigenfunctions are
\begin{eqnarray}\label{S:Order11}
&&  S^{(1)}_{1,1}=C_0\,P_{\ell }^m(X) ,\nonumber\\
&& S^{(2)}_{1,1}= 
\frac{i X P_{\ell }^m(X)}{2 \sqrt{2} \left(1-X^2\right) z_+ \left(1+z_+^2\right) \ell ^2 (\ell +1)}{\biggl [}4 \sqrt{2} A_2 m (\ell +1) (\ell +2) (2 \ell +1)  \nonumber\\
&&  \hspace{2cm} -\frac{2 i}{z_+ \left(1+z_+^2\right)} {\biggl (} K_2 m z_+^2 \left(1+z_+^2\right)^2 \sqrt{\ell  (\ell +1)} \left(\ell ^2+5 \ell +2\right) \nonumber\\
&&  \hspace{4.6cm}-\ell  (\ell +1) {\bigl [}\sqrt{2} B_0 (\ell -1) (\ell +2)  +2 i C_0 z_+ \left(1+z_+^2\right) \sqrt{\ell  (\ell +1)}{\bigr ]} {\biggr )}{\biggr ]}
   \nonumber\\
&& \hspace{1.2cm} 
-\frac{i (\ell +1-m) P_{\ell +1}^m(X)}{2 \sqrt{2} \left(1-X^2\right) z_+^2 \left(1+z_+^2\right){}^2 \ell ^2 (\ell +1)^2} {\biggl (} 4 \sqrt{2} A_2 m z_+ \left(1+z_+^2\right) \left(5 \ell ^2+5 \ell +2\right) \nonumber\\
&& \hspace{2cm} -2 i \left[K_2 m z_+^2 \left(1+z_+^2\right)^2 \sqrt{\ell  (\ell +1)} \left(\ell ^2+\ell +2\right)-\sqrt{2} B_0 (\ell -1) \ell  (\ell +1) (\ell +2)\right] \nonumber\\
&&  \hspace{2cm}-4 C_0 z_+ \left(z_+^2+1\right) [\ell  (\ell +1)]^{3/2}{\biggr )},
   \nonumber\\
   && 
S^{(3)}_{1,1}= \frac{P_{\ell }^m(X)}{2 z_+^2 \left(1+z_+^2\right)^2 \ell ^2 (\ell +1)^2} {\biggl [}-4 X^2 z_+ \left(1+z_+^2\right) \ell  (\ell +1)^2\nonumber\\
&&  \hspace{2cm} \times \left[2 A_2 \left(\ell ^2+\ell +1\right)-i \sqrt{2} K_2 z_+ \left(1+z_+^2\right) \sqrt{\ell  (\ell +1)}\right]\nonumber\\
&&  \hspace{2cm}
-A_2 z_+ \left(1+z_+^2\right) \left[4 m^2 \left(5 \ell ^2+5 \ell +2\right)+\left(1-X^2\right) \ell ^2 (\ell +1)^2 \left(\ell ^2+\ell -14\right)\right] \nonumber\\
&&  \hspace{2cm} +2 m \ell  (\ell +1) \left[\sqrt{2} C_0 z_+ \left(1+z_+^2\right) \sqrt{\ell  (\ell +1)}-i B_0 (\ell -1) (\ell +2)\right]\nonumber\\
&&  \hspace{2cm}  +i \sqrt{2} K_2 z_+^2 \left(1+z_+^2\right)^2 \sqrt{\ell  (\ell +1)} \left[m^2 \left(\ell ^2+\ell +2\right)-\left(1-X^2\right) \ell ^2 (\ell +1)^2\right]{\biggr ]}
\nonumber\\
&&  \hspace{1.2cm}+ \frac{2(\ell +1-m) X\,P_{\ell +1}^m(X)}{z_+ \left(1+z_+^2\right) \ell  (\ell +1)}\left[2 A_2 \left(\ell ^2+\ell +1\right)-i \sqrt{2} K_2 z_+ \left(1+z_+^2\right) \sqrt{\ell  (\ell +1)}\right],
\end{eqnarray}
where $C_0$ is a new arbitrary amplitude, and the frequency contribution $\omega_{1,1}$ can be found in  \eqref{CFTfreqScalar}. 

Finally,  to get the frequency correction at order ${\mathcal O}\left( \eta, a^2\right)$ we use again the integration by parts procedure that we already described to get the  ${\mathcal O}\left( \eta^0, a^2\right)$ contribution. Going through this procedure we find the frequency $\omega_{1,2}$ that can be read from  \eqref{CFTfreqScalar} and we omit here the associated long expressions for  $S_{1,2}^{(f)}$. The frequency contribution $\omega_{1,3}$ written in \eqref{CFTfreqScalar} is computed in a similar way. 

\subsection{Vector modes  \label{sec:HydroVectors}}
Consider now the vector modes. These distinguish from the scalar modes because at leading order in the viscosity and rotation they have vanishing pressure perturbation and vanishing frequency: $V_{0,0}^{(0)}=0$ and $\omega_{0,0}=0$ (for scalar modes we use the notation $ V_{j,i}^{(f)}\equiv Q_{j,i}^{(f)}$ ).. In these conditions it follows from the EoM that (as it could not be otherwise) the velocity perturbation of these modes can be expanded in terms of the Kodama-Ishisbashi vector harmonics ${\bf V}_i$, $i=X,\Phi$ (which can themselves be expressed as a function of the associated Legendre polynomials $P_{\ell }^m$; see e.g. section 4.1 of \cite{DS}). Altogether, at leading  order ${\mathcal O}\left( \eta^0, a^0\right)$, the vector hydrodynamic modes have eigenfunctions
\begin{eqnarray}\label{V:Order00}
&& V_{0,0}^{(1)}=0, \nonumber\\
&& V_{0,0}^{(2)}=\frac{i m A_3}{1-X^2} \,P_{\ell }^m(X)\,,\nonumber\\
&& V_{0,0}^{(3)}=A_3 \,(1-X^2) \,P_{\ell }^m(X)^{\prime} ,
\end{eqnarray}
and $\omega_{0,0}=0$.

The EoM at  ${\mathcal O}\left( \eta^0, a^1\right)$ and  ${\mathcal O}\left( \eta^0, a^2\right)$ combined give the frequency corrections $\omega_{0,1}$ and $\omega_{0,2}=0$.\footnote{To be more clarifying, the first order EoM determine $\omega_{0,1}=0$ but leave $V^{(2)}_{0,1}$ undetermined and $V_{0,1}^{(3)}$ is left just as a function of $V^{(2)}_{0,1}$ and its derivative. The explicit expressions for  $V^{(2)}_{0,1}$ and $V^{(3)}_{0,1}$,  as written in \eqref{V:Order01}, are found only at second order where we also determine $\omega_{0,2}$. Ultimately, this technical property of the vector modes is due to the fact that the frequency contribution of rotational odd powers vanish when $\eta=0$.}
The  eigenfunctions at order ${\mathcal O}\left( \eta^0, a^1\right)$ are
\begin{eqnarray}\label{V:Order01}
&&  V^{(1)}_{0,1}=\frac{2 A_3 z_+ \left(z_+^2+1\right)}{\ell  (\ell +1)} {\bigl [} (\ell +1-m) P_{\ell +1}^m(X)- (\ell +1)^2 X P_{\ell }^m(X) {\bigr ]},\nonumber\\
&& V^{(2)}_{0,1}= \frac{i\,m B_0}{1-X^2} P_{\ell }^m(X),   \nonumber\\
&& 
V_{0,1}^{(3)}=-B_0\, {\bigl [} (\ell +1)X \,P_{\ell }^m(X)-(\ell +1-m) P_{\ell +1}^m(X) {\bigr ]}, 
\end{eqnarray}
where $B_0$ is a new arbitrary amplitude  and  $\omega_{0,1}$ is given in \eqref{CFTfreqVector}.

To find the frequency contribution at order ${\mathcal O}\left( \eta^0, a^3\right)$ we use two of the EoM at the previous order  ${\mathcal O}\left( \eta^0, a^2\right)$  to find  $V^{(1)}_{0,2}$ explicitly and an algebraic relation for $V^{(3)}_{0,2}$ as a function of $V^{(2)}_{0,2}$ and its derivative. Then, one of the EoM at order ${\mathcal O}\left( \eta^0, a^3\right)$ is a second order ODE that only involves the unknown $V_{0,2}^{(2)} $ and its first and second derivatives, in addition to two source contributions proportional to  the associated Legendre polynomial and its derivative. It is used to find the frequency contribution  $\omega_{0,3}$ as given in \eqref{CFTfreqVector}, after using several integrations by parts as exemplified in the previous scalar mode treatment. The long relations associated to this discussion are omitted here.

We now consider the viscosity contributions.
Using the EoM at order  ${\mathcal O}\left( \eta, a^0\right)$ and ${\mathcal O}\left( \eta, a^1\right)$ we find  the eigenfunctions
\begin{eqnarray}\label{V:Order10}
&& V_{1,0}^{(1)}=0, \nonumber\\
&& V_{1,0}^{(2)}=\frac{i m K_2 }{1-X^2}P_{\ell }^m(X),\nonumber\\
&& 
V _{1,0}^{(3)}=K_2 [(\ell +1-m) P_{\ell +1}^m(X)-(\ell +1)X P_{\ell }^m(X)],
\end{eqnarray}
where $K_2$ is an arbitrary amplitude, and fix the frequency contribution $\omega_{1,0}$ as written in  \eqref{CFTfreqVector} and $\omega_{1,1}=0$. 

The  EoM at order ${\mathcal O}\left( \eta, a^1\right)$ also determine $V_{1,1}^{(1)}$ and $V _{1,1}^{(3)}$ that we write below, while  EoM at order ${\mathcal O}\left( \eta, a^21\right)$ find algebraic relations for $V_{1,2}^{(1)}$ and $V _{1,2}^{(3)}$ (that we do not write here) and a second order ODE for $V_{1,1}^{(2)}$. The relations just described are
\begin{eqnarray}\label{V:Order11}
&&V_{1,1}^{(1)}=\frac{2 K_2 z_+\left(1+z_+^2\right)}{\ell  (\ell +1)}\left[(\ell +1)^2X P_{\ell }^m(X)-(\ell +1-m) P_{\ell +1}^m(X)\right], \nonumber\\
&& 
V_{1,1}^{(2)}(X)^{\prime\prime}-\frac{6 X}{1-X^2}V_{1,1}^{(2)}(X)^\prime+\frac{(\ell -1) (\ell +2)-m^2-(\ell -2) (\ell +3)X^2}{\left(1-X^2\right)^2}V_{1,1}^{(2)}(X) 
 \nonumber\\
&& \hspace{1.4cm} +\frac{2 A_3 \left(8-\ell  (\ell +1) \left(\ell  (\ell +1) \left(\ell ^2+\ell -5\right)+14\right)\right)}{\left(1-X^2\right) z_+ \left(z_+^2+1\right) \ell  (\ell +1)}\,X\,P_{\ell }^m(X)' 
\nonumber\\
&& \hspace{1.4cm}-\frac{A_3 \,P_{\ell }^m(X)}{2 \left(1-X^2\right)^2}   {\biggl [} 
\frac{2 m^2 \left(3 \ell ^8+12 \ell ^7+16 \ell ^6+6 \ell ^5-25 \ell ^4-46 \ell ^3+58 \ell ^2+80 \ell -24\right)}{z_+ \left(z_+^2+1\right) \ell  (\ell +1) (2 \ell -1) (2 \ell +3)}\nonumber\\
&&\hspace{3.7cm}
-\frac{X^2 \left(\ell  (\ell +1) \left(\ell ^2 (\ell +1)^2-28\right)+32\right)}{z_+ \left(z_+^2+1\right)} 
+ \frac{(\ell -1) \ell ^2 (\ell +1)^2 (\ell +2)}{z_+^3 \left(z_+^2+1\right)}\nonumber\\
&&\hspace{3.7cm}
-\frac{2 \ell  (\ell +1) \left(-3 \ell ^6-9 \ell ^5+15 \ell ^3+10 \ell ^2+\ell +6\right)}{z_+ \left(z_+^2+1\right) (2 \ell -1) (2 \ell +3)}+i \omega _{1,2}L \ell ^2 (\ell +1)^2
  {\biggr ]} =0 \nonumber\\
&& 
V _{1,1}^{(3)}=\frac{i \left(1-X^2\right)}{m}{\biggl [} \left(1-X^2\right) V_{1,1}^{(2)}(X)'-2 X V_{1,1}^{(2)}(X)\nonumber\\
&&\hspace{3cm} +\frac{A_3 (\ell -1) (\ell +2) }{z_+\left(1+z_+^2\right)}\left(4 X P_{\ell }^m(X)-\frac{\ell ^2+\ell -4 }{\ell  (\ell +1)}\left(1-X^2\right) P_{\ell }^m(X)'\right){\biggr ]},
\end{eqnarray}
We use the ODE for $V_{1,1}^{(2)}$ to determine the frequency contribution $ \omega _{1,2}$, explicitly written in \eqref{CFTfreqVector}, after several integration by parts.

\section{QNMs and superradiance: a perturbative analytical analysis  ($D=4$) \label{sec:AnalyticalQNMAppendix}}

In this  appendix  we  give details of the perturbative matched asymptotic expansion that leads to the frequency quantization \eqref{FreqMatchingl2} and that we compare with the numerical results in Section \ref{sec:CompareNumAnal}. This perturbative approach was introduced to study  perturbations of a scalar field in the Kerr black hole by Starobinsky \cite{Starobinsky}, Unruh \cite{Unruh} and Detweiller \cite{detweiler}, and later used successfully to study scalar and gravitational perturbations in rotating black holes in \cite{Maldacena:1997ih,Cardoso:2004nk,Cardoso:2004hs,Cardoso:2006wa,Dias:2006zv}. 
In particular, the superradiant timescales of a scalar field in the Kerr-AdS black hole computed with this method \cite{Cardoso:2004hs} were confirmed to be accurate by the numerical analysis of  \cite{Uchikata:2009zz}.\footnote{Here we do not follow the alternative perturbative analysis of  \cite{Basu:2010uz,Bhattacharyya:2010yg,Dias:2011at,Dias:2011tj} to find QNM and superradiant frequencies because, in the present system, it requires going to a high order in perturbation theory $-$ the imaginary part of the frequency appears only at order ${\cal O}(r_+/L)^6$ $-$ where the source terms make it difficult to solve analytically the equations.}
  
The matched asymptotic expansion procedure allows to solve perturbatively the angular \eqref{MasterAng} and radial \eqref{MasterRad} equations, and yields an approximate analytical solution for the QNM and superradiant instability frequency spectra.

This analysis starts with the observation is that if we work in a regime of parameters where $\frac{a}{L} \ll 1$ and  $a \tilde{\omega} \ll 1$ the angular equation for the spin-weighted AdS-spheroidal harmonics reduces approximately to the standard equation for the spin-weighted spherical harmonics   \cite{Breuer,Berti:2005gp}. In particular it is independent of the mass parameter $M$ and cosmological radius $L$, and its regular solutions can be found analytically (see e.g. \cite{Dias:2013sdc} for a detailed construction). Here, it is important to highlight that regularity of these eigenfunctions requires that the angular eigenvalues and quantum numbers are quantized as
\begin{equation}\label{condAngN}
\lambda=(\ell -1) (\ell +2) -\frac{2 m}{\ell }\,\frac{\ell ^2+\ell +4}{\ell +1}\,a\tilde{\omega}  +{\cal O}\left( a^2\tilde{\omega}^2, \frac{a^2}{L^2} \right)\,, \qquad \hbox{with}\quad \ell=2,3,4,\cdots\,,\quad  |m|\leq \ell\,,
\end{equation}
where the azimuthal quantum number $m$ is an integer, and we have introduced the quantum number $\ell$ with properties discussed after  \eqref{KrKchi}. This fixes the angular eigenvalue spectrum and we just need to solve now the radial equation in a regime of parameters that is consistent with the approximation where  \eqref{condAngN} is valid.

We follow a standard matching asymptotic expansion analysis whereby we divide the exterior spacetime of the Kerr-AdS black hole into two regions; a near-region where $r-r_+\ll \frac{1}{\tilde{\omega} }$ and a far-region where $r-r_+\gg r_+$. In each of these regions, some of the terms in radial equation make a sub-dominant contribution and can be consistently discarded. We will find that if we further require  $\frac{r_+}{L} \ll 1$, this procedure yields an equation with an (approximate) analytical solution in both spacetime regions. The next important step is to restrict our attention to  the  regime $r_+\tilde{\omega} \ll 1$. In this regime, the far and near regions have an overlaping zone,  $r_+\ll r-r_+\ll \frac{1}{\tilde{\omega} }$, where the far and near region solutions are simultaneously valid. In this matching region, we can then match/relate the set of independent parameters that are generated in each of the two regions.  
We will also find that if we further restrict our analysis to the regime $\frac{a}{r_+}\ll 1$,  it is sufficient to work only with the the leading order contribution for the angular eigenvalues in \eqref{condAngN}, $\lambda\sim  (\ell -1) (\ell +2) $.

The regime of validity of the matching analysis can be expressed in a much simplified form.
Indeed, the rotation parameter is constrained by the extremity condition $a\leq r_+\sqrt{\frac{ 3 r_+^2+L^2}{L^2-r_+^2}}$ for $r_+<\sqrt{3} L$ and by $a<L$ for $r_+>\sqrt{3} L$ (see e.g.~\cite{Dias:2010ma}). For the regime we are interested,   $\frac{r_+}{L} \ll 1$, we thus have $a\leq r_+\sqrt{\frac{ 3 r_+^2+L^2}{L^2-r_+^2}}=r_+ +\mathcal{O}(r_+^3)$. Thus $\frac{r_+}{L} \ll 1$ automatically implies $\frac{a}{L} \ll 1$. Moreover, for $\frac{r_+}{L} \ll 1$ the (real part) of the QNM frequencies of the BH do not differ much from the normal mode frequencies of global AdS that are order $\tilde{\omega} L \sim \mathcal{O}(1)$. Therefore $\frac{r_+}{L} \ll 1$ also implies $r_+ \tilde{\omega} \ll 1$ and $a \tilde{\omega} \ll 1$.
To sum, our analysis will be valid in the regime of parameters \eqref{AnalyticApprox}. We discuss the different regions separately and discuss how to match the solutions obtained next.

\subsection{\label{sec:BH Near region}Near-region equation and regular solution at the horizon}
The near-region is defined by $r-r_+\ll \frac{1}{\tilde{\omega} }$. Introducing the wave function $\Phi=\Delta _r R_{\ell  \tilde{\omega}  m}^{(-2)}$, the radial equation \eqref{MasterRad} reads
\begin{eqnarray}
\Delta _r \Phi ''-\Delta _r' \Phi '+\left(\frac{6 r^2}{L^2}+4 i \,K_r'+\frac{K_r^2-2i \, K_r \Delta _r'}{\Delta _r}-\lambda \right)\Phi =0 \,.
\label{newRadialEq}
\end{eqnarray}
If we further restrict to the regime $\frac{r_+}{L} \ll 1$, the cosmological constant contribution can be 
neglected. Specifically, in the radial equation \eqref{newRadialEq} the following approximations are valid 
\begin{eqnarray} \label{near wave eqDelta}
&& \Delta_r {\bigl |}_{ r\sim r_+} \simeq r^2+a^2-2Mr +\cdots \simeq(r-r_+)(r-r_-) \,, \quad \hbox{with}\quad  r_- \simeq \frac{a^2}{r_+}\,, \nonumber \\
&& \left(\frac{6 r^2}{L^2}+4 i K_r'\right){\bigl |}_{ r\sim r_+}\simeq \frac{6 r_+^2}{L^2}-8 i r_+ \tilde{\omega}  \left(1-\frac{a^2}{L^2}\right)\sim -8 i r_+ \tilde{\omega} +\cdots\,,  \\
&& \frac{K_r^2-2i\,K_r \Delta _r'}{\Delta _r} {\bigl |}_{ r\sim r_+}  \simeq \frac{\Xi ^2\left(r_+^2+a^2\right)^2\left(4\pi  T_H\right)^2\varpi  (\varpi +2 i)}{\Delta _r}+8 i r_+ \tilde{\omega}+\cdots\nonumber \\
&& \hspace{3.42cm}  \simeq  \frac{\left(r_+ -r_-\right)^2 \varpi  (\varpi +2 i)} {(r-r_+)(r-r_-)}+8 i r_+ \tilde{\omega}+\cdots\,, \nonumber 
\end{eqnarray}
where $\Omega_H,T_H$ are the angular velocity and temperature defined in \eqref{OmegaT}, and motivated by the BC \eqref{bcH} we have introduced the superradiant factor,
\begin{eqnarray}
\varpi \equiv \frac{\tilde{\omega} -m \Omega _H}{4\pi  T_H} \simeq  (\tilde{\omega}-m\Omega_{\rm H})\frac{r_+^2+a^2}{r_+-r_-} \,,
\end{eqnarray}
With these near-region approximations the radial equation \eqref{newRadialEq} is then 
\begin{eqnarray}
\Delta _r \Phi^{\prime\prime}(r)-\Delta _r' \Phi^\prime(r)+\left(\frac{\left(r_+ -r_-\right)^2 \varpi  (\varpi +2 i)} {(r-r_+)(r-r_-)}-(\ell -1) (\ell +2) \right)\Phi(r) \simeq 0  \,,
\label{near wave eq}
\end{eqnarray}
where, in the approximation regime \eqref{AnalyticApprox}, we replaced the eigenvalue $\lambda$ by its leading contribution in \eqref{condAngN} (the requirement $a/r_+\ll 1$ is fundamental here since we neglect a contribution proportional to $m s a/r_+$ when compared with $\lambda\sim   (\ell -1) (\ell +2) $).
Introducing a new radial coordinate $z$ and wavefunction $F$ defined as
\begin{eqnarray}
z=\frac{r-r_+}{r-r_-}\, , \qquad 0\leq z \leq 1\,; \qquad \Phi=z^{i \,\varpi} (1-z)^{\ell-1}\,F\,, 
 \label{newCoordWf}
\end{eqnarray}
the near-region radial wave equation takes the form
\begin{eqnarray}
& &  \hspace{-0.5cm} z(1\!-\!z)\partial_z^2 F+ {\biggl [} (-1+i\, 2\varpi)-\left [ 1+2\ell+ i\, 2\varpi \right
]\,z {\biggr ]}
\partial_z F -(\ell+1) \left [ \ell-1+ i \,2\varpi\right ] F=0\,.
 \label{near wave hypergeometric}
\end{eqnarray}
This  is a standard hypergeometric equation \cite{abramowitz}, $z(1\!-\!z)\partial_z^2
F+[c-(a+b+1)z]\partial_z F-ab F=0$, whose most general solution in the neighborhood of $z=0$ is $A_{in}\, z^{1-c} F(a-c+1,b-c+1,2-c,z)+A_{out}\,
F(a,b,c,z)$. Using \eqref{newCoordWf}, one finds that the most general solution of the near-region
equation is therefore
\begin{eqnarray}\label{hypergeometric solution}
\Phi = A_{in}\, z^{2-i\,\varpi}(1-z)^{\ell-1}
F(a-c+1,b-c+1,2-c,z) +A_{out} \,z^{i\,\varpi}(1-z)^{\ell-1}  F(a,b,c,z) \,,
\end{eqnarray}
with
\begin{equation}
a=\ell-1+i\,2\varpi \,,  \qquad b=\ell+1 \,, \qquad c=-1+ i\,2\varpi \,.
\label{hypergeometric parameters}
\end{equation}
The first term represents an ingoing wave at the horizon $z=0$, while the second term in (\ref{hypergeometric solution}) represents an outgoing wave 
which we set to zero, $A_{out}=0$, to guarantee that no perturbations come off the horizon. 

For the matching we need the large $r$ (i.e $z\rightarrow 1$) behavior of the ingoing near-region solution. To get this, we use the $z \rightarrow 1-z$ transformation law for the hypergeometric function
\cite{abramowitz},
\begin{eqnarray}
& \hspace{-0.5cm}  F(a\!-\!c\!+\!1,b\!-\!c\!+\!1,2\!-\!c,z)=
(1\!-\!z)^{c-a-b}
\frac{\Gamma(2-c)\Gamma(a+b-c)}{\Gamma(a-c+1)\Gamma(b-c+1)}
 \,  F(1\!-\!a,1\!-\!b,c\!-\!a\!-\!b\!+\!1,1\!-\!z) & \nonumber \\
&  \hspace{3.5cm}+ \frac{\Gamma(2-c)\Gamma(c-a-b)}{\Gamma(1-a)\Gamma(1-b)}
 \,  F(a\!-\!c\!+\!1,b\!-\!c\!+\!1,-c\!+\!a\!+\!b\!+\!1,1\!-\!z)\,,\nonumber
\end{eqnarray}
and the property $ F(a,b,c,0)=1$. Finally, noting that when $r\rightarrow \infty$ one has $1-z = (r_+-r_-)/r$,
one obtains the large $r$ behavior of the near-region wave solution that is regular at the horizon,
\begin{eqnarray}
 \hspace{-1.2cm} && \Phi \sim A_{in}\, \Gamma (3-2 i \varpi ) \left[
\frac{ \left(r_+-r_-\right)^{-\ell -2} \Gamma (2 \ell +1)}{\Gamma (\ell +1) \Gamma (3+\ell -2 i \varpi )}r^{\ell +2}+\frac{ \left(r_+-r_-\right)^{\ell -1} \Gamma (-2 \ell -1)}{\Gamma (-\ell ) \Gamma (2-\ell -2 i \varpi )}r^{1-\ell } \right].
 \label{largeNear}
\end{eqnarray}
%

\subsection{\label{sec:BH Far region}Far-region wave equation
and global AdS solution}

The far-region is defined by $r-r_+\gg r_+$. In this region the effects induced by the black hole mass and angular momentum can be neglected to leading approximation. The far-region background where the gravitational perturbation propagate is then simply global AdS spacetime.
Our approximations then yield $\Delta _r \simeq r^2\left(1+\frac{r^2}{L^2}\right)$ and, in the regime where the eigenvalue $\lambda$ is given by the leading contribution in \eqref{condAngN}, the radial equation \eqref{MasterRad} boils down to
\begin{equation}\label{far wave eq}
\partial _r\left(\Delta _r\partial _r R_{\ell \tilde{\omega} m}^{(-2)}\right)+\left[\frac{\left(\tilde{\omega}  r^2+i \Delta _r'\right){}^2}{\Delta _r}+2 \left(\frac{9 r^2}{L^2}+1\right)-8 i r \tilde{\omega} -(\ell -1) (\ell +2) \right]R_{\ell \tilde{\omega} m}^{(-2)} \simeq 0\,.
\end{equation}
This is again a hypergeometric equation in whose most general solution is 
\begin{eqnarray}\label{far sol}
&& R_{\ell \tilde{\omega} m}^{(-2)}=\frac{L }{r}\left(\frac{L}{r}+i\right)^{\frac{1}{2} (L \tilde{\omega} -2)} \left(\frac{L}{r}-i\right)^{-\frac{1}{2} (L \tilde{\omega} +2\ell )} {\biggl [} B_0\:  F\left(\ell -1,\ell +1+L \tilde{\omega} ;2 (\ell +1);\frac{2 r}{r+i L}\right)\nonumber\\
&& \hspace{1.4cm}+B_1 (-2i)^{-(2\ell +1)}\,\left(\frac{L}{r}-i\right)^{2 \ell +1} \, F\left(-\ell -2,L \tilde{\omega} -\ell ;-2 \ell ;\frac{2 r}{r+i L}\right){\biggr ]}\,,
\end{eqnarray}
where $B_0, B_1$ are at this point arbitrary amplitudes whose ratio will be constrained by the asymptotic global AdS BC.\footnote{\label{foot}If we were working exactly in global  AdS ($a=0$ and $M=0$ everywhere) this  solution  would  be  exact and extending all the way down to the origin where regularity would require setting $B_1=0$. Then the asymptotically global AdS BC imposed below instead of constraining the ratio $B_1/B_0$  would instead quantize the frequencies that can propagate in global AdS. Indeed we can explicitly check that the expression for $B_1/B_0$ that we get when we do the procedure described below \eqref{TheBCsVa0} vanishes when we insert the global AdS frequencies for scalar, $ \tilde{\omega} L=1+\ell+2p$, or vector modes, $\tilde{\omega} L=2+\ell+2p$ (integer $p$ is the radial overtone).}

Asymptotically the solution decays as 
\begin{eqnarray}\label{far solAsymp}
&& 
\hspace{-0.6cm}R_{\ell \tilde{\omega} m}^{(-2)}|_{r\to  \infty }\simeq e^{i\frac{\pi }{2} (L \tilde{\omega} +\ell )}
\times 
\nonumber\\
&& {\biggl \{} -i \, \frac{L}{r}\,{\biggl [} \,B_0 \,F{\bigl (}\ell-1,\ell+1+L \tilde{\omega};2 (\ell +1);2{\bigr )} + \,2^{-(2\ell +1)} B_1\, F{\bigl (}-\ell -2,L \tilde{\omega} -\ell ;-2 \ell ;2{\bigr )} {\biggr]}
\nonumber\\
&& 
\hspace{0.2cm} +\frac{L^2}{r^2}\,{\biggl [} \frac{1}{2}\frac{B_0}{\ell +1}  {\biggl (}\left[2 \left(L^2 \tilde{\omega} ^2+1\right)+\ell  (L \tilde{\omega} -1)-\ell ^2\right] \, F{\bigl (}\ell ,\ell+2 +L \tilde{\omega};2 \ell +3;2{\bigr )}\nonumber\\
&& \hspace{3.6cm} +\ell  (\ell -1+L \tilde{\omega} ) \, F{\bigl (}\ell +1,\ell +2+L \tilde{\omega} ;2 \ell +3;2{\bigr )}{\biggr )}\nonumber\\
&& \hspace{1.4cm}-2^{-(2\ell +1)} B_1  {\biggl (}L \tilde{\omega}  \, F{\bigl (}-\ell -2,L \tilde{\omega} -\ell ;-2 \ell ;2{\bigr )}\nonumber\\
&&
\hspace{3.8cm}-(\ell +2) \, F{\bigl (}-\ell -1,L \tilde{\omega} -\ell ;-2 \ell ;2{\bigr )}{\biggr )}{\biggr ]}{\biggr \}}+O\left(\frac{L^3}{r^3}\right).
\end{eqnarray}
To have an asymptotically global AdS perturbation we need to match this decay with  \eqref{FrobRinf}, namely, $R_{\tilde{\omega} \ell m}^{(- 2)}{\bigl |}_{r\to \infty} \! \sim \! B_{+}^{(- 2)}\,\frac{L}{r}+ B_{-}^{(-2)}\,\frac{L^2}{r^2}+O\left(\frac{L^3}{r^3}\right)$ and impose the BC \eqref{TheBCs}, $B_-^{(-2)} \! = \!  i \,\beta B_+^{(-2)}$. In the regime we are working one has $a\simeq 0$ and the BC expressions \eqref{TheBCsS}-\eqref{TheBCaux} for $\beta$ simplify considerably reducing to
\begin{eqnarray}
&&  1) \quad \beta=\beta_{\rm \bf s}= -L \tilde{\omega}  \left(1+\frac{\lambda }{\lambda -2\left(L^2 \tilde{\omega} ^2-1\right)}\right), \label{TheBCsSa0}  \\ 
&&  2)  \quad  \beta=\beta_{\rm \bf v}=\frac{\lambda }{2 L \tilde{\omega} }-L \tilde{\omega} \,, \label{TheBCsVa0}
\end{eqnarray}
for scalar and vector modes, respectively. Here, $ \lambda=(\ell -1) (\ell +2) $.
Going through this asymptotic matching we find how the amplitudes  $B_0 \left( B_+^{(-2)},\beta \right)$ and  $B_1 \left( B_+^{(-2)},\beta \right)$ must be related to the BC parameters  $B_+^{(-2)} $ and $\beta$  for the perturbation to be asymptotically global AdS. 

For a later matching with the near-region solution we will need the small $r$ behaviour of the far-region solution
$\Phi=\Delta _r R_{\ell \tilde{\omega} m}^{(-2)}$   with  $R_{\ell \tilde{\omega} m}^{(-2)}$ given by \eqref{far sol}. This is 
\begin{equation}\label{smallFar}
\Phi \sim B_{+}^{(- 2)} \frac{1}{\alpha_D}\left[ i \,e^{-i \, \frac{\pi }{2}(\ell +L \tilde{\omega} )} L^{-\ell }(\ell +1)\,\alpha_N\, r^{\ell +2} 
- e^{i\,\frac{\pi }{2} (\ell -L \tilde{\omega} )}L^{\ell +1}\ell \,\beta_N\, r^{1-\ell }\right],
 \end{equation}
where we  defined
\begin{eqnarray}\label{smallFaraux}
&& \alpha_D=(\ell +1) (\ell +2) (\ell -L \tilde{\omega} )  F{\bigl (} -\ell -1,L \tilde{\omega} -\ell +1,1-2 \ell ,2{\bigl )}  F{\bigl (} \ell -1,L \tilde{\omega} +\ell +1,2 (1+\ell ),2{\bigl )}  \nonumber\\
&& \hspace{1cm} +  \ell  \, F{\bigl (} -\ell -2,L \tilde{\omega} -\ell ,-2 \ell ,2{\bigl )} {\biggl [}(\ell +1) (2 \ell +1)  F{\bigl (} \ell -1,L \tilde{\omega} +\ell +1,2 (\ell +1),2{\bigl )},   \nonumber\\
&&  \hspace{6.2cm}+(\ell -1) (\ell +1+L \tilde{\omega} )  F{\bigl (} \ell ,L \tilde{\omega} +\ell +2,2 \ell +3,2{\bigl )}{\biggl ]},
 \nonumber\\
&& \alpha_N= \ell  (\ell +2-\beta -L \tilde{\omega} ) F{\bigl (}-\ell -2,L \tilde{\omega} -\ell ,-2 \ell ,2{\bigl )}\nonumber\\
&& \hspace{1cm}+(\ell +2) (\ell -L \tilde{\omega} ) F{\bigl (} -\ell -1,L \tilde{\omega} -\ell +1,1-2 \ell ,2{\bigl )},  \nonumber\\
&& \beta_N=(\ell +1) (L \tilde{\omega} +\ell -1+\beta )\,F{\bigl (} \ell -1,L \tilde{\omega} +\ell +1,2 (\ell +1),2{\bigl )}\nonumber\\
&& \hspace{1cm}+(\ell -1) (L \tilde{\omega} +\ell +1) \,F{\bigl (} \ell ,L \tilde{\omega} +\ell +2,2 \ell +3,2{\bigl )}.
 \end{eqnarray}
 
\subsection{\label{sec:BH Match}Matching. QNM and superradiant  frequencies}
 
In the regime $r_+\tilde{\omega} \ll 1$, the near and far regions have an overlaping zone, $r_+\ll r-r_+\ll \frac{1}{\tilde{\omega} }$, where both are simultaneously valid. The requirement that the solutions can be matched across the overlapping zones related
the amplitudes $A_{in},\,B_+^{(-2)} $ and quantizes the frequency $\tilde{\omega}$. In particular, the frequencies that are allowed to propagate in the Kerr-AdS black hole are found matching the large $r$ behavior \eqref{largeNear} of the near-region solution with the small $r$ behaviour \eqref{smallFar} of the far-region solution. This yields two conditions, one following from the matching of the $r^{\ell+2}$ coefficients and the other from the matching of the  $r^{1-\ell}$ coefficients. One of these constraints is used to find the ratio between the near and far region amplitudes $A_0/B_+^{(-2)}$ that is then inserted in the other constraint to finally yield the matching condition that quantizes the frequency spectrum:\footnote{To get this result, as observed in a similar context in \cite{Pani:2012bp}, we should keep in mind that the angular eigenvalue is an integer strictly only in the limit of zero rotation. Therefore the ratio of gamma functions that appears in our computation should be taken as
\begin{equation}
\lim_{\epsilon\to 0}\frac{\Gamma (-\ell-\epsilon )}{\Gamma (-2 \ell -2\epsilon)}= 
\frac{4(2\ell -1)!}{(-1)^{\ell}(\ell -1)!} \,,
\end{equation}
after using the gamma function property $ \Gamma(-n+\epsilon)\sim (-1)^n/(n!\epsilon)$, for $\epsilon \ll 1$ and integer $n$ (assuming at the starting point that $\epsilon=0$  gives a result that differs from the correct one by a factor of $2$).}
\begin{eqnarray}\label{FreqMatching}
&& \hspace{-2.5cm} 
\frac{\ell !}{(2 \ell -1)!}\frac{ i (\ell +1) \Gamma\left(\ell +1\right) \Gamma\left(\ell +3-2 i \varpi \right) } {4 L^{2 \ell +1} \ell ^2 \Gamma \left[2( \ell +1)\right]\Gamma\left(2-\ell -2 i \varpi \right) }\left(r_+-\frac{a}{r_+^2}\right)^{2 \ell +1} \times
\nonumber\\
&& \hspace{0.8cm} {\biggl [} \ell  (\ell +2-L \tilde{\omega} -\beta )\,F {\bigl (}-\ell -2,L \tilde{\omega} -\ell ,-2 \ell ,2{\bigr )} \nonumber\\
&& \hspace{1cm}   +(\ell +2) (\ell -L \tilde{\omega} )\,F {\bigl (}-\ell -1,L \tilde{\omega} -\ell +1,1-2 \ell ,2{\bigr )}   {\biggl ]} \nonumber\\
&&=(\ell +1) (\ell -1+L \tilde{\omega} +\beta )\,F{\bigl (}\ell -1,L \tilde{\omega} +\ell +1,2 (1+\ell ),2{\bigr )} \nonumber\\
&& \hspace{0.6cm} +(\ell -1) (\ell +1+L \tilde{\omega} )\,F{\bigl (}\ell ,L \tilde{\omega} +\ell +2,2 \ell +3,2{\bigr )},
\end{eqnarray}
where the superradiant factor $\varpi$ was introduced in \eqref{superradiance factor}, and the asymptotic BC parameter $\beta$ is given by \eqref{TheBCsSa0} for scalar, and by \eqref{TheBCsVa0} for vector perturbations. Recall that this expression is valid in the approximation regime \eqref{AnalyticApprox}.

This frequency quantization condition simplifies considerably when we choose a particular harmonic $\ell$.
In particular, for the lowest harmonic, $\ell=2$, it reduces to \eqref{FreqMatchingl2}. 
We leave the detailed discussion of the solution of this frequency quantization condition and the  comparison with the associated exact numerical results to subsection \ref{sec:CompareNumAnal}.

\end{appendix}


\end{document}